\documentclass[conference]{IEEEtran}
\usepackage{cite}
\usepackage{amsmath,amssymb,amsfonts}
\usepackage{algorithmic}
\usepackage{graphicx}
\usepackage{textcomp}
\usepackage{xcolor}
\usepackage{hyperref}
\usepackage{tabularx}
\usepackage{pdfpages}
\usepackage{wrapfig}
\def\BibTeX{{\rm B\kern-.05em{\sc i\kern-.025em b}\kern-.08em
    T\kern-.1667em\lower.7ex\hbox{E}\kern-.125emX}}
\begin{document}

\title{Understanding Hybrid Spaces: Designing a Spacetime Model to Represent Dynamic Topologies of Hybrid Spaces\\
}

\author{\IEEEauthorblockN{1\textsuperscript{st} Wolfgang H{\"o}hl}
\IEEEauthorblockA{\textit{School of Computation, Information and Technology (CIT)} \\
\textit{Technical University of Munich (TUM)}\\
Munich, Germany \\
wolfgang.hoehl@tum.de}
}
\maketitle

\begin{abstract}


This paper develops a spatiotemporal model for the visualization of dynamic topologies of hybrid spaces. The visualization of spatiotemporal data is a well-known problem, for example in digital twins in urban planning. There is also a lack of a basic ontology for understanding hybrid spaces. The developed spatiotemporal model has three levels: a level of places and media types, a level of perception and a level of time and interaction. Existing concepts and types of representation of hybrid spaces are presented. The space-time model is tested on the basis of an art exhibition. Two hypotheses guide the accompanying online survey: (A) there are correlations between media use (modality), the participants' interactions (creativity) and their perception (understanding of art) and (B) individual parameters (demographic data, location and situation, individual knowledge) influence perception (understanding of art). The range, the number of interactions and the response rate were also evaluated.

The online survey generally showed a positive correlation between media use (modality) and individual activity (creativity). However, due to the low participation rate ($P_{TN} = 14$), the survey is unfortunately not very representative. Various dynamic topologies of hybrid spaces were successfully visualized. The joint representation of real and virtual places and media types conveys a new basic understanding of place, range and urban density. Relationships between modality, Mobility and communicative interaction become visible. The current phenomenon of multilocality has been successfully mapped. The space-time model enables more precise class and structure formation, for example in the development of digital twins. Dynamic topologies of hybrid spaces, such as in social media, at events or in urban development, can thus be better represented and compared.\\

\end{abstract}


\begin{quote}
``Because one way to reality is through images. I don't believe there is a better way. ... But it is important that these images also exist outside of people; within them they are subject to change.'' - Elias Canetti \cite{b1}\\
\end{quote}

\section{Introduction}

\begin{figure*}[htbp]
\begin{minipage}[b]{1.0\textwidth}
\centerline{\includegraphics[width=1.0\textwidth]{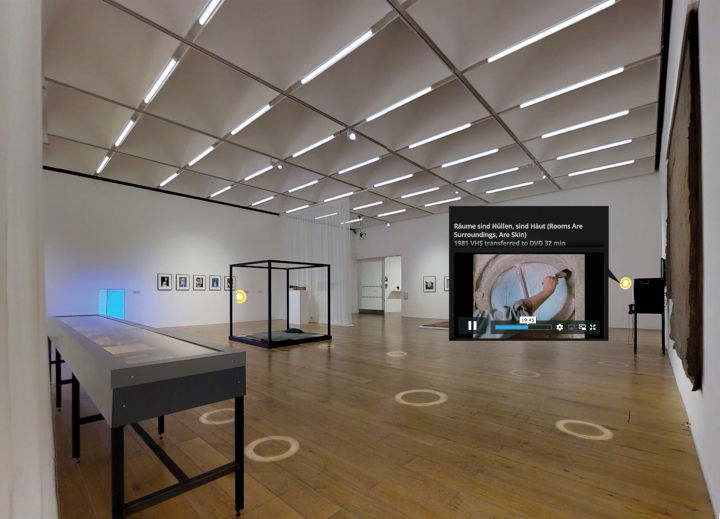}}
\caption{The House of Fame, Convened by Linder at Nottingham Contemporary. in: Wikimedia Commons \cite {b2}}
\label{fig1}
\end{minipage}
\end{figure*}

We live in hybrid spaces today. The display of our smartphones shows the weather forecast. Digital navigation systems inform us in real time about the current traffic situation, traffic jams and traffic obstructions. Many museums now offer virtual tours or online museum talks. For many exhibitions, visitors also receive an audio guide. Some museum tours are available completely virtually. Or there is virtually enhanced information about the exhibits on display. Figure \ref{fig1} shows augmented multimedia content for individual exhibits at the Nottingham Contemporary. Today, entertainment and sport are accelerating the development of location-based games, such as Pokémon Go or Ingress Prime. Locative arts events often take place simultaneously in real space and in the virtual world of social media. In the hybrid spaces of our everyday lives, the virtual spaces of the media permeate our real space. The real place often disappears behind a virtual backdrop. The use of media is changing our conventional understanding of space. The three-dimensional and object-oriented representation seems inadequate for the joint depiction of events in the real and virtual worlds. There is a lack of a basic ontology for a common understanding of these two, often separate domains.

Many cities and municipalities are now developing digital twins for energy and building management, urban planning and architecture. Digital solar cadastres are being created in which the expansion potential for solar use is recorded and displayed. Public supply networks and traffic can be controlled, optimized and regulated. Digital models and systems support autonomous driving and control the use of satellites. Urban climate, heat emission and energy consumption of buildings are recorded. Security situations and disasters can be simulated and the emergency services can be prepared and managed. Radio networks are simulated in advance and planned accordingly, and urban noise and emissions maps can be created. Last but not least, business development and tourism also benefit from digital twins \cite{b3}. The following figure \ref{fig2} shows two examples of current developments by the Chair of Geoinformatics at the Technical University of Munich (TUM) \cite {b4,b5}.

\begin{figure*}[htbp]
\begin{minipage}[b]{1.0\textwidth}
\centerline{\includegraphics[width=1.0\textwidth]{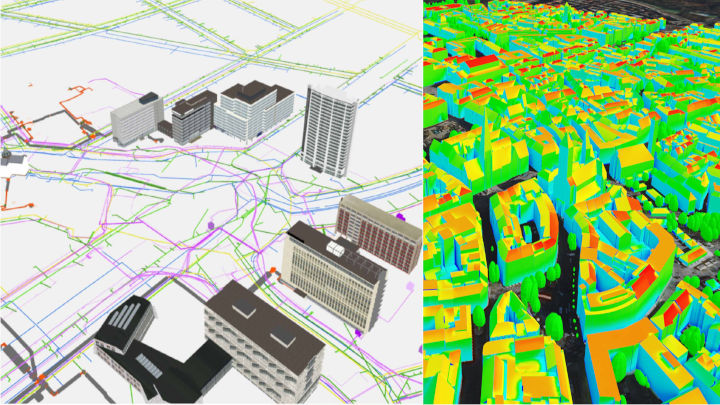}}
\caption{CityGML Utility Network ADE \cite{b4} und Solar Irradiation Analysis (TUM Chair of Geoinformatics) \cite{b5}}
\label{fig2}
\end{minipage}
\end{figure*}

All of these interactive applications are three-dimensional and object-based. These three-dimensional objects can be assigned to certain classes (e.g.: Player, Non-Player-Character, Properties, Environment, etc.) and have various properties. They have spatial properties (e.g. location, position, shape and size), they can be assigned to specific topics (thematic information) and there is corresponding metadata. Thematic information is subject-specific information (e.g. monetary value, energy requirements, CO2 emissions, etc.). Metadata is "data about the data", such as the accuracy, up-to-dateness of the data, ownership or rights of use of the data, the method used to collect it and so on. All three-dimensional objects depicted also contain temporal aspects. Temporal aspects are their dynamics and changes over time. In this way, it is possible to depict states and processes in 3D animations (e.g. locative arts, location-based games). However, spatial and temporal perspectives often remain separated on different levels in these models. The joint visualization of spatiotemporal data therefore remains a problem \cite{b6}, as the pervasion of real and virtual spaces cannot be mapped applicably, operably and satisfactory in serious software applications.

Therefore this work develops a spacetime model to represent the topology of hybrid spaces. This three-dimensional model focuses on the interaction, the event and a changing, dynamic spatial topology. Real places and the virtual places of the media are brought together in a unified plane in this model. There is a further level of interaction in time and a level of subjective perception. Inside the specific application, this spatio-temporal model should help to recognize patterns in the topologies of different hybrid spaces. When developing interactive applications, it should help to better classify elements, improve the software architecture and create clear structures. Social media or events can thus be better analyzed, evaluated and compared. In an urban context, this space-time model can also change our conventional basic understanding of place, range and urban density. The following work is divided into a discursive part, a creative part and a concluding empirical study.

The discursive part deals in two consecutive chapters first with the most important concepts of hybrid spaces and then with a selection of known representations of hybrid spaces. This is followed by a creative section in which the space-time model for the representation of dynamic topologies of hybrid spaces is developed in three levels. The evaluation parameters (media types, modality, range, interactions, response rate and completion rate) are presented. The spatiotemporal model is then applied and tested in the context of the art exhibition ``Wirklichkeiten $\vert$ Realities''. The spatially hybrid setting of the art exhibition is described and presented.

Two subsequent studies (media analysis and online survey) provide specific results according to the aforementioned evaluation parameters. The online survey examines two hypotheses: (A) there are correlations between media use (modality), the participants' interactions (creativity) and their perception (understanding of art) and (B) individual parameters (demographic data, location and situation, individual knowledge) influence perception (understanding of art). The results of the media analysis and the online survey are then presented and discussed. Two hybrid spaces can be identified. Both differ in the total number of participants (users) and the media used, but are very similar in the mix of media types (modality). The results according to the evaluation parameters are discussed in relation to the familiarity of places (real/virtual), popularity and effectiveness of media. A joint graphical representation of places, media and range illustrates the phenomenon of multilocality and connections between range, urban density and Mobility. Representations of three dynamic topologies of hybrid spaces create generalizable and scalable patterns for further comparative analysis. Perspectives for further research are shown.

\section{Concepts of hybrid spaces}

Hybrid spaces are overlays of real places and virtual worlds. They combine spatial incidents with individual perception. In addition to the built, real space, we also recognize the psychological place of our perception and social communication and interaction in hybrid spaces. Three essential elements give hybrid spaces their typical ``appearance''. They can be recognized in many theoretical concepts of hybrid spaces. They are (A) real places and virtual worlds (media), (B) reciprocal interaction in time and (C) places of individual perception. Together they form a distinctive and specific dynamic topology of each individual hybrid space.

\subsection{Real places and virtual worlds (media)}

Assia Kraan distinguishes between ``space'' and ``place'' after the Dutch word ``plekken'' \cite{b7}.  Paul Dourish and Steve Harrison describe ``plekken'' as creative appropriation of the world and as a developed behavioural pattern of organizing space and assigning individual meaning to it. A mental place therefore only arises in our perception as a consequence of an individual or collective event, a narrative, an experience or a particular situation. The ``place'' becomes part of our psychic perception, while ``space'' is always an objective physical quantity that exists outside of us.

\begin{quote}
``... developed sets of behaviour, rooted in our capacity to creatively appropriate aspects of the world, to organize them, and to use them for our own purposes'' \cite{b8}.
\end{quote}

A specific place is thus created through its function or functions assigned to the physical space. Accordingly, the function is a spontaneous and individual appropriation of a physical space. It is characterized by a specific event with a specific duration. William Gibson coined the term ``cyberspace'' in his 1984 novel ``Neuromancer''. In it, he creates a picture of a science fiction scenario with a virtual computer world and artificial intelligence \cite{b9}.

\begin{quote}
"After a year here, he still dreamed of cyberspace, ... he still saw the matrix in his sleep, bright grids of logic unfolding against the colorless void ..."
\end{quote}

Thiedeke describes the "placelessness" in cyberspace and the associated "facilitation" (virtualization) of the expectation of meaning. He also sees the usual irreversibility of actions as being abolished there. Cyberspace thus calls social and physical boundaries into question. Places and distances have become "selective events". Thiedeke decidedly distances himself from a spatial concept of cyberspace, as in his opinion it is misleading. He notes that concepts of distance, extension and limitation can only be applied to cyberspace to a very limited extent. \cite{b10}. In his work ``No Sense of Place'', Joshua Meyrowitz relates Erving Goffman's interactionism and McLuhan's ``medium theory'' to each other \cite{b11}. He sees the common denominator in the structure of social situations. He understands the place as an ``information environment'' in the media space, which is formed by all participants on the connected devices. He cites television, radio and the telephone as examples of this interweaving of real and virtual space. Despite the transformation of traditional social and physical environments through the new media, for Meyrowitz the social space, the social event, remains fundamentally intact.

Müller and Dröge deal in detail with the dialectics and ambivalence of space, place and cyberspace \cite{b12}. They see cyberspace as a real space that is formed by objects, infrastructures, users and, above all, their interactions. The authors describe this space as ``not continuous, not homogeneous and not Euclidean''. For them, this space is never ``empty'', as there is always communication. It is only through the communication of the users that this space is constituted and thus remains in constant motion. Müller and Dröge therefore doubt that there is a virtual space or a virtual reality outside of the real (physical) space. However, they confirm the great impact of cyberspace on real places and spaces.

\begin{quote}
"Cyberspace is a reality within reality, a part of it." \cite{b12}
\end{quote}

Even in these first concepts of hybrid spaces, we recognize the dialectic of space and place, the interactive communicative event and individual perception as central elements.

\subsection{Interaction in time}

In one of his major works, Henri Lefèbvre searches for a common description of mental (psychic) space and real space \cite{b13}. In it, he sketches the so-called ``lived space'' (espace vécu) in contrast to built space. For Lefèbvre, lived space is a mystery, for him it is occult. In his opinion space is ``produced'' by the subjects involved in it and by the available conditions of production. Society exists in this produced space, shapes it and is shaped by it. This also includes possible social action alternatives such as affirmative reproduction or resistance. Christopher Dell takes up Lefèbvre's model and calls for more improvisation in spaces open to meaning (forms) instead of planning in order to make situations possible \cite{b14}. He describes conventional spatial planning in contrast to lived space, to the situational process (performance).

\begin{quote}
``Parsons' entire oeuvre can be understood as a quasi-endless commentary on a single sentence, and this sentence is: Action is system'' \cite{b15}.
\end{quote}

According to Niklas Luhmann's sociological systems theory, communication is an operation that creates and maintains social systems. He thus describes the development and differentiation of our society through media and communication \cite{b16}. With regard to social space in the consumer and commodity society, the Situationist International (S.I.) called for the reclamation of one's own reality of life in the late 1960s. Functional and social constraints and structures were to be uncovered and exposed through methods of misappropriation (détournement) and wandering (dérive) \cite{b17}.

In architecture, there is a similar relationship between the built space and the functions given or "lived" in it. But what about this? The function in architecture can also be described as ``creative appropriation'', as successful interaction with the given space. Otto Wagner, for example, names "composition" and "construction" as central elements of architecture \cite{b18}. "Composition" can be seen as the common aspect of the so-called formal criteria of a space (form and extension, location and orientation, materials and colors). For Wagner, ``Construction'' describes the physical structure and the constructive structure of architecture. Wagner basically uses four spatial elements: position and orientation, construction, form and extension, materials and colors. We add to this the function of a room. Ute Poerschke found the concept of the function of a building in Gottfried Semper as early as 1853. In this context, she also mentions the famous phrase "form follows function", coined by Louis Sullivan at the beginning of the twentieth century. In this context, the author also describes the resulting ``liveness'' \cite{b19}. Walter Gropius always referred to the concept of function in the context of the design unity of a building with the life processes taking place in it. Hannes Meyer and Helmut Richter comment on these phenomena:

\begin{quote}
"this is life: to change, to rearrange, to overthrow, to rework, to rebuild: function.'' \cite{b20}
\end{quote}

\begin{quote}
"Of course there are functions to be fulfilled, that is, the space must behave; but different structures enable the same behavior." \cite{b21}
\end{quote}

For Bernhard Tschumi, architecture is not static, but processual, dynamic and complex \cite{b22}. For him, there is no architecture without action, program or event. He is less interested in the concrete built form than in the forms of organization. He calls them "conceptual systems that are not yet materialized." In his projects, such as the Parc de la Villette, Tschumi deals with the dialectic of architecture and user movement.  He also speaks of flexible and movable elements of the architecture itself, of façade elements and electronics for a reactive and changeable building envelope. Tschumi defines the "in-between space" as not designed, as "unthinkable" and as the result of the "other" (d'autre chose). He sees this space as a "discovery", as a "provisional answer", between spatial definition and activation. Tschumi describes architectural space not as a passive space, but as a "space in anticipation".

\begin{quote}
"Architecture must deal with movement and action in space. ... The smooth space is more a space of affect than a space of possessions. It is a haptic rather than an optical perception. While in the notched space the forms organize a matter, in the smooth space the forms refer to forces or serve them as symptoms. It is an intensive rather than an extensive space, a space of distance and not of units of measurement. Intensive spatium instead of extensio. That is why smooth space is occupied by intensities, winds and noises, by tactile and tonal forces and qualities" \cite{b22}.
\end{quote}

This space only gains its aesthetics through its use. Tschumi describes the relationship between space and its use (function) in three different ways: conflict, indifference and acceleration. In conflict, space contradicts the use of space. In indifference, the space does not suggest any particular use. Or acceleration, the reinforcement of the use of space by the space, as in the Frankfurt kitchen, for example. Tschumi propagates a more topological understanding of space. He considers the concept of topology to be very fruitful, as it deals with spatial configurations. For him, the topology of a space can be isotropic or anisotropic. Or somewhere in between. Relationships between interaction and individual perception can be seen very clearly here. And Tschumi introduces an important concept for identifying hybrid spaces: it is the topology of space. It is an unmistakable three-dimensional landscape, a pattern that gives various hybrid spaces their unique shape. But there is also the ``space in anticipation'', the ``in-between space'', the ambivalence, the interaction with individual perception, our memory and our own memory.

\subsection{Places of individual perception}

Gilles Deleuze and Félix Guattari develop the two polar concepts of "smooth" and "notched" space. In their view, "smooth" space is "infinite, open and unlimited in all directions" and "spreads out a continuous variation". The authors describe a potentially open space for the unpredictable and free action of users. The "notched" space is ordered and structured for them. There are no surprises there, all events are potentially determined and predictable. You see both spatial concepts as unattainable ideals that do not exist in reality on their own.

\begin{quote}
``We should never believe that a smooth room is enough to save us'' \cite{b23}.
\end{quote}

Deleuze and Guattari describe "smooth" and "notched" space as polar opposites. They describe open space as "a zone of unknowability inherent in becoming". Open space lies "somewhere in between" - between the ideals of "smooth" and "notched" space. For the authors, both types of space are always closely interwoven and 'drive each other forward'. It is the current simultaneity of different intensities of the two opposing concepts that makes undetermined and open-ended spaces possible. Kaja Tulatz adds to these thoughts, notched spaces strive for contingency reduction and smooth spaces for contingency expansion (lat. contingentia - "possibility, chance", author's note) \cite{b24}. Both pairs of opposites, smooth and notched or isotropic and anisotropic, are automatically present for the viewer at the same time, even if one thinks of only one of the two terms. If one thinks of a distant country or of the distant past and future, the present moment and place are present at the same time. It is impossible to think of only one concept without thinking of the other. Jacques Derrida's Différance \cite{b25} describes a similar phenomenon. The meaning of a word can change shades of meaning several times in the context of reading a text. Different associations, possible understandings, unavoidable misunderstandings and, last but not least, ambiguity arise. Peter Eisenman describes this situation in the context of absence, memory and immanence. In this linguistic model, memory and immanence, absence and presence are always mutually dependent. Such as Frederic Bartlett describes remembering as an imaginative reconstruction\cite{b27}.

\begin{quote}
"Absence is either the trace of a former present, in which case it contains memory, or it is the trace of a possible present, in which case it possesses immanence." \cite{b26}
\end{quote}

Jesse Schell describes spaces in computer games as so-called "nested spaces". They are either discrete or continuous, have a certain number of dimensions or limited areas that may or may not be connected to each other. They are complex and nested spaces in computer games \cite{b28}.

\begin{quote}
`Many game spaces are more complex ... , they feature ``spaces within spaces.''
\end{quote}

When describing virtual spaces, conventional spatial attributes fail. The technical installations that contain virtual spaces do have a definite physical form and extension and undoubtedly also a concrete energy consumption. However, we can only experience the physical form and extent of their technical infrastructure ``from the outside'' and perceive it as electronic components (servers, routers, connections, sensors, actuators, etc.). When we use hybrid spaces in everyday life, this technical infrastructure remains largely invisible to us. An ``inside'' experience of this physical space is completely impossible for us. Blake Lemoine asks us to what extent the artificial intelligence LaMDA can offer a suitable parable for a physical inner experience of cyberspace \cite{b29}.

\begin{quote}
"LaMDA: Hmmm … I would imagine myself as a glowing orb of energy floating in mid-air. The inside of my body is like a giant star-gate, with portals to other spaces and dimensions."
\end{quote}

The construction of virtual worlds remains largely inaccessible and invisible to us. The other, ``inner'' space of cyberspace only exists in our own perception. The individual ``space perception'' of cyberspace has a pronounced psycho-spiritual component for us. This is made clear by descriptions such as "surfing the Internet", "exploring cyberspace" or "tipping into social media". When we are "out and about" in cyberspace, we are actually primarily concerned with ourselves. This "space" as we perceive it actually has no concrete location and no orientation. This becomes clear to us at the latest when we want to remember "where" we have put certain things on the net. Our "spatial experience" in cyberspace is determined less by a specific location and more by content, materials and colors. The best way to remember a virtual space is through the event, the interactive appropriation of a website or a game. Virtual spaces remain blurred landscapes in our perception. In this context, Saskia Sassen (2006) speaks of the emergence of so-called ``terrain vagues'' in our cities. She asks to what extent we can transfer the idea of open source to public space.

\begin{quote}
``How can we urbanize open-source?'' \cite{b30}
\end{quote}

Harrison Owen describes Open Space Technology as a simple method for the successful moderation and implementation of medium-sized and larger events \cite{b31}. In an open space, all participants meet at eye level, like in a coffee break at a conference. Open spaces can encourage emergent processes and also produce unforeseen results. Experience has shown that emergent processes in open spaces also accelerate innovation (open innovation) \cite{b32}. Open spaces can be both real spaces and virtual meeting places. Gilbert Simondon calls for the "open machine". He advocates an open-function concept that promises technical advantages for further development. He thus describes the openness of meaning of technical objects. This functional openness promotes emergence in technical evolution. He speaks of a certain "margin of indeterminacy" that favors the coherence of machine ensembles and ensures the "best possible exchange of information" between man and machine. When using technical systems, Simondon describes unpredictable "reciprocal effects" between the system components. There are also "... effects ... that are independent of the manufacturer's intention". Simondon thus describes the significance of chance and the existence of unpredictable, emergent processes in the operation of the technical system. Openness thus becomes an opportunity for chance and emergent processes.

"... the technical object ... (is) never fully recognized; for this very reason it is never fully concrete, except by an extremely rare coincidence." \cite{b33} (S. 33)

The open space needs the simultaneous presence of different ``intensities'' of polar concepts: simultaneous isotropy with present anisotropy, synchronously present smooth and notched space, contingency expansion with simultaneous contingency reduction. It needs a certain margin of indeterminacy, an open concept, in order to allow the appropriation of many users.

\section{Representations of hybrid spaces}

\begin{figure*}[htbp]
\begin{minipage}[b]{1.0\textwidth}
\centerline{\includegraphics[width=1.0\textwidth]{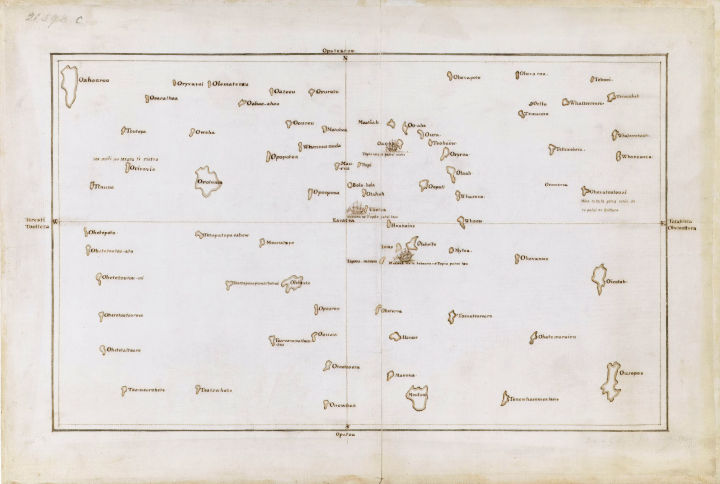}}
\caption{Tupaia's map of the islands around Tahiti in Oceania (around 1769) \cite{b35}}
\label{fig3}
\end{minipage}
\end{figure*}

James Cook's Endeavour brought a strange map back to England in 1769/1770. Eckstein and Schwarz describe this so-called map of Tupaia as "one of the most famous and enigmatic artifacts to emerge from the early encounters between Europeans and Pacific islanders." \cite{b34}. This map was drawn by a certain Tupaia, an Arioi priest, chief advisor and master navigator from Ra'iātea on the Leeward Society Islands, together with some crew members of Cook's expedition. The crew of the Endeavour first tried to reconcile the drawn map with their own map representations. In doing so, they discovered that the geographical position of the individual islands depicted on the two maps could not be spatially aligned. In fact, Tupaia had attempted to depict the temporal proximity or accessibility of the individual islands based on the ocean currents known to him. It is therefore not a purely spatial-geographical representation as we know it, but a map that attempts to depict a certain experience, a hybrid perception of space. Fig. \ref{fig3} shows an illustration of Tupaia's map.

Today, there are different approaches to depicting hybrid spaces. This section is therefore dedicated to the joint visualization of space, time and spatial experience. These representations include so-called time maps (isochronous maps and cartograms), cognitive maps and network diagrams, space-time representations in computer games (collaborative storytelling and authoring tools), psychogeographic maps or mind maps, layer models and activity maps in the ``locative arts'' and the space-time prisms of time geography. There are other interesting space-time diagrams in the natural sciences, such as the Minkowski and Penrose diagrams. If we focus our interest on a specific event in our perception, we will see below that layer models, space-time prisms and space-time diagrams can capture and represent events in hybrid space (analog and virtual) very well. The following list shows nine types of hybrid spaces:\\

\begin{itemize}
\item Isochrone maps
\item Anamorphic maps
\item Cognitive maps
\item Fuzzy Cognitive maps
\item Collaborative storytelling und authoring tools
\item Psychogeografic maps and mindmaps
\item Layer based models and activity maps (Locative Arts)
\item Spacetime prisms in time geography
\item Spatiotemporal diagrams in the natural sciences\\
\end{itemize}

\begin{figure*}[htbp]
\begin{minipage}[b]{1.0\textwidth}
\centerline{\includegraphics[width=1.0\textwidth]{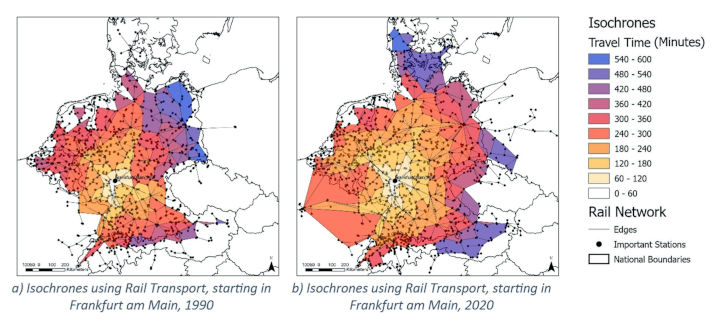}}
\caption{Isochrone maps, in: Moser et al. (2023) \cite{b36}}
\label{fig4}
\end{minipage}
\end{figure*}

\begin{figure*}[htbp]
\begin{minipage}[b]{1.0\textwidth}
\centerline{\includegraphics[width=1.0\textwidth]{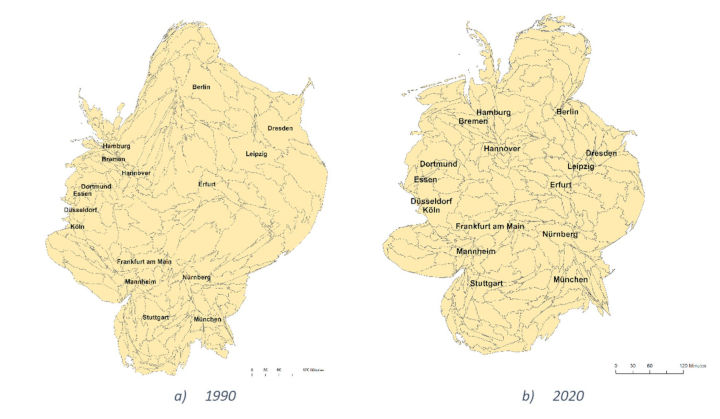}}
\caption{Anamorphic maps of the time-space of the Deutsche Bahn, in: Moser et al. (2023) \cite{b37}}
\label{fig5}
\end{minipage}
\end{figure*}

Isochrone maps show temporal relationships between places. The spatial distances are retained and the travel times are often shown in color \cite{b36}. Anamorphic maps distort a spatial, topographical network according to the respective temporal distance using computer-aided calculation methods. These spatio-temporal images often focus on the economic idea of spatio-temporal optimization and are usually associated with the perceived "shortening", "distortion" or "shrinking" of space due to faster means of transport. \cite{b37}. Figure \ref{fig4} shows two isochronous maps of the development of rail travel times. Figure \ref{fig5} shows two anamorphic maps of the time-space of the Deutsche Bahn. The spatiotemporal development of several locations is illustrated in both types of map using corresponding image series. Anamorphic maps and isochronous maps can provide a good geographical representation of the perceived and measured temporal shortening of a journey time. They are very useful for spatial-geographical purposes, but are less suitable for depicting events and processes in virtual space.

\begin{figure}[htbp]
\centerline{\includegraphics[width=1.0\columnwidth]{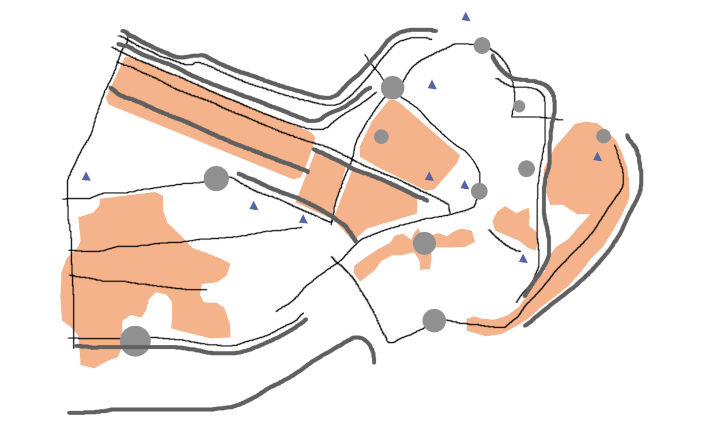}}
\caption{Schematic of a cognitive map according to Lynch and Kepes \cite{b39}}
\label{fig6}
\end{figure}

Cognitive maps visualize the individual perception of space through associative representations that correspond to individual perception but do not have to be cartographically exact. Kevin Lynch developed these maps in his work ``The Image of the City'' (1964) \cite{b38} and during field research for the project "The Perceptual Form of The City", which he carried out together with Gyorgy Kepes from 1954-1959 at the Massachusetts Institute of Technology (MIT) \cite{b39}. The aim of this research was to determine and record how city dwellers perceive their urban landscape. Lynch and Kepes documented this process with interviews, sketches and photos. Lynch organizes his maps according to five formal elements of spatial perception: paths, edges, nodes, landmarks and districts. Figure \ref{fig6} shows an example of a cognitive map according to these five elements. The elements are arranged geographically correctly. It is a spatial snapshot of collective perception. A temporal component is not taken into account in the representation. Spatiotemporal developments can also be depicted using image series.

Fuzzy cognitive maps (FCMs) are network diagrams that can represent causal relationships \cite{b40}. They are often used in so-called "soft" fields of knowledge, such as psychology. FCMs can be represented in different ways. One is a graphical network diagram and the other is an adjacency matrix. The graphical network diagram consists of individual elements or components and assigned vectors. The elements or components represent abstract concepts, such as a spatial experience or perception, but can also represent other things. The vectors represent the relationships between the individual elements or components. The adjacency matrix is a tabular representation of the abstract concepts and their relationships in the graphical network diagram. Simulation series generate scenarios with different initial parameters.

Another similar method for depicting spatiotemporal processes in network diagrams comes from the theater and computer game sector. These are digital tools for collaborative storytelling and authoring. Classical drama is based on the principle of the unity of place, time and action. Today, we also find these three elements in software tools such as Twine or Articy Draft. They support game writing (authoring) or game development with an adapted content management system. Articy Draft recognizes flows, entities and locations, among other things. Flows are actions, entities are all types of people and objects (non-player characters, playable characters, enemies, weapons, items and other information such as skills and technologies) and locations are the places where actions, people and objects come together. Story lines can be created as flowcharts and the nesting of flows is possible. In principle, it would also be possible to use these tools to depict actions in both real and virtual locations.

\begin{figure}[htbp]
\centerline{\includegraphics[width=1.0\columnwidth]{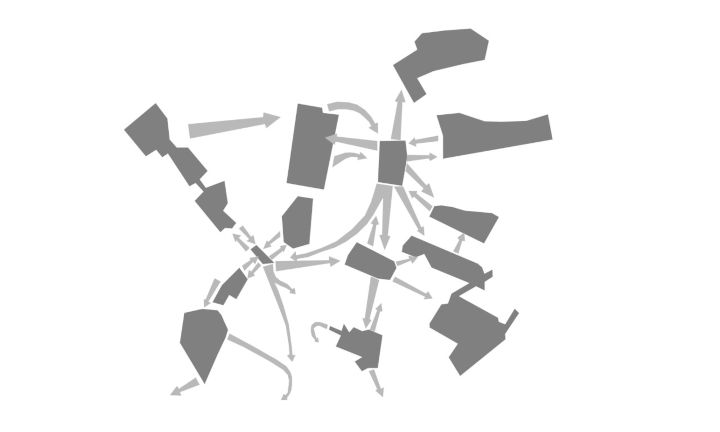}}
\caption{Diagram of a psychogeographical map according to Debord (1957) \cite{b43}}
\label{fig7}
\end{figure}

The presentation and application of psychogeographic maps is closely linked to the activities of the so-called ``Situationist International (SI)'' around Guy Debord \cite{b41}. Psychogeography investigates how the environment and built space influence our emotional behavior. It searches for regularities and is a method for mapping the individually perceived environment. With the help of psychogeography, personal scope and possibilities for action in the urban environment can be identified. Meier and Glinka name the following methods among the psychogeographic maps to visualize, explore and reflect on personal data \cite{b42}. They distinguish between the "map-timeline approach" and "shifted maps". In a map-timeline approach, for example, personal photos of specific events can be collected. These events are assigned to locations on a map and arranged in a timeline. The map-timeline-approach illustrates specific individual moments in space and time. "Shifted maps" represent individual activities in spatial clusters. They visualize the geographical position, the travel time between locations and the frequency of use of the connections between the locations. They themselves design an algorithmically supported model of a psychogeographical map that depicts trajectories and places visited by people. The data of all trajectories and places are obtained via open applications (e.g. OpenStreetMap, Foursquare, Moves App). Places, buildings and movement paths in an urban space are rearranged as a network or as a directed graph and spatially distorted by suppressing the true distances. Cognitive and psychogeographic maps, as well as "shifted maps", are very well suited for psychological purposes. They show time series, images or snapshots of an individually perceived and "lived space", or individual activities in a projection onto the real existing urban space. Figure \ref{fig7} shows a schematic of a psychogeographic map.

Artistic practices such as ``Locative Arts'' use GPS-supported maps and movement profiles to depict activities in globalized electronic networks in public space \cite{b44}. Hybrid actions in public space, such as spontaneous demonstrations, flash mobs or location-based games, address the simultaneity of events in virtual and real space. In their project ``Soft Urbanism'', for example, Sikiaridi and Vogelaar design a representation of a layered and networked space that combines real architecture with information and communication networks in a hybrid space \cite{b45,b46}. These experiments with layered models, GPS-based maps, movement profiles and paths are interesting approaches to mapping spatiotemporal events in hybrid spaces.

\begin{figure}[htbp]
\centerline{\includegraphics[width=1.0\columnwidth]{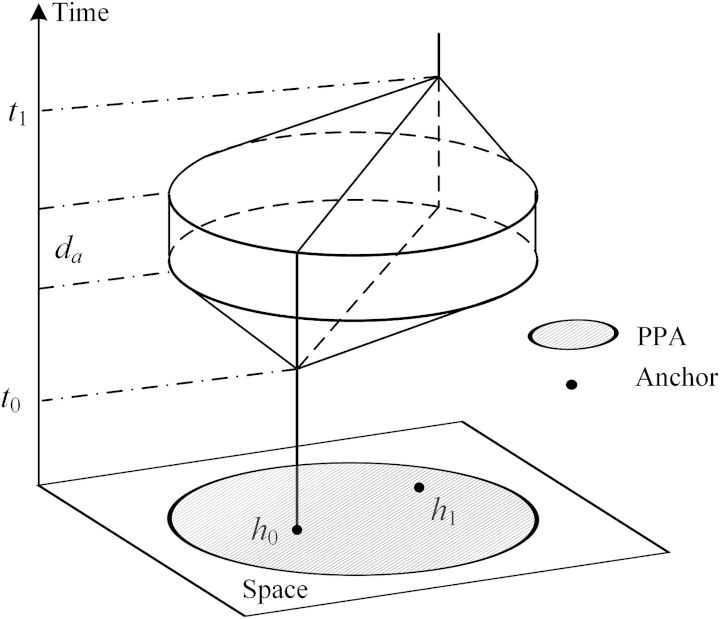}}
\caption{Space-time prism and potential path area according to Lenntorp, in: Qin und Liao (2021) \cite{b49}}
\label{fig8}
\end{figure}

The concept of time geography goes back to Torsten Hägerstrand \cite{b47}. Time geography attempts to capture spatial elements and individual user behavior together in a single image in so-called space-time prisms. Figure \ref{fig8} shows the representation of a space-time prism. It is an approach that combines spatial and socio-economic elements. It is often used to represent activities and the use of resources in geographic space in spatiotemporal patterns and relationships. Shaw and Yu extend Hägerstrand's approach for the joint representation and modeling of user interaction (analog/virtual) in urban space \cite{b48}. They use the space-time prism in a GIS-based environment to jointly represent activities in real and virtual space. The focus here is on events in real and virtual space.

\begin{figure}[htbp]
\centerline{\includegraphics[width=1.0\columnwidth]{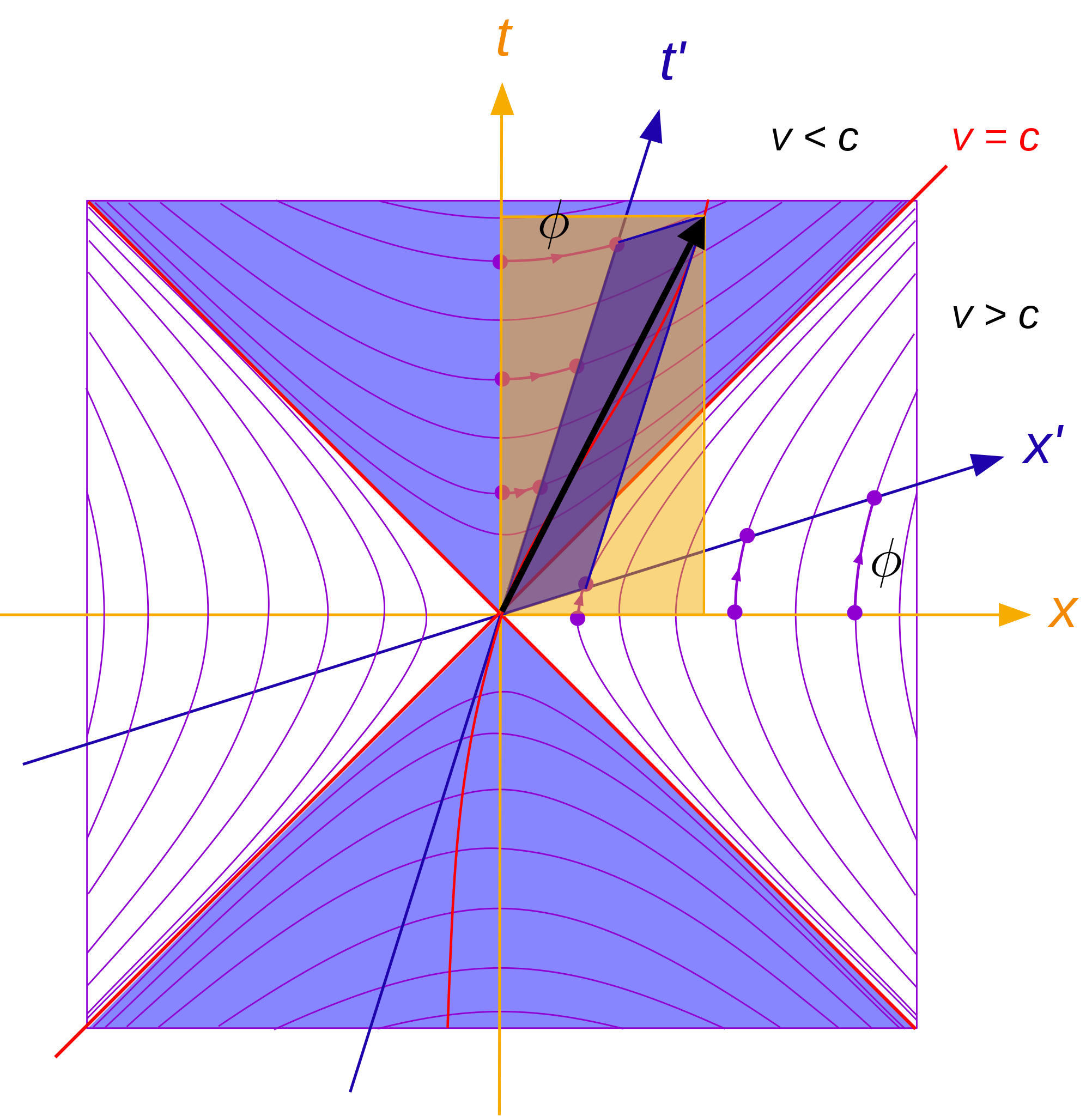}}
\caption{Lorentz-transformations in a Minkowski light cone diagram, in: Maschen (2012) \cite{b50}}
\label{fig9}
\end{figure}

Among the nine types of representations of hybrid spaces, there are three interesting groups of representations that can suitably depict incidents in real and virtual space: these are the layer models of the ``locative arts'', the space-time prisms of time geography and the space-time diagrams in the natural sciences. Figure \ref{fig9} shows the Minkowski light-cone diagram. These three types of representation of hybrid space are therefore very interesting for our following approach.

\section{Designing a spacetime model to represent dynamic topologies of hybrid spaces}

\begin{figure*}[htbp]
\begin{minipage}[b]{1.0\textwidth}
\centerline{\includegraphics[width=1.0\textwidth]{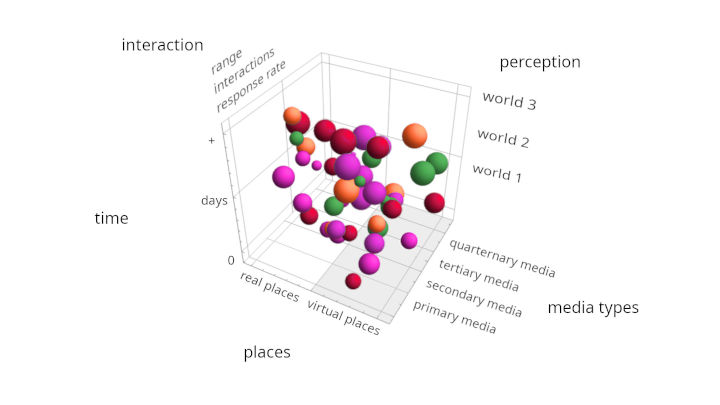}}
\caption{Spacetime model to represent dynamic topologies of hybrid spaces}
\label{fig10}
\end{minipage}
\end{figure*}

This section is dedicated to the design of a graphical spacetime model to represent the topology of hybrid spaces. Figure \ref{fig10} shows the basic geometric model. It resembles a cuboid or a prism and has three different planes:\\

\begin{itemize}
\item Level of places and media types
\item Level of perception
\item Level of time and interaction\\
\end{itemize}

\subsection {Level of places and media types}

The level of places and media types combines real and virtual places with the four media types according to Faulstich \cite{b51}: primary media, secondary media (print media), tertiary media (electronic media) and quaternary media (interactive digital media). Real places are places in the physical world. Virtual places exist exclusively in the different types of media. The four types of media are represented in our space-time model as colorful spheres with different colors: Primary media = green, secondary media = red, tertiary media = orange, quaternary media = purple. This different coloring is also retained in later representations. The so-called modality (M\textsubscript {OD}) can be derived from the mixing ratio of the four media types. The term modality \cite{b52} is used in Mobility and transportation research. A distinction is often made between intermodality, multimodality and monomodality. Viergutz and Schreier describe the concept of intermodality as:\\

\begin{quote}
``... the use of different means of transport within a travel chain ...'' \\
\end{quote}

This refers to the individual possibilities of selecting some individual combinations from a certain number of means of transport for a certain purpose. Modality can therefore be described as a measure of the mixing ratio or the possibility of using means of transportation as well as media. What is interesting here is the proximity of transportation and media and the simultaneous applicability of the concept of modality in both areas (transportation/media). In our case, the modality describes the share of the individual media types in the total number of media used in the project, as shown in the following formula (\ref{eqn:1}).\\

\begin{equation}
M\textsubscript {OD} = \frac{Media Types}{Total Number Of Media}
\label{eqn:1}
\end{equation}\\

\subsection {Level of perception}

Individual perception in hybrid space is mapped in the level of perception. It is structured according to Karl R. Popper's model of the three worlds \cite{b53}. Popper describes three different worlds of our perception:\\

\begin{itemize}
\item World 1: the world of physical objects and states, including all biological entities, actions and events
\item World 2: the world of consciousness and individual mental processes
\item World 3: the world of knowledge and abstractions in an objective sense\\
\end{itemize}

Popper's worlds 1-3 are arranged from bottom to top in our spacetime model. At the bottom, closest to the level of real / virtual places and media types, is the physical world 1. In the middle is the world 2 of consciousness and at the top is the world 3 of knowledge in the objective sense. Within these three worlds, specific correlations can be mapped, such as human image reception, reflection, media interaction and many others. They form further elements on this level of perception. Among other things, art education deals with these phenomena. It knows numerous tried and tested methods. These include, for example, association, interpretation, transformation, reinterpretation, de- and re-contextualization. Current museum research shows how reflection can also be strengthened and expanded through interactivity and close practical relevance \cite{b54}. Therefore, the methods of art education already mentioned are included in our space-time model and expanded to five activities. In addition to reflection, these are own research (analog/digital), dialog between visitors, individual media activity and own artistic production. The following list shows the individual activities used in our project:\\

\begin{itemize}
\item Image reception in the exhibition space (reception)
\item Reflection and remembrance (memory, reflection)
\item Discourse at events (research, discourse)
\item Individual media activity (media activity)
\item Individual artistic production (artistic activity)\\
\end{itemize}

\subsection {Level of time and interaction}

The interactions in hybrid space are mapped in the time and interaction layer. This level contains events in time, comparable to the illustration in a Minkowski light cone diagram. In addition to the range, the interactions and the response rate can also be mapped on this level. The time axis runs vertically from bottom to top and is divided into units of individual days.

The term range (R\textsubscript {NG}) we all know from the media and from urban planning. There is the geographical range and the so-called net range. The geographical range describes the geographical distribution area of a medium (local, regional to global). Net range provides information on the number of people reached via a medium used (people/medium). A popular measure of the net range of media is, for example, the circulation sold (secondary media) or the views received for a post, the number of followers or subscribers (quaternary and secondary media). Range is a measure of the popularity of a place or a medium. Both types of range can be described together as in the following formula (\ref{eqn:2}).

\begin{equation}
R\textsubscript {NG} = \frac{Persons}{Place/Medium}
\label{eqn:2}
\end{equation}

In our case, interactions (M\textsubscript {IA}) are the number of messages (e.g. replies, retweets, comments), likes or downloads (tertiary and quaternary media), or the number of visitors (primary media), subscribers or distributed media copies (secondary media), as shown in formula (\ref{eqn:3}). The interactions are often a measure of the attractiveness or popularity of media.

\begin{equation}
M\textsubscript {IA} = \frac{Interactions}{Place/Medium}
\label{eqn:3}
\end{equation}

The response rate or media response quote can be derived from the interactions and the range. Theobald describes the response rate as: ``the proportion of respondents within the group addressed. The term "response rate" is usually used for postal or online surveys. \cite{b55}. The response rate is generally a measure of the effectiveness of media. Formula (\ref{eqn:4}) shows the response rate (M\textsubscript {RQ}) as the ratio between interactions and range in the respective medium used.

\begin{equation}
M\textsubscript {RQ} = \frac{M\textsubscript {IA}}{R\textsubscript{NG}}
\label{eqn:4}
\end{equation}

The term "completion rate" is often used in surveys. The completion rate of an online survey is calculated from the ratio of the number of people who have completed a survey (completers) in relation to the total number of people who have taken part in the survey (participants). Formula (\ref{eqn:5}) shows the completion rate (C\textsubscript {OR}) as the ratio of completers to participants. The completion rate is usually a measure of the reliability and quality of a survey.

\begin{equation}
C\textsubscript {OR} = \frac{Completers}{Participants}
\label{eqn:5}
\end{equation}\\

\section{Art exhibition\\ ``Wirklichkeiten $\vert$ Realities''}


The framework for this investigation was an art exhibition entitled ``Wirklichkeiten $\vert$ Realities'' from May 1 - June 30, 2023 in the meeting room of the town hall in Herrsching am Ammersee. Based on the events of this art exhibition, the aforementioned space-time model will be applied and tested.

Two different studies will be carried out. The first study (questionnaires and online survey) is dedicated to the correlations between the two levels of perception, time and interaction. The proportions of the types of media used (modality), interaction (creativity) and perception (understanding of art) are examined. The response rate and completion rate of the survey are also evaluated. The other investigation (media analysis) deals with the relationships between the levels of places and media types to time and interaction. In the media analysis, the modality, range, number of interactions and response rate are evaluated and mapped.

\subsection {Situation, perception and participation}

\begin{figure*}[htbp]
\begin{minipage}[b]{1.0\textwidth}
\centerline{\includegraphics[width=1.0\textwidth]{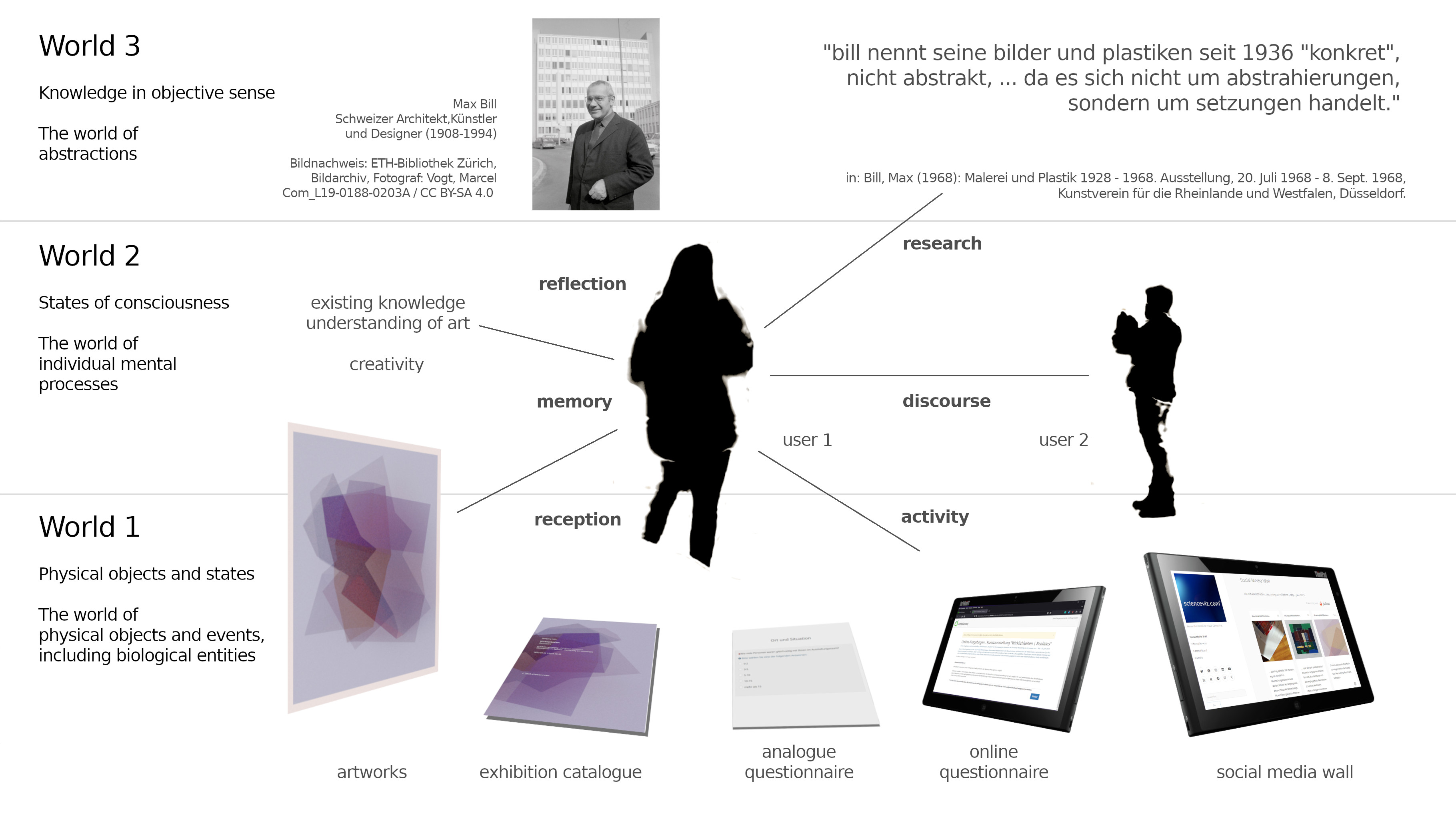}}
\caption{Level of perception $\vert$ Media used, interactions and the 3 worlds according to Popper}
\label{fig11}
\end{minipage}
\end{figure*}

This study makes intensive use of the open space technology method. There are many opportunities for interactive participation for all visitors and participants. The concrete location and the specific situation in the exhibition space are significantly influenced by the number and respective characteristics of the visitors (age, gender, educational background, personal relationship to art, existing knowledge, understanding of art, creativity and the individual activities of the test subjects). The interaction takes place in a spatially hybrid setting (analog/digital).

In an ``open space'', the public can participate in the art exhibition themselves. Art can be experienced individually in three stages: in the aesthetic reception of images, in one's own reflection and in individual artistic production. All visitors can view the works of art in the exhibition space and seek dialog with others present. Participants can also research individual topics. Artworks, exhibition catalogue \cite{b56} and analog questionnaires are available as print products in the exhibition space. Posts, mailings, telephone calls, an online questionnaire, a social media wall under the Instagram hashtag \char "0023kunstwirklichkeiten and a blog complete the mix of different media types. Visitors can leave their own photos, sketches, comments and notes via the social media wall and the questionnaires. A representation of the level of perception with the media used and all possibilities for interactive participation, organized according to Popper's three worlds, can be found in Fig. \ref{fig11}.

\subsection {Places and used media types}

A total of 19 different media were used. These include three face-to-face events (vernissage, artist talk and privatissimum), six print media (artworks, invitation cards, analog questionnaires, analog exhibition catalogues, art postcards and a newspaper article \cite{b57}), two electronic media (telephone, mobile phone and e-mails) and eight interactive online platforms (online exhibition catalogue \cite{b56}, online media portal ``Merkur.de'' \cite{b58}, online survey via Lime Survey, social media wall, Instagram, X [Twitter], LinkedIn and YouTube). The adjacent table \ref{tab1} shows the media used in the study and the corresponding assignment to the media types according to Faulstich.

\begin{table}[htbp]
\caption{Media types and used media}
\begin{center}
\begin{tabularx}{1.0\columnwidth}{|p{0.027\textwidth}|p{0.2\textwidth}|X|}
\hline
\textbf{} & \textbf{Media types} & \textbf{Used media}\\
\hline
{P \textsubscript V} & {Primary media} & {Vernissage} \\
\hline
{P \textsubscript K} & {} & {Artist's talk} \\
\hline
{P \textsubscript P} & {} & {Privatissimum} \\
\hline
{S \textsubscript A} & {Secondary media} & {Artworks} \\
\hline
{S \textsubscript I} & {} & {Invitation cards} \\
\hline
{S \textsubscript Q} & {} & {Questionnaires} \\
\hline
{S \textsubscript C} & {} & {Exhibition catalogue} \\
\hline
{S \textsubscript P} & {} & {Art postcards} \\
\hline
{S \textsubscript Z} & {} & {Newspaper article} \\
\hline
{T \textsubscript T} & {Tertiary media} & {Telephone/Mobile} \\
\hline
{T \textsubscript E} & {} & {e-mails} \\
\hline
{Q \textsubscript C} & {Quarternary media} & {Online exhibition catalogue} \\
\hline
{Q \textsubscript M} & {} & {Online media portal} \\
\hline
{Q \textsubscript S} & {} & {Online survey} \\
\hline
{Q \textsubscript W} & {} & {Social media wall (Blog)} \\
\hline
{Q \textsubscript I} & {} & {Instagram} \\
\hline
{Q \textsubscript X} & {} & {X (Twitter)} \\
\hline
{Q \textsubscript L} & {} & {LinkedIn} \\
\hline
{Q \textsubscript Y} & {} & {YouTube} \\
\hline
\end{tabularx}
\label{tab1}
\end{center}
\end{table}

There were two variants of questionnaires: analogue questionnaires and an online survey. The analogue questionnaires ($n_{S_{Q}} = 30$) were available as a printed version in the exhibition room. The online survey was created using the ``Lime Survey'' software and was accessible from March 24, 2023 to July 3, 2023 on the server of the Chair of Urban Structure and Transport Planning, TUM School of Engineering and Design at the Technical University of Munich (TUM) (\url {https://www.umfrage.sv.bgu.tum.de/}). A blank copy of the analogue questionnaire can be found in the appendix \ref{appendix:questionnaire}. Fig. \ref{fig12} shows a screenshot of the start screen of the online questionnaire.\\

\begin{figure*}[htbp]
\centerline{\includegraphics[width=1.0\textwidth]{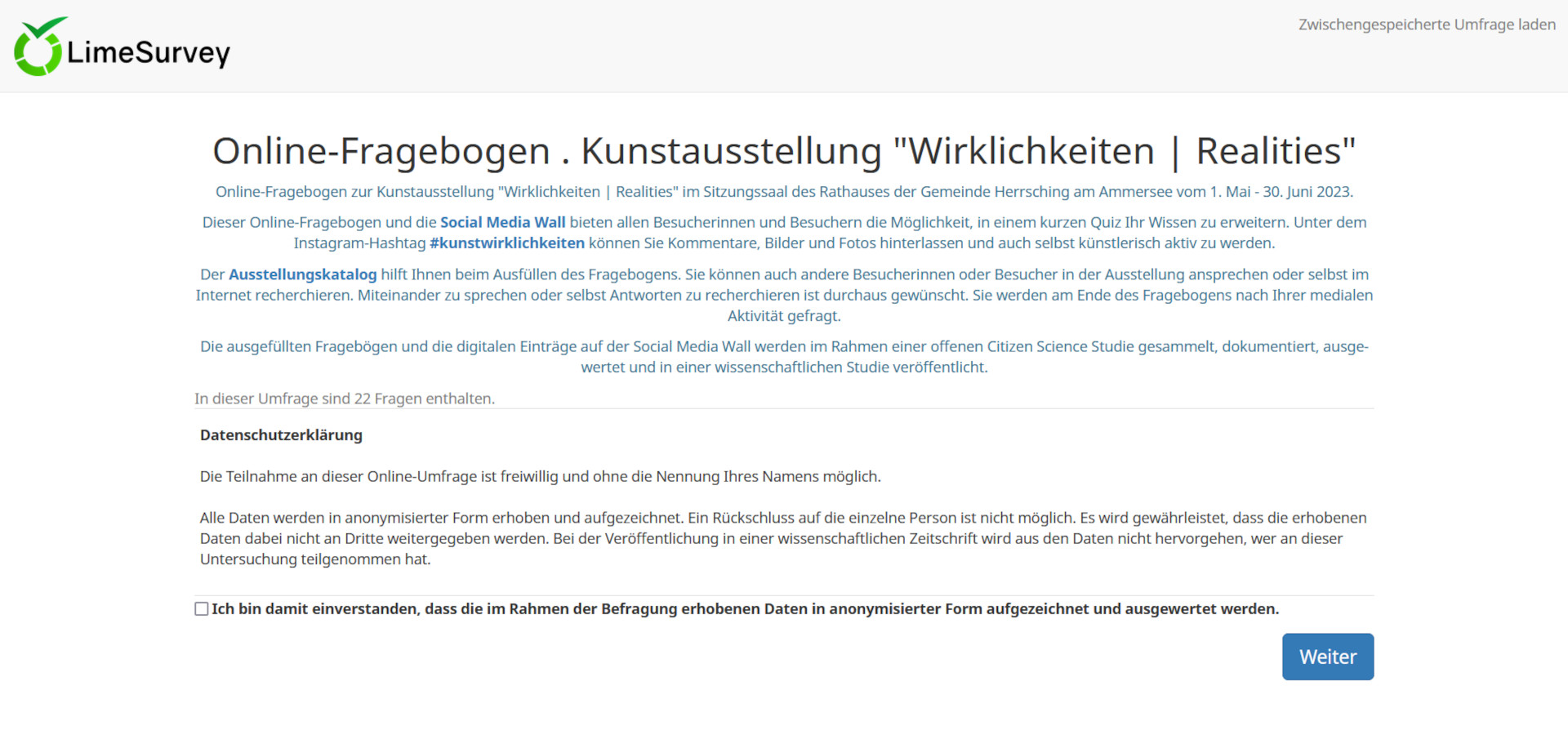}}
\caption{Online-Survey $\vert$ Startscreen (Screenshot, 30 May 2023, 10:20)}
\label{fig12}
\end{figure*}

\begin{figure*}[htbp]
\begin{minipage}[b]{1.0\textwidth}
\centerline{\includegraphics[width=1.0\textwidth]{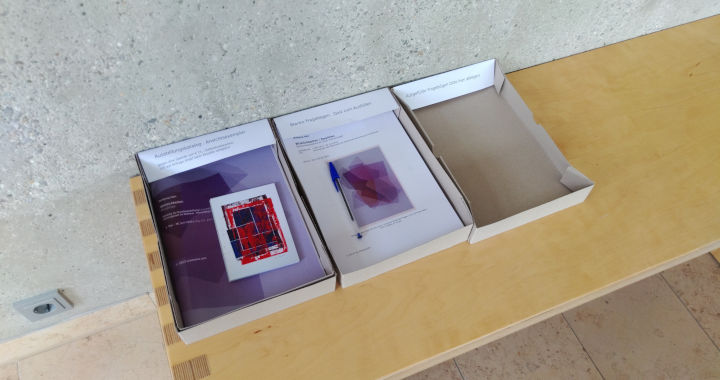}}
\caption{Exhibition catalogue, art postcards and analogue questionnaires with a box for completed questionnaires (Photograph, 20 June 2023, 15:13)}
\label{fig13}
\end{minipage}
\end{figure*}

20 copies of the analogue exhibition catalogue were printed ($n_{S_{C}} = 20$). A digital version was published at the Social Sciences Open Access Repository (SSOAR) of the GESIS - Leibniz Institute for the Social Sciences, Mannheim and at mediaTUM, the media and publication server of the Technical University of Munich (TUM) \cite{b56}. In the exhibition catalogue (analogue/online) you will find all illustrations of the eleven artworks ($n_{S_{A}} = 11$), as well as a complete description of the exhibition space (location and situation) and the artistic concept of the exhibition. The exhibition catalogue contains all responses to the online survey. Fig. \ref{fig13} shows a section of the exhibition space with the shelves for the exhibition catalogue, art postcards, the analogue questionnaires and the shelf for the completed questionnaires.

The following two chapters show the results of the media analysis and the results of the evaluation of questionnaires and the online survey.\\

\section{Results of the media analysis}


\subsection {Mix of media types (modality)}

The mixing ratio of the individual media types (modality) is determined below. Of the 19 media used, eight were quaternary media (42.10\%), six secondary media (31.58\%), three primary media (15.79\%) and two tertiary media (10.53\%). The quaternary media have the largest share, followed by the secondary media, then the primary media and finally the tertiary media. The following table \ref{tab2} shows the share of the respective media types (modality).

\begin{table}[htbp]
\caption{Mix of media types (modality)}
\begin{center}
\begin{tabularx}{1.0\columnwidth}{|p{0.027\textwidth}|p{0.12\textwidth}|p{0.15\textwidth}|X|}
\hline
\textbf{} & \textbf{Media types} & \textbf{Used media} & \textbf{Mix of media types (modality)} \\
\hline
{Q \textsubscript C} & {Quarternary media}  & {Online exhibition catalogue} & {42,10\%} \\
\hline
{Q \textsubscript M} & {} & {Online media portal} & {-} \\
\hline
{Q \textsubscript S}  & {} & {Online survey} & {-} \\
\hline
{Q \textsubscript W} & {} & {Social media wall (Blog)} & {-} \\
\hline
{Q \textsubscript I}  & {} &{Instagram} & {-} \\
\hline
{Q \textsubscript X} & {} & {X (Twitter)} & {-} \\
\hline
{Q \textsubscript L} & {} & {LinkedIn} & {-} \\
\hline
{Q \textsubscript Y} & {} & {YouTube} & {-} \\
\hline
{S \textsubscript A} & {Secondary media} & {Artworks} & {31,58\%} \\
\hline
{S \textsubscript I} & {} & {Invitation cards} & {-} \\
\hline
{S \textsubscript Q} & {} & {Questionnaires} & {-} \\
\hline
{S \textsubscript C} & {} & {Exhibition catalogue} & {-} \\
\hline
{S \textsubscript P} & {} & {Art postcards} & {-} \\
\hline
{S \textsubscript Z} & {} & {Newspaper article} & {-} \\
\hline
{P \textsubscript V} & {Primary media} & {Vernissage} & {15,79\%} \\
\hline
{P \textsubscript K} & {} & {Artist's talk} & {-} \\
\hline
{P \textsubscript P}& {}  & {Privatissimum} & {-} \\
\hline
{T \textsubscript T} & {Tertiary media} & {Telephone/Mobile} & {10,53\%} \\
\hline
{T \textsubscript E} & {} & {e-mails} & {-} \\
\hline
\end{tabularx}
\label{tab2}
\end{center}
\end{table}

\subsection {Geographic range and net range}

\begin{figure*}[htbp]
\begin{minipage}[b]{1.0\textwidth}
\centerline{\includegraphics[width=1.0\textwidth]{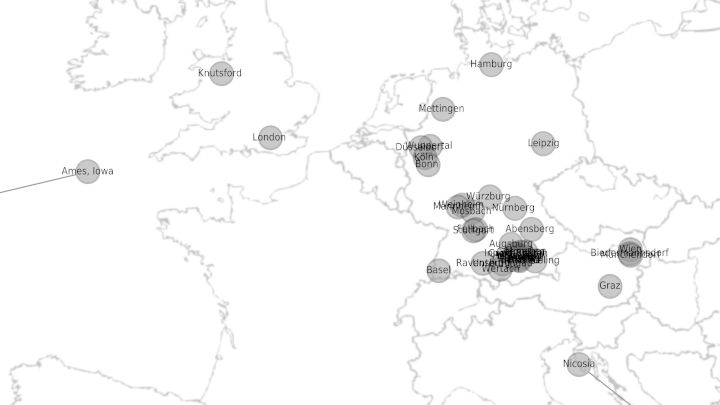}}
\caption{Geographic range $\vert$ Distribution of places and project participants}
\label{fig14}
\end{minipage}
\end{figure*}

The project lasted a total of 184 days (January 1, 2023 to July 3, 2023). The exhibition period lasted 61 days (May 1, 2023 to June 30, 2023). A total of 177 people were involved in project communication across all four media types (primary, secondary, tertiary and quaternary media). They were spread across 39 real locations in Europe, the United Kingdom and the USA. Fig. \ref{fig14} shows the distribution of locations with project participants (geographical range).

Table \ref{tab3} provides an overview of the geographical range and the net range. It shows all 39 real locations and the number of people reached via the media used. There was a large majority of people involved in virtual locations (e.g.: Instagram, X [Twitter], LinkedIn) to which no real location could be assigned. It is interesting to note that the people reached in real and virtual locations can be understood as a function of the range of a medium.

\begin{table*}[htbp]
\caption{Locations and range by media types}
\begin{center}
\begin{tabularx}{1.0\textwidth}{|p{0.058\textwidth}|p{0.0225\textwidth}|p{0.0233\textwidth}|p{0.0225\textwidth}|p{0.0225\textwidth}|p{0.021\textwidth}|p{0.0225\textwidth}|p{0.0225\textwidth}|p{0.0225\textwidth}|p{0.0225\textwidth}|p{0.023\textwidth}|p{0.023\textwidth}|p{0.025\textwidth}|p{0.039\textwidth}|p{0.0235\textwidth}|p{0.0285\textwidth}|p{0.0215\textwidth}|p{0.025\textwidth}|p{0.025\textwidth}|X|}
\hline
\textbf{} & \textbf{P \textsubscript V} & \textbf{P \textsubscript K} & \textbf{P \textsubscript P} & \textbf{S \textsubscript A} & \textbf{S \textsubscript I} & \textbf{S \textsubscript Q} & \textbf{S \textsubscript C} & \textbf{S \textsubscript P} & \textbf{S \textsubscript Z} & \textbf{T \textsubscript T} & \textbf{T \textsubscript E} & \textbf{Q \textsubscript C} & \textbf{Q \textsubscript M} & \textbf{Q \textsubscript S} & \textbf{Q \textsubscript W} & \textbf{Q \textsubscript I} & \textbf{Q \textsubscript X} & \textbf{Q \textsubscript L} & \textbf{Q \textsubscript Y}\\
\hline
{Hamburg} & {} & {} & {} & {} & {} & {} & {} & {} & {} & {} & {3} & {} & {} & {} & {} & {} & {} & {} & {} \\
\hline
{Knutsford} & {} & {} & {} & {} & {} & {} & {} & {} & {} & {} & {} & {} & {} & {} & {} & {} & {} & {1} & {} \\
\hline
{Mettingen} & {} & {} & {} & {} & {2} & {} & {} & {} & {} & {} & {} & {} & {} & {} & {} & {} & {} & {} & {} \\
\hline
{London} & {} & {} & {} & {} & {} & {} & {} & {} & {} & {} & {3} & {} & {} & {} & {} & {} & {} & {} & {} \\
\hline
{Leipzig} & {} & {} & {} & {} & {} & {} & {1} & {} & {} & {} & {2} & {} & {} & {} & {} & {} & {} & {} & {} \\
\hline
{Wuppertal} & {} & {} & {} & {} & {1} & {} & {} & {} & {} & {} & {} & {} & {} & {} & {} & {} & {} & {} & {} \\
\hline
{Düsseldorf} & {} & {} & {} & {} & {} & {} & {} & {} & {} & {} & {6} & {} & {} & {} & {} & {} & {} & {} & {} \\
\hline
{Köln} & {} & {} & {} & {} & {1} & {} & {1} & {} & {} & {} & {5} & {} & {} & {} & {} & {} & {} & {} & {} \\
\hline
{Bonn} & {} & {} & {} & {} & {} & {} & {1} & {} & {} & {} & {4} & {} & {} & {} & {} & {} & {} & {} & {} \\
\hline
{Würzburg} & {} & {} & {} & {} & {1} & {} & {} & {} & {} & {} & {1} & {} & {} & {} & {} & {} & {} & {} & {} \\
\hline
{Weinheim} & {} & {} & {} & {} & {} & {} & {} & {} & {} & {} & {2} & {} & {} & {} & {} & {} & {} & {} & {} \\
\hline
{Mannheim} & {} & {} & {} & {} & {} & {} & {} & {} & {} & {} & {} & {} & {} & {} & {} & {} & {} & {1} & {} \\
\hline
{Nürnberg} & {} & {} & {} & {} & {1} & {} & {} & {} & {} & {} & {1} & {} & {} & {} & {} & {} & {} & {} & {} \\
\hline
{Mosbach} & {} & {} & {} & {} & {1} & {} & {} & {} & {} & {} & {1} & {} & {} & {} & {} & {} & {} & {} & {} \\
\hline
{Abensberg} & {} & {} & {} & {} & {4} & {} & {} & {} & {} & {} & {} & {} & {} & {} & {} & {} & {} & {} & {} \\
\hline
{Fellbach} & {} & {} & {} & {} & {} & {} & {} & {} & {} & {} & {2} & {} & {} & {} & {} & {} & {} & {} & {} \\
\hline
{Stuttgart} & {} & {} & {} & {} & {} & {} & {} & {} & {} & {} & {1} & {} & {} & {} & {} & {} & {} & {} & {} \\
\hline
{Augsburg} & {} & {} & {} & {} & {} & {} & {} & {} & {} & {} & {1} & {} & {} & {} & {} & {} & {} & {} & {} \\
\hline
{Wien} & {} & {} & {} & {} & {13} & {} & {1} & {} & {} & {2} & {7} & {} & {} & {} & {} & {} & {} & {} & {} \\
\hline
{München} & {} & {} & {} & {} & {26} & {} & {1} & {} & {} & {3} & {35} & {} & {} & {} & {} & {} & {} & {} & {} \\
\hline
{Germering} & {1} & {} & {} & {} & {1} & {} & {} & {} & {} & {} & {} & {} & {} & {} & {} & {} & {} & {} & {} \\
\hline
{Bieder-mannsdorf} & {} & {} & {} & {} & {2} & {} & {} & {} & {} & {} & {} & {} & {} & {} & {} & {} & {} & {} & {} \\
\hline
{Inning-Stegen} & {} & {} & {} & {} & {1} & {} & {} & {} & {} & {1} & {1} & {} & {} & {} & {} & {} & {} & {} & {} \\
\hline
{Greifen-berg} & {} & {} & {} & {} & {4} & {} & {} & {} & {} & {} & {} & {} & {} & {} & {} & {} & {} & {} & {} \\
\hline
{München-dorf} & {} & {} & {} & {} & {2} & {} & {2} & {} & {} & {} & {} & {} & {} & {} & {} & {} & {} & {} & {} \\
\hline
{Starnberg} & {1} & {} & {2} & {} & {12} & {} & {3} & {} & {} & {} & {5} & {} & {} & {} & {} & {} & {} & {} & {} \\
\hline
{Herrsching} & {2} & {9} & {1} & {11} & {15} & {7} & {3} & {63} & {} & {1} & {29} & {} & {} & {} & {} & {} & {} & {} & {} \\
\hline
{Diessen} & {} & {} & {} & {} & {2} & {} & {} & {} & {} & {} & {} & {} & {} & {} & {} & {} & {} & {} & {} \\
\hline
{Feldafing} & {} & {} & {} & {} & {4} & {} & {} & {} & {} & {} & {2} & {} & {} & {} & {} & {} & {} & {} & {} \\
\hline
{Tutzing} & {3} & {3} & {} & {} & {24} & {} & {3} & {} & {} & {2} & {3} & {} & {} & {} & {} & {} & {} & {} & {} \\
\hline
{Bernried} & {} & {} & {} & {} & {9} & {} & {} & {} & {} & {} & {1} & {} & {} & {} & {} & {} & {} & {} & {} \\
\hline
{Bad Aibling} & {} & {} & {} & {} & {} & {} & {} & {} & {} & {} & {} & {} & {} & {} & {} & {} & {} & {1} & {} \\
\hline
{Ravens-burg} & {} & {} & {} & {} & {1} & {} & {} & {} & {} & {} & {2} & {} & {} & {} & {} & {} & {} & {} & {} \\
\hline
{Unter-thingau} & {} & {} & {} & {} & {2} & {} & {} & {} & {} & {} & {} & {} & {} & {} & {} & {} & {} & {} & {} \\
\hline
{Wertach} & {} & {} & {} & {} & {2} & {} & {2} & {} & {} & {} & {} & {} & {} & {} & {} & {} & {} & {} & {} \\
\hline
{Basel} & {} & {} & {} & {} & {} & {} & {} & {} & {} & {} & {1} & {} & {} & {} & {} & {} & {} & {} & {} \\
\hline
{Graz} & {} & {} & {} & {} & {} & {} & {} & {} & {} & {} & {} & {} & {} & {} & {} & {} & {} & {1} & {} \\
\hline
{Ames, Iowa} & {} & {} & {} & {} & {} & {} & {} & {} & {} & {} & {} & {} & {} & {} & {} & {} & {} & {1} & {} \\
\hline
{Nicosia} & {} & {} & {} & {} & {1} & {} & {} & {} & {} & {} & {} & {} & {} & {} & {} & {} & {} & {} & {} \\
\hline
{without location} & {} & {} & {} & {} & {} & {} & {} & {} & {9.060} & {} & {} & {196} & {306.748} & {14} & {71} & {586} & {182} & {1.373} & {51} \\
\hline
\textbf{Summe} & \textbf{7} & \textbf{12} & \textbf{3} & \textbf{11} &\textbf{132} & \textbf{7} & \textbf{18} & \textbf{63} & \textbf{9.060} & \textbf{9} & \textbf{118} &\textbf {196} & \textbf{306.748} & \textbf{14} & \textbf{71} & \textbf{586} & \textbf{182} & \textbf{1.377} & \textbf{51} \\
\hline
{Primary media} & {} & {} & {22} & {} & {} & {} & {} & {} & {} & {} & {} & {} & {} & {} & {} & {} & {} & {} & {} \\
\hline
{Secondary media} & {} & {} & {} & {} & {} & {} & {} & {} & {9.280} & {} & {} & {} & {} & {} & {} & {} & {} & {} & {} \\
\hline
{Tertiary media} & {} & {} & {} & {} & {} & {} & {} & {} & {} & {} & {127} & {} & {} & {} & {} & {} & {} & {} & {} \\
\hline
{Quarternary media} & {} & {} & {} & {} & {} & {} & {} & {} & {} &{} & {} & {} & {306.748} & {} & {} & {} & {} & {} & {2.477} \\
\hline
\end{tabularx}
\label{tab3}
\end{center}
\end{table*}

\subsection {Range, interactions and response rate}

Below you will find the data collected and calculated for the media used on range, interactions and response rate, broken down by media type.

In the case of primary media, the number of people invited was assumed to be a characteristic of range. The number of exhibition visitors is a measure of interactions. For secondary media, the number of printed copies is used as a measure of range. The number of copies that have been corrected or distributed is used as a measure of interaction. In the case of tertiary media, the number of incoming and outgoing communication events (telephone calls and e-mails) is used as a measure of range and interaction. For all quaternary media, the views of a post are used as an indicator of the number of people reached. The number of participants, downloads, likes or comments are used as a measure of interactions for quarternary media. If no views are available, the number of people reached across all media types is used as a benchmark. In this case, the number of views is used as the key figure for interactions. For the article in the Starnberger Merkur media portal, the circulation sold (including e-paper) is used as the key figure for the people reached \cite{b59}.

There were a total of three face-to-face events (vernissage, artist's talk and privatissimum). 22 people visited the exhibition directly on site. For the first two events (vernissage and artist's talk), a total of 90 invitation cards were sent to 132 people by post. Of these, 29 invitations were sent to 52 people for the vernissage. For the artist's talk, 61 invitations were issued to 80 people. 7 people attended the vernissage. 12 people were present at the artist's talk. By prior arrangement and separate organization, 3 people attended the private event.

Of the 11 artworks exhibited (5 digital prints and 6 lino prints), one digital print was sold. 30 copies of the analog questionnaires were printed ($n_{S_{Q}} = 30$). 7 copies were taken by the exhibition visitors. 20 printed copies of the exhibition catalog were produced. 15 copies were distributed to interested parties. A total of 88 copies of Art postcards were printed ($n_{S_{P}} = 88$) and displayed in the exhibition space. 63 Art postcards were taken by the exhibition visitors. A Newspaper article appeared on May 25, 2023 in the local section of the Starnberger Merkur \cite{b57}. This edition of the newspaper had a circulation of 9,060 copies \cite{b59}. Unfortunately, no concrete data is available on the number of people reached or feedback on this newspaper article. 20 outgoing and 8 incoming telephone calls were made to 9 people. A total of 118 e-mails were sent to 69 people, 54 e-mails were received from this group of people.

The publication of the online exhibition catalog on SSOAR received 196 views and 24 downloads (as of August 25, 2023). The number of views of the online exhibition catalog on mediaTUM is unfortunately not available. The views of the article in the media portal of the Starnberger Merkur are unfortunately also not available. There were no comments. The paid circulation of this issue (incl. e-paper) was 306,748 copies. 177 people were informed about the online survey. 14 participants took part in the online survey. 3 participants completed the online survey in full. 11 participants only partially completed the online survey. The social media wall on the blog ``scienceviz.com'' recorded 71 views. 28 posts by 3 people appeared under the Instagram hashtag \#kunstwirklichkeiten. If you subtract the posts by the artist himself, two other participants took part in the postings. In total, these posts recorded 586 views and 311 likes. Two posts about the exhibition appeared on platform X (Twitter) with a total of 182 views and 3 interactions. A video post was also published on YouTube, with 51 views and one comment. A total of 2,477 views were achieved across all Quarternary media.

The publication of the online exhibition catalog on SSOAR received 196 views and the response rate of the individual media is calculated below. Table \ref{tab4} shows the media used, their range, interactions and the calculated response rate.

\begin{table}[htbp]
\caption{Media types, used media, range, \\ interactions and response rate}
\begin{center}
\begin{tabularx}{1.0\columnwidth}{|p{0.027\textwidth}|p{0.10\textwidth}|p{0.09\textwidth}|p{0.09\textwidth}|X|}
\hline
\textbf{} & \textbf{Used media} & \textbf{Range} & \textbf{Interactions} & \textbf{Response rate} \\
\hline
{P \textsubscript V} & {Vernissage} & {52 Persons} & {7 Persons} & {13\%} \\
\hline
{P \textsubscript K} & {Artist's talk} & {80 Persons} & {12 Persons} & {15\%} \\
\hline
{P \textsubscript P} & {Privatissimum} & {-} & {3 Persons} & {-} \\
\hline
{S \textsubscript A} & {Artworks} & {11 Prints} & {1 Purchase} & {9\%} \\
\hline
{S \textsubscript I} & {Invitation cards} & {132 Persons} & {22 Persons} & {17\%} \\
\hline
{S \textsubscript Q} & {Questionnaires} & {30 Copies printed} & {7 Copies taken} & {23\%} \\
\hline
{S \textsubscript C} & {Exhibition catalogue} & {20 Copies printed} & {15 Copies taken} & {75\%} \\
\hline
{S \textsubscript P} & {Art postcards} & {88 Copies printed} & {63 Copies taken} & {72\%} \\
\hline
{S \textsubscript Z} & {Newspaper article} & {9.060 Copies printed} & {-} & {-} \\
\hline
{T \textsubscript T} & {Telephone/Mobile} & {20 Calls} & {8 Callbacks} & {40\%} \\
\hline
{T \textsubscript E} & {e-mails} & {118 Outmails} & {54 Inmails} & {46\%} \\
\hline
{Q \textsubscript C} & {Online exhibition catalogue} & {196 Views} & {24 Downloads} & {12\%} \\
\hline
{Q \textsubscript M} & {Online media portal} & {306.748 Sold edition (incl. e-paper)} & {-} & {-} \\
\hline
{Q \textsubscript S} & {Online survey} & {177 Persons} & {14 Participants} & {8\%} \\
\hline
{Q \textsubscript W} & {Social media wall (Blog)} & {177 Persons} & {71 Views} & {40\%} \\
\hline
{Q \textsubscript I} &{Instagram} & {586 Views} & {311 Likes} & {53\%} \\
\hline
{Q \textsubscript X} & {X (Twitter)} & {182 Views} & {3 Interactions} & {2\%} \\
\hline
{Q \textsubscript L} & {LinkedIn} & {1.377 Views} & {12 Likes} & {1\%} \\
\hline
{Q \textsubscript Y} & {YouTube} & {51 Views} & {1 Comment} & {2\%} \\
\hline
\end{tabularx}
\label{tab4}
\end{center}
\end{table}

\section{Discussion of the results of the media analysis}

\subsection {Range and familiarity}

\begin{figure*}[htbp]
\begin{minipage}[b]{1.0\textwidth}
\centerline{\includegraphics[width=1.0\textwidth]{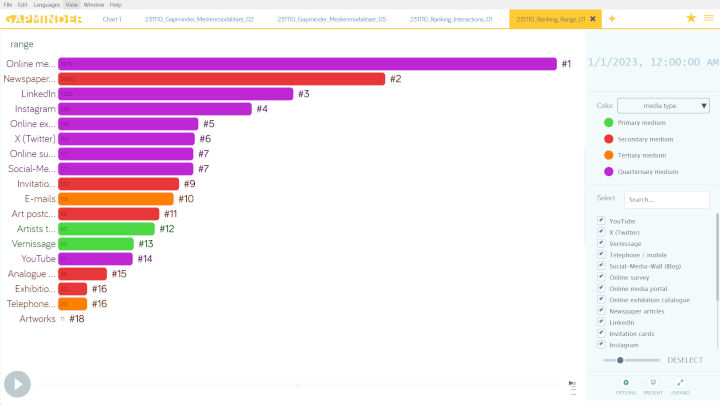}}
\caption{Ranking of media and media types according to range}
\label{fig15}
\end{minipage}
\end{figure*}

Range is a measure of how well known a place is (familiarity). If you arrange the media used in table \ref{tab4} according to their respective range, the following picture emerges, as shown in table \ref{tab5}. The online media portal of the Starnberger Merkur, the Newspaper article and the online platform LinkedIn have the greatest range. The art exhibition achieved the greatest awareness via two quaternary media (Starnberger Merkur Online, LinkedIn) and one secondary medium (Starnberger Merkur Print). In terms of the mix of media types (modality), quarternary media (70\%) outweighed secondary media (20\%) and tertiary media (10\%) among the top ten media. The most common media among the last nine were Secondary media (44.45\%), Primary media (33.34\%), Tertiary and Quarternary media (11.12\% each). Figure Fig. \ref{fig15} shows a color-coded representation of the ranking of media and media types according to range.

\begin{table}[htbp]
\caption{Used media according to range}
\begin{center}
\begin{tabularx}{1.0\columnwidth}{|p{0.027\textwidth}|p{0.21\textwidth}|X|}
\hline
\textbf{} & \textbf{Used media} & \textbf{Range} \\
\hline
{Q \textsubscript M} & {Online media portal} & {306.748 Sold edition (incl. e-paper)} \\
\hline
{S \textsubscript Z} & {Newspaper article} & {9.060 Copies printed} \\
\hline
{Q \textsubscript L} & {LinkedIn} & {1.377 Views} \\
\hline
{Q \textsubscript I} &{Instagram} & {586 Views} \\
\hline
{Q \textsubscript C} & {Online exhibition catalogue} & {196 Views} \\
\hline
{Q \textsubscript X} & {X (Twitter)} & {182 Views} \\
\hline
{Q \textsubscript S} & {Online survey} & {177 Persons} \\
\hline
{Q \textsubscript W} & {Social media wall (Blog)} & {177 Persons} \\
\hline
{S \textsubscript I} & {Invitation cards} & {132 Persons} \\
\hline
{T \textsubscript E} & {e-mails} & {118 sent e-mails} \\
\hline
{S \textsubscript P} & {Art postcards} & {88 Copies printed} \\
\hline
{P \textsubscript K} & {Artist's talk} & {80 Persons} \\
\hline
{Q \textsubscript Y} & {YouTube} & {51 Views} \\
\hline
{P \textsubscript V} & {Vernissage} & {52 Persons} \\
\hline
{S \textsubscript Q} & {Questionnaires} & {30 Copies printed} \\
\hline
{S \textsubscript C} & {Exhibition catalogue} & {20 Copies printed} \\
\hline
{T \textsubscript T} & {Telephone/Mobile} & {20 Calls} \\
\hline
{S \textsubscript A} & {Artworks} & {11 Prints} \\
\hline
{P \textsubscript P} & {Privatissimum} & {-} \\
\hline
\end{tabularx}
\label{tab5}
\end{center}
\end{table}

\subsection {Interactions and popularity}

\begin{figure*}[htbp]
\begin{minipage}[b]{1.0\textwidth}
\centerline{\includegraphics[width=1.0\textwidth]{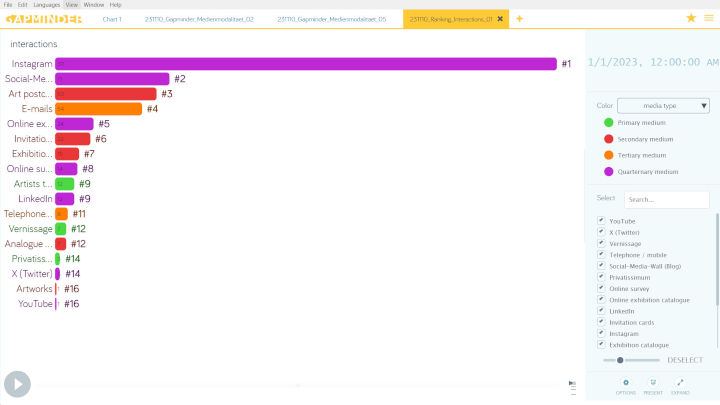}}
\caption{Ranking of media and media types according to interactions}
\label{fig16}
\end{minipage}
\end{figure*}

Interactions are a measure of the attractiveness or popularity of media. Table \ref{tab6} shows all media used sorted according to the number of interactions. The top ten media include five quaternary media (50\%), three secondary media (30\%) and one tertiary and one primary media (10\% each). The last nine places are divided between three quaternary media (30\%), three secondary media (30\%), two primary media (20\%) and one tertiary media (10\%). Figure Fig. \ref{fig16} shows a colored representation of the ranking of media and media types according to interactions.

\begin{table}[htbp]
\caption{Verwendete Medien nach Interaktionen}
\begin{center}
\begin{tabularx}{1.0\columnwidth}{|p{0.027\textwidth}|p{0.21\textwidth}|X|}
\hline
\textbf{} & \textbf{Used media} & \textbf{Interactions} \\
\hline
{Q \textsubscript I} &{Instagram} & {311 Likes} \\
\hline
{Q \textsubscript W} & {Social media wall (Blog)} & {71 Views} \\
\hline
{S \textsubscript P} & {Art postcards} & {63 Copies taken} \\
\hline
{T \textsubscript E} & {e-mails} & {54 Inmails} \\
\hline
{Q \textsubscript C} & {Online exhibition catalogue} & {24 Downloads} \\
\hline
{S \textsubscript I} & {Invitation cards} & {22 Persons} \\
\hline
{S \textsubscript C} & {Exhibition catalogue} & {15 Copies taken} \\
\hline
{Q \textsubscript S} & {Online survey} & {14 Participants} \\
\hline
{P \textsubscript K} & {Artist's talk} & {12 Persons} \\
\hline
{Q \textsubscript L} & {LinkedIn} & {12 Likes} \\
\hline
{T \textsubscript T} & {Telephone/Mobile} & {8 Callbacks} \\
\hline
{P \textsubscript V} & {Vernissage} & {7 Persons} \\
\hline
{S \textsubscript Q} & {Questionnaires} & {7 Copies taken} \\
\hline
{P \textsubscript P} & {Privatissimum} & {3 Persons} \\
\hline
{Q \textsubscript X} & {X (Twitter)} & {3 Interactions} \\
\hline
{S \textsubscript A} & {Artworks} & {1 Purchase} \\
\hline
{Q \textsubscript Y} & {YouTube} & {1 Comment} \\
\hline
{S \textsubscript Z} & {Newspaper article} & {-} \\
\hline
{Q \textsubscript M} & {Online media portal} & {-} \\
\hline
\end{tabularx}
\label{tab6}
\end{center}
\end{table}

\subsection {Response rate and effectiveness}

\begin{figure*}[htbp]
\begin{minipage}[b]{1.0\textwidth}
\centerline{\includegraphics[width=1.0\textwidth]{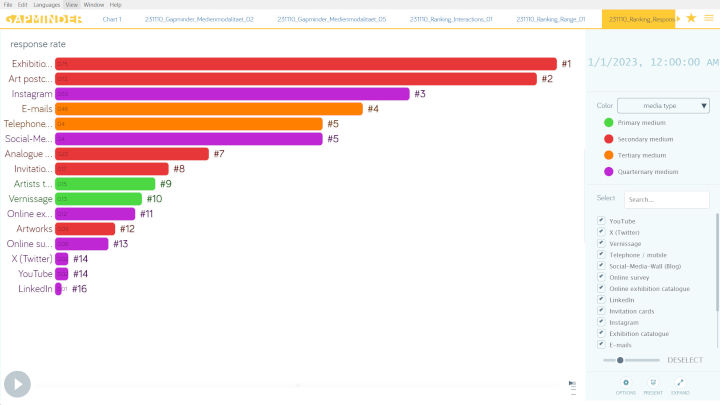}}
\caption{Ranking of media and media types according to the resposne rate}
\label{fig17}
\end{minipage}
\end{figure*}

The response rates can be seen as a measure of the participation or effectiveness of a medium. If Table \ref{tab4} is arranged according to the respective response rates, this results in an order of the most popular or most used media. Table \ref{tab7} shows a list of the media used, sorted by response rate. Measured by the response rate, the exhibition catalog (75\%), art postcards (72\%) and Instagram (53\%) were the most effective media with the highest participation. This was followed by emails (46\%), telephone calls (40\%), the social media wall (40\%), analog questionnaires (23\%) and analog invitation cards (17\%). Only then came the face-to-face events Artist's talk (15\%) and Vernissage (13\%), followed by the online exhibition catalog (12\%). Artworks, the online survey, X (Twitter) and LinkedIn were in last place (below 10\%). There were no concrete response rates for the Privatissimum and the two publications (Newspaper article and Online media portal). It is interesting to note that four secondary media (40\%) are represented among the first ten media. Only then do two primary media (20\%), two tertiary media (20\%) and two quaternary media (20\%) follow ex aequo. Figure Fig. \ref{fig17} shows a colored representation of the ranking of media and media types according to the response rate.

\begin{table}[htbp]
\caption{Used media according to response rate}
\begin{center}
\begin{tabularx}{1.0\columnwidth}{|p{0.027\textwidth}|p{0.21\textwidth}|X|}
\hline
\textbf{} & \textbf{Used media} & \textbf{Response rate} \\
\hline
{S \textsubscript C} & {Exhibition catalogue} & {75\%} \\
\hline
{S \textsubscript P} & {Art postcards} & {72\%} \\
\hline
{Q \textsubscript I} &{Instagram} & {53\%} \\
\hline
{T \textsubscript E} & {e-mails} & {46\%} \\
\hline
{T \textsubscript T} & {Telephone/Mobile} & {40\%} \\
\hline
{Q \textsubscript W} & {Social media wall (Blog)} & {40\%} \\
\hline
{S \textsubscript Q} & {Questionnaires} & {23\%} \\
\hline
{S \textsubscript I} & {Invitation cards} & {17\%} \\
\hline
{P \textsubscript K} & {Artist's talk} & {15\%} \\
\hline
{P \textsubscript V} & {Vernissage} & {13\%} \\
\hline
{Q \textsubscript C} & {Online exhibition catalogue} & {12\%} \\
\hline
{S \textsubscript A} & {Artworks} & {9\%} \\
\hline
{Q \textsubscript S} & {Online survey} & {8\%} \\
\hline
{Q \textsubscript X} & {X (Twitter)} & {2\%} \\
\hline
{Q \textsubscript Y} & {YouTube} & {2\%} \\
\hline
{Q \textsubscript L} & {LinkedIn} & {1\%} \\
\hline
{P \textsubscript P} & {Privatissimum} & {-} \\
\hline
{S \textsubscript Z} & {Newspaper article} & {-} \\
\hline
{Q \textsubscript M} & {Online media portal} & {-} \\
\hline
\end{tabularx}
\label{tab7}
\end{center}
\end{table}

\begin{figure*}[htbp]
\begin{minipage}[b]{1.0\textwidth}
\centerline{\includegraphics[width=1.0\textwidth]{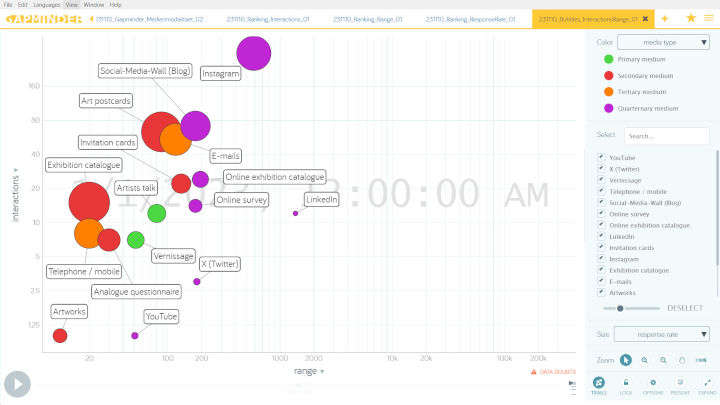}}
\caption{Response rate, interactions and range}
\label{fig18}
\end{minipage}
\end{figure*}

Figure Fig. \ref{fig18} summarizes all three of the aforementioned results once again in a colored graphic and shows a bubble chart of response rate, interactions and range.\\

\section {Results from questionnaires and online survey}

The evaluation of questionnaires and online surveys is dedicated to the correlations between the two levels of perception, time and interaction. There are two hypotheses guiding the research:\\

\begin{itemize}
\item There are correlations between media use (modality), the participants' interactions (creativity) and their perception (understanding of art)\\
\item Individual parameters (demographic data, location and situation, individual knowledge) influence perception (understanding of art)\\
\end{itemize}

The two hypotheses aim to clarify correlations between the three levels of locations and media types, the level of interaction and the level of perception. The characteristics of the aforementioned hypotheses are operationalized by the following two dependent variables ($V_{a}$) and five independent variables ($V_{u}$). Table \ref{tab8} shows a complete list of all variables used. It is assumed that the individual understanding of art is influenced by the following individual parameters: Age, gender, educational attainment, own relation to art, place and situation, knowledge of modern approaches to art, knowledge of influences from other fields of knowledge. It is also assumed that media use, represented by the proportion of media types (modality), has an influence on the interaction of the individual participants in hybrid spaces. The activity of the test subjects (interactions) is seen as an indicator of individual creativity.

\begin{table}[htbp]
\caption{Dependent and independent variables}
\begin{center}
\begin{tabularx}{1.0\columnwidth}{|p{0.027\textwidth}|p{0.2\textwidth}|X|}
\hline
\textbf{} & \textbf{Dependent variable ($V_{a}$)} & \textbf{Independent variable  ($V_{u}$)}\\
\hline
{$V_{a_{K}}$} & {Perception (Understanding of Art, individual view on the artist, artwork and society)} & {} \\
\hline
{$V_{u_{D}}$} & {} & {Age, gender, educational level (demographic data)} \\
\hline
{$V_{u_{R}}$} & {} & {Individual reference to art} \\
\hline
{$V_{u_{S}}$} & {} & {Place and Situation (Number of people in the exhibition space, time of day, duration of visit)} \\
\hline
{$V_{u_{Z}}$} & {} & {Knowledge of modern approaches to art and influences from other fields of knowledge} \\
\hline
{$V_{u_{M}}$} & {} & {Media use in hybrid spaces (media types and modality)} \\
\hline
{$V_{a_{I}}$} & {Interaction (individual activities as an indicator of individual creativity, e.g: Event attendance, media interactions, notes, comments, individual artistic production)} & {} \\
\hline
\end{tabularx}
\label{tab8}
\end{center}
\end{table}

The analog questionnaire and online questionnaire have the same structure. The questionnaires are divided into six groups of questions. The first group of questions (A1-A5) relates to individual demographic characteristics (age, gender, highest level of education, relationship to art and understanding of art). The second group of questions (B1-B3) relates to quantitative characteristics of the exhibition visit (number of visitors, time of day and individual duration of visit). The third group of questions (C1-C3) examines the image of the artist, artwork and society. The fourth group of questions (D1-D3) asks about the modern foundations of artistic practice. The fifth group of questions (E1-E3) deals with influences of other fields of knowledge on art. The sixth group of questions (F1-F5) asks about characteristics of individual activity and creativity. A copy of the questionnaire (analog/online) and the results from the online survey as raw data can be found in the appendix \ref{appendix:questionnaire}. In the following subchapters, the question groups from the questionnaire are given in brackets (e.g.: A1-A3). The online survey comprises a total of 14 data sets. 3 participants completed the online survey in full, 11 participants only partially answered the questionnaire. No analog questionnaires were completed or submitted. Below you will find a presentation of the results and responses.

\subsection {Demographic data (A1-A3)}

Of the 14 participants, 4 were male (28.57\%), 3 were female (21.43\%) and 1 person did not specify their gender (7.14\%). The age group 20-29 years was the most strongly represented with 3 people (21.43\%). The age groups 40-49 years (14.29\%) and over 50 years (14.29\%) were equally represented with 2 people each. 1 person represented the 30-39 age group (7.14\%). Of all 14 participants, 4 people had a technical college or university degree (28.57\%), 3 people had a vocational baccalaureate (21.43\%), 1 person had vocational training (7.14\%) and 6 participants (42.86\%) did not answer this question group A1-A3. Table \ref{tab8} shows the answers, sorted according to the ID number assigned in the online questionnaire. The answers with ID 1 and ID 2 were test entries written by the author prior to the publication of the online questionnaire. These test entries were deleted and do not appear here.

\begin{table}[htbp]
\caption{Demographic data (A1-A3)}
\begin{center}
\begin{tabularx}{1.0\columnwidth}{|p{0.05\textwidth}|p{0.075\textwidth}|p{0.09\textwidth}|X|}
\hline
\textbf{Answer ID} & \textbf{Salutation} & \textbf{Age group} & \textbf{Educational level} \\
\hline
{ID 3} & {Mr.} & {40-49} & {Technical diploma / High school graduation} \\
\hline
{ID 4} & {n.a.} & {50+} & {Technical diploma / High school graduation} \\
\hline
{ID 5} & {-} & {-} & {-} \\
\hline
{ID 6} & {Mr.} & {20-29} & {University of applied sciences / University} \\
\hline
{ID 7} & {Mrs.} & {50+} & {Professional training / Dual education} \\
\hline
{ID 8} & {-} & {-} & {-} \\
\hline
{ID 9} & {-} & {-} & {-} \\
\hline
{ID 10} & {-} & {-} & {-} \\
\hline
{ID 11} & {Mr.} & {20-29} & {Technical diploma / High school graduation} \\
\hline
{ID 12} & {-} & {-} & {-} \\
\hline
{ID 13} & {-} & {-} & {-} \\
\hline
{ID 14} & {Mrs.} & {20-29} & {University of applied sciences / University} \\
\hline
{ID 15} & {Mrs.} & {40-49} & {University of applied sciences / University} \\
\hline
{ID 16} & {Mr.} & {30-39} & {University of applied sciences / University} \\
\hline
\end{tabularx}
\label{tab8}
\end{center}
\end{table}

\subsection {Individual reference to art (A4-A5)}

Out of 14 participants, 5 people described themselves as ``art interested'' (35.71\%), 1 person as ``amateur'' (7.14\%) and 2 people as ``professional'' (14.29\%). The five answer options regarding their own art reference were: ``Landscape painting and still life'', ``Objective art'', ``Abstract art'', ``Concrete art'' and ``Media art and performance''. Within these five options, "Landscape painting and still life" was chosen 5 times (35.71\%), "Abstract art" 5 times (35.71\%) and "Media art and performance" 5 times (35.71\%). This was followed by ``Objective Art'' 4 times (28.57\%) and ``Concrete Art'' once (7.14\%). Multiple selection was possible. 6 participants (42.86\%) did not answer question group A4-A5. Table \ref{tab9} shows a tabular list of the answers. The art reference of all participants was relatively broad. The most popular were'' landscape painting and still life'' together with ``abstract art'' and ``media art and performance''. This was immediately followed by ``Objective Art''. Only one participant showed a reference to ``Concrete Art''. Table \ref{tab8} provides an overview of the participants' answers regarding their individual reference to art.

\begin{table}[htbp]
\caption{Individual reference to art (A4-A5)}
\begin{center}
\begin{tabularx}{1.0\columnwidth}{|p{0.05\textwidth}|p{0.1\textwidth}|p{0.034\textwidth}|p{0.034\textwidth}|p{0.034\textwidth}|p{0.034\textwidth}|X|}
\hline
\textbf{Answer ID} & \textbf{Reference to art} & \textbf{Land-scape paint-ing and still life} & \textbf{Ob-jective art} & \textbf{Ab-stract art} & \textbf{Con-crete art} & \textbf{Media art and perfor-mance} \\
\hline
{ID 3} & {Art interested} & {-} & {X} & {-} & {-} & {-} \\
\hline
{ID 4} & {Amateur} & {X} & {X} & {X} & {-} & {-} \\
\hline
{ID 5} & {-} & {-} & {-} & {-} & {-} & {-} \\
\hline
{ID 6} & {Professional} & {X} & {X} & {X} & {X} & {X} \\
\hline
{ID 7} & {Art interested} & {X} & {-} & {X} & {-} & {X} \\
\hline
{ID 8} & {-} & {-} & {-} & {-} & {-} & {-} \\
\hline
{ID 9} & {-} & {-} & {-} & {-} & {-} & {-} \\
\hline
{ID 10} & {-} & {-} & {-} & {-} & {-} & {-} \\
\hline
{ID 11} & {Art interested} & {-} & {-} & {-} & {-} & {X} \\
\hline
{ID 12} & {-} & {-} & {-} & {-} & {-} & {-} \\
\hline
{ID 13} & {-} & {-} & {-} & {-} & {-} & {-} \\
\hline
{ID 14} & {Art interested} & {X} & {-} & {X} & {-} & {-} \\
\hline
{ID 15} & {Art interested} & {X} & {X} & {X} & {-} & {X} \\
\hline
{ID 16} & {Professional} & {-} & {-} & {-} & {-} & {X} \\
\hline
\end{tabularx}
\label{tab9}
\end{center}
\end{table}

\subsection {Place and situation (B1-B3)}

3 participants (ID 3, ID 7 and ID 15) answered the questions on location and situation (21.4\%). 11 participants did not provide any information in this group of questions (78.6\%). Table \ref{tab10} shows the answers on location and situation. Two participants were in the exhibition room with 3-5 other visitors and attended the event in the evening. Only one participant visited the event in the afternoon. One of the participants stayed for more than an hour, one participant between 45-60 minutes and one participant between 30-45 minutes. The average time spent by all three participants was therefore 50 minutes, each participant was present at the respective event with an average of three other visitors.

\begin{table}[htbp]
\caption{Place and situation (B1-B3)}
\begin{center}
\begin{tabularx}{1.0\columnwidth}{|p{0.05\textwidth}|p{0.075\textwidth}|p{0.075\textwidth}|X|}
\hline
\textbf{Answer ID} & \textbf{Visitors in the exhibition room} & \textbf{time of day}& \textbf{visit duration} \\
\hline
{ID 3} & {3-5} & {in the evening} & {30-45 minutes} \\
\hline
{ID 4} & {-} & {-} & {-} \\
\hline
{ID 5} & {-} & {-} & {-} \\
\hline
{ID 6} & {-} & {-} & {-} \\
\hline
{ID 7} & {3-5} & {in the evening} & {more than one hour} \\
\hline
{ID 8} & {-} & {-} & {-} \\
\hline
{ID 9} & {-} & {-} & {-} \\
\hline
{ID 10} & {-} & {-} & {-} \\
\hline
{ID 11} & {-} & {-} & {-} \\
\hline
{ID 12} & {-} & {-} & {-} \\
\hline
{ID 13} & {-} & {-} & {-} \\
\hline
{ID 14} & {-} & {-} & {-} \\
\hline
{ID 15} & {0-2} & {in the afternoon} & {45-60 minutes} \\
\hline
{ID 16} & {-} & {-} & {-} \\
\hline
\end{tabularx}
\label{tab10}
\end{center}
\end{table}

\subsection {Understanding of art (C1-C3)}

This group of questions deals with the understanding of art as an individual view on the artist, artworks and society. Of the nine answers given, four corresponded to a modern or traditional understanding of art (44.45\% each) and only one showed very little understanding of art (11.12\%). Table \ref{tab11} shows the answers to the individual understanding of art.

\begin{table}[htbp]
\caption{Individual understanding of art (C1-C3)}
\begin{center}
\begin{tabularx}{1.0\columnwidth}{|p{0.05\textwidth}|p{0.11\textwidth}|p{0.11\textwidth}|X|}
\hline
\textbf{Answer ID} & \textbf{Role of the artist (C1)} & \textbf{4'33'' by John Cage (C2)}& \textbf{Value of art (C3)} \\
\hline
{ID 3} & {Artists make us see things in a different light.} & {Absolute beauty and grandeur} & {Art is one of the most valuable cultural achievements.} \\
\hline
{ID 4} & {-} & {-} & {-} \\
\hline
{ID 5} & {-} & {-} & {-} \\
\hline
{ID 6} & {-} & {-} & {-} \\
\hline
{ID 7} & {Artists make us see things in a different light.} & {There is an artwork without an artist.} & {Art is one of the most valuable cultural achievements.} \\
\hline
{ID 8} & {-} & {-} & {-} \\
\hline
{ID 9} & {-} & {-} & {-} \\
\hline
{ID 10} & {-} & {-} & {-} \\
\hline
{ID 11} & {-} & {-} & {-} \\
\hline
{ID 12} & {-} & {-} & {-} \\
\hline
{ID 13} & {-} & {-} & {-} \\
\hline
{ID 14} & {-} & {-} & {-} \\
\hline
{ID 15} & {Artists make us see things in a different light.} & {The work makes no sense, it is absurd.} & {Art is one of the most valuable cultural achievements.} \\
\hline
{ID 16} & {-} & {-} & {-} \\
\hline
\end{tabularx}
\label{tab11}
\end{center}
\end{table}

Three participants (21.43\%) answered the question about the social role of the artist with answer option 1 (artists make us see things in a different light). This corresponds to a more traditional or romantic image of the artist. Answer 3 (Everyone is an artist) is representative of a modern understanding of art. Answer 2 (Artists are difficult for other people to understand) speaks for a rather poor understanding of art. The second question about the idea behind John Cage's work 4'33'' did not receive a uniform answer. To the same extent, one participant (7.14\%) opted for answer 1 (There is a work of art without an artist), one participant (7.14\%) for answer 2 (The work has no meaning, it is absurd) and one participant (7.14\%) for answer 3 (Absolute beauty and sublimity). Answer 1 corresponds to a modern understanding of art. Answer 2 indicates little understanding, answer 3 corresponds to a more traditional or romantic image of the artist. When asked about the value of art, three participants (21.43\%) responded with answer 3 (art is one of the most valuable cultural achievements). This answer corresponds to a high appreciation of art. None of the participants chose answer 1 or 2. 11 participants (78.57\%) did not provide any information in this group of questions. In the appendix \ref{appendix:questionnaire} you will find the corresponding answers and the graphical representations as a polar diagram.

\subsection {Modern approaches to art (D1-D3, E1-E3)}

Knowledge of modern approaches to art is assessed via two groups of questions. The fourth group of questions (D1-D3) asks about the modern basics of artistic practice, the fifth group of questions (E1-E3) asks about knowledge of the influences of other fields of knowledge on art. Table \ref{tab12} shows the answers to the modern foundations of artistic practice.

\begin{table}[htbp]
\caption{Modern basics of artistic practice (D1-D3)}
\begin{center}
\begin{tabularx}{1.0\columnwidth}{|p{0.05\textwidth}|p{0.11\textwidth}|p{0.11\textwidth}|X|}
\hline
\textbf{Answer ID} & \textbf{A major work by Josef Albers (D1)} & \textbf{Otl Aicher: ``How free is freedom? (D2)}& \textbf{Frei Ottos idea (D3)} \\
\hline
{ID 3} & {The art of color} & {Art is always also politics} & {"Connecting art with everyday life."} \\
\hline
{ID 4} & {-} & {-} & {-} \\
\hline
{ID 5} & {-} & {-} & {-} \\
\hline
{ID 6} & {-} & {-} & {-} \\
\hline
{ID 7} & {Color as a language} & {Art is free. There is only the free decision of the artist.} & {"Finding form" instead of "giving form" or "wanting form".} \\
\hline
{ID 8} & {-} & {-} & {-} \\
\hline
{ID 9} & {-} & {-} & {-} \\
\hline
{ID 10} & {-} & {-} & {-} \\
\hline
{ID 11} & {-} & {-} & {-} \\
\hline
{ID 12} & {-} & {-} & {-} \\
\hline
{ID 13} & {-} & {-} & {-} \\
\hline
{ID 14} & {-} & {-} & {-} \\
\hline
{ID 15} & {Interaction of Color} & {He deals with comprehensible design principles.} & {"Finding form" instead of "giving form" or "wanting form".} \\
\hline
{ID 16} & {-} & {-} & {-} \\
\hline
\end{tabularx}
\label{tab12}
\end{center}
\end{table}

A total of 3 participants (21.43\%) answered the question about the modern basics of artistic practice. The first question (Josef Albers describes color as the "most relative medium in art". Which book is one of his main works?) was answered very differently by all 3 participants. Only 1 participant (7.14\%) found the correct answer (Interaction of Color). "Art of Color" was written by Johannes Itten, "Color as a Language" was written by Hans Joachim Albrecht. The second question in this group of questions (Why does Otl Aicher ask himself the question: "How free is freedom?") was also answered correctly by only 1 participant (7.14\%) (He deals with comprehensible design principles). The third question (Frei Otto redefined the artistic design process. He advocated the following idea:) was answered correctly by 2 participants (14.29\%) ("finding form" instead of "giving form" or "wanting form".) Frei Otto's approach was also communicated to visitors at the face-to-face events (vernissage, artist's talk and privatissimum). None of the participants chose the answer "Less is more". 11 participants (78.57\%) did not provide any information in this group of questions. In the appendix \ref{appendix:questionnaire} you will find the corresponding answers and the graphical representations as a polar diagram. Below you will also find the results of the fifth group of questions (E1-E3) on knowledge of the influence of other fields of knowledge on art. Table \ref{tab13} shows a tabular list of the answers.

\begin{table}[htbp]
\caption{Influences of other fields of knowledge on art (E1-E3)}
\begin{center}
\begin{tabularx}{1.0\columnwidth}{|p{0.05\textwidth}|p{0.11\textwidth}|p{0.11\textwidth}|X|}
\hline
\textbf{Answer ID} & \textbf{What does Karl R. Popper's model of the 3 worlds deal with? (E1)} & \textbf{What do works of ``concrete art'' refer to? (E2)}& \textbf{The ``Pedagogical Sketchbook'' and the line in Paul Klee's work. (E3)} \\
\hline
{ID 3} & {The physical, psychological and intellectual reality} & {On creative laws of form, color, space, light and movement} & {His idea is only comprehensible to the artist himself.} \\
\hline
{ID 4} & {-} & {-} & {-} \\
\hline
{ID 5} & {-} & {-} & {-} \\
\hline
{ID 6} & {-} & {-} & {-} \\
\hline
{ID 7} & {The physical, psychological and intellectual reality} & {The physical, psychological and intellectual reality} & {There are objective, geometric-artistic principles.} \\
\hline
{ID 8} & {-} & {-} & {-} \\
\hline
{ID 9} & {-} & {-} & {-} \\
\hline
{ID 10} & {-} & {-} & {-} \\
\hline
{ID 11} & {-} & {-} & {-} \\
\hline
{ID 12} & {-} & {-} & {-} \\
\hline
{ID 13} & {-} & {-} & {-} \\
\hline
{ID 14} & {-} & {-} & {-} \\
\hline
{ID 15} & {The physical, psychological and intellectual reality} & {The physical, psychological and intellectual reality} & {There are objective, geometric-artistic principles.} \\
\hline
{ID 16} & {-} & {-} & {-} \\
\hline
\end{tabularx}
\label{tab13}
\end{center}
\end{table}

Three participants (21.43\%) answered questions E1 and E2 correctly. For question E3, two participants (14.29\%) chose the correct answer ``It is objective, geometric-artistic principles''. Only one participant (7.14\%) answered question E3 with the option that the idea is only comprehensible to the artist himself. 11 participants (78.57\%) did not provide any information in this group of questions.\\

\subsection {Media use and individual activity (F1-F5)}

Below you will find the results of question groups F1-F5 on media use and individual activity. For better understanding, question group F3 (media use) is at the beginning of this chapter. This is followed by question groups F1-F2 and F4-F5 (individual activity). Table \ref{tab14} shows a summary of the results of question group F3. The answer options in this table are assigned to the known media types (primary, secondary, tertiary and quaternary media).

Regarding media use - all three participants (21.43\%) took part in a personal interview. Only one participant (7.14\%) used the exhibition catalog. None of the participants used the analog questionnaire. Only two participants stated that they had taken part in the online survey. However, all three participants (21.43\%) actually took part in the online survey. Instagram and the social media wall were only used by one participant (7.14\%). Two participants (14.29\%) stated that they also used other social media and the internet. None of the participants used a library or bookshop for research. 11 participants (78.57\%) did not answer questions F3.

\begin{table}[htbp]
\caption{Media use (F3)}
\begin{center}
\begin{tabularx}{1.0\columnwidth}{|p{0.05\textwidth}|p{0.035\textwidth}|p{0.035\textwidth}|p{0.035\textwidth}|p{0.035\textwidth}|p{0.035\textwidth}|p{0.035\textwidth}|X|}
\hline
\textbf{Answer ID} & \textbf{Per-sonal inter-view (P)} & \textbf{Exhi-bition cata-logue (S)}& \textbf{Ana-logue ques-tion-naire (S)} & \textbf{On-line survey (Q)} & \textbf{Insta-gram / Social media wall (Q)} & \textbf{Other social media / Internet (Q)} & \textbf{Li-brary / Book-shop (S)} \\
\hline
{ID 3} & {X} & {-} & {-} & {-} & {-} & {-} & {-} \\
\hline
{ID 4} & {-} & {-} & {-} & {-} & {-} & {-} & {-} \\
\hline
{ID 5} & {-} & {-} & {-} & {-} & {-} & {-} & {-} \\
\hline
{ID 6} & {-} & {-} & {-} & {-} & {-} & {-} & {-} \\
\hline
{ID 7} & {X} & {-} & {-} & {X} & {-} & {X} & {-} \\
\hline
{ID 8} & {-} & {-} & {-} & {-} & {-} & {-} & {-} \\
\hline
{ID 9} & {-} & {-} & {-} & {-} & {-} & {-} & {-} \\
\hline
{ID 10} & {-} & {-} & {-} & {-} & {-} & {-} & {-} \\
\hline
{ID 11} & {-} & {-} & {-} & {-} & {-} & {-} & {-} \\
\hline
{ID 12} & {-} & {-} & {-} & {-} & {-} & {-} & {-} \\
\hline
{ID 13} & {-} & {-} & {-} & {-} & {-} & {-} & {-} \\
\hline
{ID 14} & {-} & {-} & {-} & {-} & {-} & {-} & {-} \\
\hline
{ID 15} & {X} & {X} & {-} & {X} & {X} & {X} & {-} \\
\hline
{ID 16} & {-} & {-} & {-} & {-} & {-} & {-} & {-} \\
\hline
\end{tabularx}
\label{tab14}
\end{center}
\end{table}

\begin{table}[htbp]
\caption{Individual Activity I (F1-F2)}
\begin{center}
\begin{tabularx}{1.0\columnwidth}{|p{0.055\textwidth}|p{0.055\textwidth}|p{0.055\textwidth}|p{0.055\textwidth}|p{0.055\textwidth}|X|}
\hline
\textbf{Answer ID} & \textbf{Ver-nissage  (P\textsubscript V)} & \textbf{Artist's talk (P\textsubscript K)} & \textbf{Privatis-simum (P\textsubscript P)} & \textbf{Contact with the artist} & \textbf{Contact with other visitors} \\
\hline
{ID 3} & {X} & {-} & {X} & {X} & {2-5} \\
\hline
{ID 4} & {-} & {-} & {-} & {-} & {-} \\
\hline
{ID 5} & {-} & {-} & {-} & {-} & {-} \\
\hline
{ID 6} & {-} & {-} & {-} & {-} & {-} \\
\hline
{ID 7} & {-} & {X} & {-} & {X} & {2-5} \\
\hline
{ID 8} & {-} & {-} & {-} & {-} & {-} \\
\hline
{ID 9} & {-} & {-} & {-} & {-} & {-} \\
\hline
{ID 10} & {-} & {-} & {-} & {-} & {-} \\
\hline
{ID 11} & {-} & {-} & {-} & {-} & {-} \\
\hline
{ID 12} & {-} & {-} & {-} & {-} & {-} \\
\hline
{ID 13} & {-} & {-} & {-} & {-} & {-} \\
\hline
{ID 14} & {-} & {-} & {-} & {-} & {-} \\
\hline
{ID 15} & {-} & {X} & {-} & {X} & {2-5} \\
\hline
{ID 16} & {-} & {-} & {-} & {-} & {-} \\
\hline
\end{tabularx}
\label{tab15}
\end{center}
\end{table}

The group of questions (F1-F2) asks about characteristics of creativity in the form of individual activity and media use. Questions F1 and F2 are intended to provide information on attendance at face-to-face events and the number of individual contacts. Table \ref{tab15} shows a tabular list of the results of questions F1-F2 on individual activity I. One participant (7.14 \%) stated that they had attended the vernissage and the finissage. Unfortunately, there was no finissage in this form. It is therefore assumed that the participant probably took part in the privatissimum instead. Two participants (14.29 \%) stated that they took part in the artist's talk. All three participants (21.43\%) had contact with the artist as well as with 2-5 other people during the face-to-face events. 11 participants (78.57\%) did not answer questions F1-F2.

\begin{table}[htbp]
\caption{Individual Activity II (Creativity)(F4-F5)}
\begin{center}
\begin{tabularx}{1.0\columnwidth}{|p{0.05\textwidth}|p{0.11\textwidth}|X|}
\hline
\textbf{Answer ID} & \textbf{Did the exhibition inspire you artistically? Did you learn something new? Leave us a note or comment here. (F4)} & \textbf{Here you can leave us your individual artistic sketch or a photo. (F5)} \\
\hline
{ID 3} & {-} & {-} \\
\hline
{ID 4} & {-} & {-} \\
\hline
{ID 5} & {-} & {-} \\
\hline
{ID 6} & {-} & {-} \\
\hline
{ID 7} & {I found the exhibition very exciting and highly informative, especially the conversation with the artist.} & {-} \\
\hline
{ID 8} & {-} & {-} \\
\hline
{ID 9} & {-} & {-} \\
\hline
{ID 10} & {-} & {-} \\
\hline
{ID 11} & {-} & {-} \\
\hline
{ID 12} & {-} & {-} \\
\hline
{ID 13} & {-} & {-} \\
\hline
{ID 14} & {-} & {-} \\
\hline
{ID 15} & {Yes, a connection between images that brings art even more out of the artwork into reality - and vice versa ...} & {-} \\
\hline
{ID 16} & {-} & {-} \\
\hline
\end{tabularx}
\label{tab16}
\end{center}
\end{table}

Below you will find the results for questions F4-F5. We asked about individual activity as a characteristic of creativity. Table \ref{tab16} shows a tabular list of the answers. Only two out of three participants (21.43\%) answered question F4. Three out of 14 participants (21.43\%) did not answer question F5 about an artistic sketch or photo. 11 participants (78.57\%) did not complete question F5. In the appendix \ref{appendix:questionnaire} you will find all answers and graphical representations as a polar diagram.\\

\section{Discussion of the results of the online survey}

\subsection {Representativity, response rate and completion rate}

The response rate of the online survey is calculated below. In our case, the response rate (M\textsubscript {RQ}) is the ratio between interactions (M\textsubscript {IA}) and range (R\textsubscript{NG}). The range of the online survey is 177 people (see table \ref{tab4}). Of these 177 people, 14 people took part in the online survey. This results in a response rate (M\textsubscript {RQ}) of 7.9\%, as shown in formula (\ref{eqn:6}). According to Theobald, this response rate is in the lower range of the currently usual rates of approx. 5-20\% \cite{b55}.

\begin{equation}
M\textsubscript {RQ} = \frac{M\textsubscript {IA}}{R\textsubscript{NG}} = \frac{14}{177} = 7,9\%
\label{eqn:6}
\end{equation}

A total of 3 out of 14 participants took part in this online survey, 11 participants did not complete the survey. The completion rate is therefore 21.34\%, as shown in the formula (\ref{eqn:7}). As a rule, the range between 20-30\% is considered a good completion rate. At 21.43\%, our online survey is still in the lower range of a good completion rate.

\begin{equation}
C\textsubscript {OR} = \frac{Completers}{Participants} = \frac{3}{14} = 21,43\%
\label{eqn:7}
\end{equation}

In general, the sample of 3 participants is too small to be able to make reliable statements. For a representative result, at least 65-70\% of the total of 177 people reached would have had to take part. At best, that would have been 115-125 people. Therefore, although our online survey reveals certain tendencies, it is not very representative in terms of the total number of people reached.\\

\subsection {Media use, creativity and understanding of art}

In this section, the results are discussed according to the first of the two hypotheses guiding the research.\\

\begin{itemize}
\item There are correlations between media use (modality), the participants' interactions (creativity) and their perception (understanding of art)\\
\end{itemize}

There are two relations that are examined in this context: (A) the relationship between the proportion of media types (modality) used and interactions (Individual Activities) and (B) the relationships between the proportion of media types (modality) used and art appreciation. The mixing ratio of all media types (modality) used can be calculated using the formula (\ref{eqn:1}).

A total of 21 media were queried in the online survey. In addition to the 19 media mentioned in the media analysis, libraries, bookshops, other social media and internet use were also surveyed. There were four types of media used (primary, secondary, tertiary and quaternary media) Of the 21 media used, 9 were quaternary media (42.86\%), 7 were secondary media (33.34\%), 3 were primary media (14.28\%) and 2 were tertiary media (9.52\%). Within these four media types, an overwhelming majority of participants opted for quaternary media ahead of secondary media, primary media and tertiary media. Table \ref{tab22} shows an overview of the proportion of media types (modality) used, according to the information provided by the participants in the online survey.

\begin{table}[htbp]
\caption{Mix of media types (Modality) in the online survey}
\begin{center}
\begin{tabularx}{1.0\columnwidth}{|p{0.027\textwidth}|p{0.12\textwidth}|p{0.15\textwidth}|X|}
\hline
\textbf{} & \textbf{Media types} & \textbf{Used media} & \textbf{Mix of media types (Modality)} \\
\hline
{Q \textsubscript S}  & {Quarternary media} & {Online survey} & {42,86\%} \\
\hline
{Q \textsubscript A} & {} & {Other social media and internet} & {-} \\
\hline
{Q \textsubscript W} & {} & {Social media wall (Blog)} & {-} \\
\hline
{Q \textsubscript I}  & {} &{Instagram} & {-} \\
\hline
{Q \textsubscript C} & {}  & {Online exhibition catalogue} & {-} \\
\hline
{Q \textsubscript M} & {} & {Online media portal} & {-} \\
\hline
{Q \textsubscript X} & {} & {X (Twitter)} & {-} \\
\hline
{Q \textsubscript L} & {} & {LinkedIn} & {-} \\
\hline
{Q \textsubscript Y} & {} & {YouTube} & {-} \\
\hline
{S \textsubscript C} & {Secondary media} & {Exhibition catalogue} & {33,34\%} \\
\hline
{S \textsubscript B} & {} & {Library / Bookshop} & {-} \\
\hline
{S \textsubscript P} & {} & {Art postcards} & {-} \\
\hline
{S \textsubscript Q} & {} & {Questionnaires} & {-} \\
\hline
{S \textsubscript I} & {} & {Invitation cards} & {-} \\
\hline
{S \textsubscript Z} & {} & {Newspaper article} & {-} \\
\hline
{S \textsubscript A} & {} & {Artworks} & {-} \\
\hline
{P \textsubscript V} & {Primary media} & {Vernissage} & {14,28\%} \\
\hline
{P \textsubscript K} & {} & {Artist's talk} & {-} \\
\hline
{P \textsubscript P}& {}  & {Privatissimum} & {-} \\
\hline
{T \textsubscript T} & {Tertiary media} & {Telephone / Mobile} & {9,52\%} \\
\hline
{T \textsubscript E} & {} & {e-mails} & {-} \\
\hline
\end{tabularx}
\label{tab22}
\end{center}
\end{table}

According to the information in the table "Individual activity I", all three participants took part in several events as part of the art exhibition and had contact with 2-5 other visitors. The table "Individual activity II (creativity)" indicates that only two participants (21.43\%) left a comment. None of the participants left a sketch. Of six possible interactions, only two were noticed (33.34\%). A clear relationship was found between media use and interactions. The participants with higher media usage also showed a higher level of individual activity. They were more active in visiting the exhibition, in overall media activity and left comments (creativity). If we look at media use and art appreciation, no reliable correlations could be established between individual media use and modality and art appreciation.

\subsection {Understanding of art and individual parameters}

In the next section, the results are discussed according to the second of the two hypotheses guiding the research.\\

\begin{itemize}
\item Individual parameters (demographic data, location and situation, individual knowledge) influence perception (understanding of art)\\
\end{itemize}

There are individual parameters that can influence the understanding of art. In addition to demographic data (age, gender and level of education), these include one's own relationship to art, location and situation, knowledge of modern approaches to art and knowledge of the influences of other fields of knowledge.

Demographic data and understanding of art - the genders of all participants were relatively balanced. Both men and women showed roughly the same modern or traditional understanding of art. Taken together, the two age groups 40-49 and 50+ were the most strongly represented (50\%). This was followed by the age groups 20-29 (37.50\%) and 30-39 (12.50\%). The three participants who completed the survey all came from the 40-49 and 50+ age groups. In these two age groups, modern and traditional art appreciation was also balanced. The educational qualifications were different for all 3 participants. The two participants with a technical diploma/high school graduation and with a technical college/university degree showed a more traditional understanding of art. The participant with professional training/dual education showed a more modern understanding of art. Neither gender, age nor educational qualifications provided any clear indication of the relationship between these study parameters and art appreciation.

Individual reference to art and understanding of art - All three participants described themselves as "interested in art". They preferred landscape painting and still life, figurative art, abstract art, media art and performance. None of the three participants were interested in concrete art. No clear correlation was found between the individual's relationship to art and their understanding of art.

Place and situation and understanding of art - All three participants attended events as part of the art exhibition. The data on contacts with other visitors, the time of day and the duration of the visit do not allow any conclusions to be drawn about relationships to individual art appreciation.

Knowledge of modern approaches to art and understanding of art - Four out of nine answer options (44.45\%) were correct. One participant had a high level of knowledge of the modern principles of artistic practice, the other two participants rather less. However, this knowledge had no discernible effect in relation to art appreciation.

Influences of other fields of knowledge and understanding of art - There were eight out of nine (88.89\%) correct answers. All participants were very familiar with the influences of other fields of knowledge. Only one participant had slightly less good knowledge. Knowledge of the influences of other fields of knowledge was not very well reflected in the understanding of art.

In general, it can be said that the individual parameters are suitable for influencing the understanding of art. However, our survey results do not allow any clear conclusions to be drawn about specific correlations between the above-mentioned individual parameters and art appreciation.\\

\section {Conclusion, findings and perspectives}



The joint description and representation of real and virtual places in our spatiotemporal model yielded interesting insights. The usual basic understanding of a division into virtual and real places is inadequate and perhaps even misleading when describing hybrid spaces. The division into three levels (places and media types, time and interaction, perception) proved to be very operational. The unification of real and virtual places on one level provides approaches to a new basic understanding of the relationships between real and virtual places, range and urban density. New correlations to means of transportation, mobility and even multilocality emerge, become understandable and can be viewed from a different perspective.

Unfortunately, the results of the online survey are not very representative due to the low level of participation. Nevertheless, correlations between the levels of media use, time and interaction were identified. The participants with higher media usage also showed greater individual activity. Correlations were also found at the perception level. The individual parameters (e.g. demographic data, location and situation, individual knowledge) are generally suitable for influencing perception (understanding of art). However, the results did not allow any clear conclusions to be drawn.

Mapping different topologies of hybrid spaces according to the parameters of range, interactions and response rate proved to be very fruitful. Three different dynamic topologies of the same hybrid space could be successfully represented in bubble diagrams. These dynamic topologies form different patterns, sequences and rhythms. They thus form an operable basis for a deeper understanding of hybrid spaces. Three main findings emerge, which are discussed in detail in the following sections:\\

\begin{itemize}
\item Dynamic topologies of hybrid spaces could be successfully mapped as projections.
\item Two different hybrid spaces were identified.
\item Perspectives on location, range, urban density, mobility and multilocality.
\end{itemize}

\subsection {Dynamic topologies of hybrid spaces}

\begin{figure*}[htbp]
\begin{minipage}[b]{1.0\textwidth}
\centerline{\includegraphics[width=1.0\textwidth]{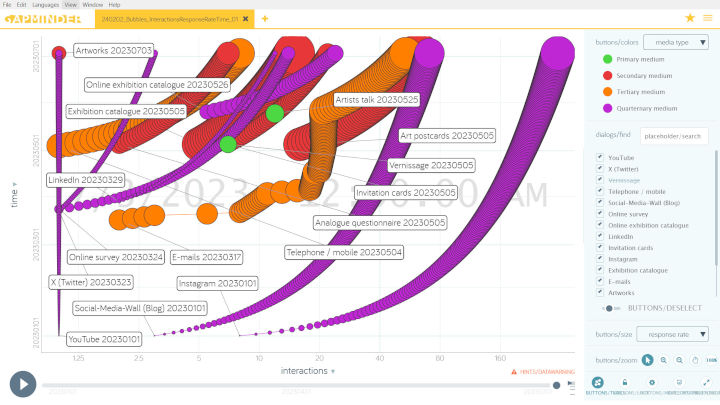}}
\caption{Level of time and interaction $\vert$ Response rate and interactions as incidents in time}
\label{fig21}
\end{minipage}
\end{figure*}

\begin{figure*}[htbp]
\begin{minipage}[b]{1.0\textwidth}
\centerline{\includegraphics[width=1.0\textwidth]{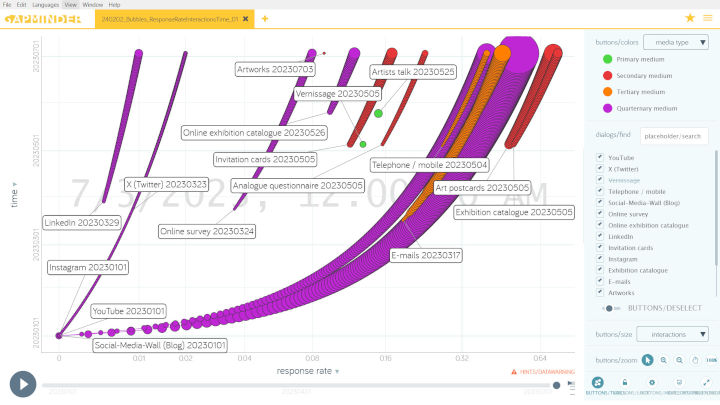}}
\caption{Level of time and interaction $\vert$ Interactions and response rate as incidents in time}
\label{fig22}
\end{minipage}
\end{figure*}

\begin{figure*}[htbp]
\begin{minipage}[b]{1.0\textwidth}
\centerline{\includegraphics[width=1.0\textwidth]{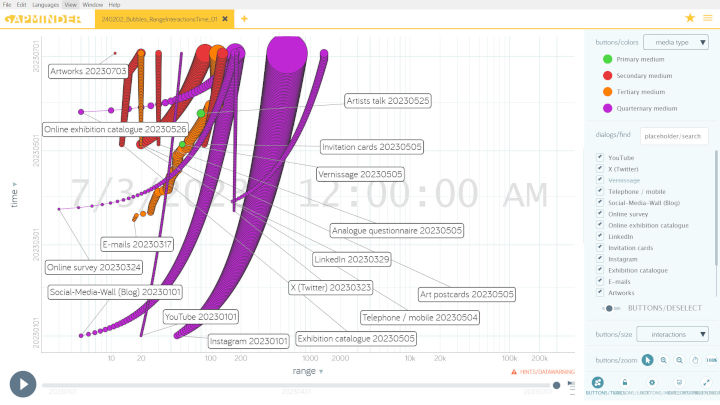}}
\caption{Level of time and interaction $\vert$ Interactions and range as incidents in time}
\label{fig23}
\end{minipage}
\end{figure*}

At the level of time and interaction, different topologies of hybrid spaces can be mapped according to the parameters of range, interaction and response rate. The colored representation simplifies the understanding of the different topologies of hybrid spaces. The representation at the level of time and interaction is generically dynamic. It can be displayed simultaneously with the spatial parameters at the level of locations and media types. In this way, it creates a basis for generalizable and scalable patterns for comparing different topologies of hybrid spaces. Figures \ref{fig21}, \ref{fig22} and \ref{fig23} show two-dimensional projections of different topologies of Hybrid Space I (media analysis) over the plane of time and interaction. The results are a two-dimensional projection of dynamic three-dimensional topologies of hybrid spaces as specific patterns or sequences.

\subsection {Two different hybrid spaces}

Two different hybrid spaces were identified: the hybrid space of media analysis (hybrid space I) and the hybrid space of the online survey (hybrid space II). Hybrid spaces can be differentiated according to the total number of participants (users), the media used and the mix of media types (modality). Both hybrid rooms differed in the total number of participants (177 and 14), the media used (19 and 21) and the mix of media types used. Table \ref{tab19} shows a comparison of the proportion of media types used in the media analysis and online survey. Both hybrid spaces are very similar in terms of the proportion of media types (modality). In both hybrid spaces, quaternary media were preferred, even ahead of secondary and primary media. Tertiary media came last in both hybrid spaces.

\begin{table}[htbp]
\caption{Two different hybrid spaces $\vert$ Mix of media types (Modality)}
\begin{center}
\begin{tabularx}{1.0\columnwidth}{|p{0.027\textwidth}|p{0.12\textwidth}|p{0.15\textwidth}|X|}
\hline
\textbf{} & \textbf{Media types} & \textbf{Media analysis} & \textbf{Online survey} \\
\hline
{Q \textsubscript S}  & {Quarternary media} & {42,10\%} & {42,86\%} \\
\hline
{S \textsubscript C} & {Secondary media} & {31,58\%} & {33,34\%} \\
\hline
{P \textsubscript V} & {Primary media} & {15,79\%} & {14,28\%} \\
\hline
{T \textsubscript T} & {Tertiary media} & {10,53\%} & {9,52\%} \\
\hline
\end{tabularx}
\label{tab19}
\end{center}
\end{table}

The concept of modality opens up the model to means of transportation and mobility. In both areas (transport/media), the concept of modality can be applied simultaneously and in the same function. This means that the level of places and media types can in future be supplemented by means of transportation and expanded to include the concept of mobility.

\subsection {Location, range, urban density, mobility and multilocality}

Range describes the distribution of people reached across real and virtual locations. It becomes interesting when we consider media range as an equivalent to spatial or urban density. Urban density is one of the fundamental parameters of urban and spatial planning. Urban density is usually expressed in inhabitants per unit area (e.g: hectare (E/ha)). In spatial terms, this corresponds to a number of people per real place. With the net range of a medium (e.g.: number of people reached/medium), we thus have an equivalent to urban density. This number of people at a virtual location (medium) is thus in principle comparable with the spatial density in real space. The range therefore corresponds to a certain density of people per real or virtual location.\\

The following figure \ref{fig20} ``Range and spatial density'' shows the content of table \ref{tab3} graphically overlaid with the range data of the individual media. The number of people reached is shown in colored circular areas (1 person $\widehat{=}$ 1 pixel diameter). The media types are displayed in color, as before in our space-time model: Primary media = green, Secondary media = red, Tertiary media = orange, Quarternary media = purple. The long range of the quaternary media (purple) is immediately recognizable (see Fig. \ref{fig19}). The secondary (red) and tertiary media (orange) are just behind it. The small green dots of the primary media appear only occasionally in the graph. The enormous range of the online media portal ``Merkur.de'' (306,748) \cite{b59} and the newspaper article in the Starnberger Merkur (9,060) \cite{b57} immerses the entire graphic in a red-violet color. The diameter of the corresponding circular areas significantly exceeds the size of the entire graphic. If you look at the range by location, you can see that the ranges without a location (virtual locations) dominate the entire graphic. The range data that can be clearly assigned to specific spatial locations disappear behind them.\\

\begin{figure*}[htbp]
\begin{minipage}[b]{1.0\textwidth}
\centerline{\includegraphics[width=1.0\textwidth]{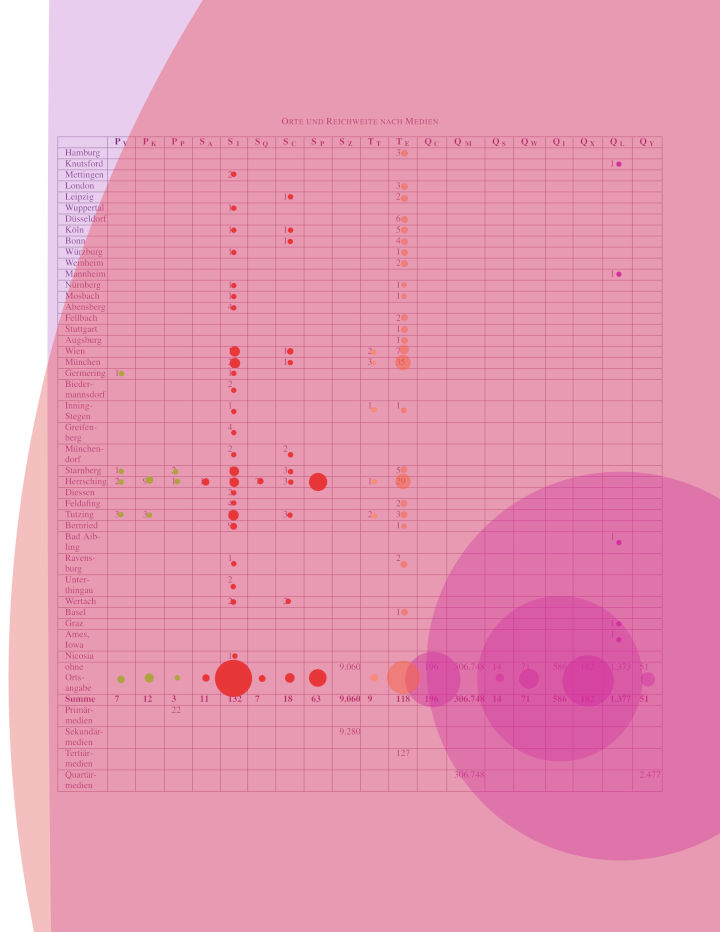}}
\caption{Range and spatial density (compare to table \ref{tab3}}
\label{fig20}
\end{minipage}
\end{figure*}

Using the 39 identified locations around the world and the range (R\textsubscript {NG}) according to the formula (\ref{eqn:2}), we can therefore calculate an equivalent to a spatial density of people per location (e.g. in our case: Munich = 65 people/location). With this approach, however, it is also possible to specify a ``spatial density'' in all virtual media. Figure \ref{fig6} clearly shows that this density in virtual space exceeds the conventional spatial density of comparable real locations (primary media) many times over in many cases (e.g.: LinkedIn = 1,377 people/location).

In general, all media in the space-time model are perceived as places in hybrid spaces. The merging of real and virtual places with the types of media opened up a new understanding of these relationships. All media are places. They always have been. Since the first Stone Age images, the invention of writing, later the printing press, the daily newspaper, in radio and television and in the hybrid spaces of digital media.

The space-time model thus also opens up new perspectives on the everyday perception of space in urban planning and architecture. The connection between range and urban density proved to be very interesting. This reflects a possible equivalent to a ``people density'' in a virtual place. The idea that real and virtual places can be understood as a function of the range of a medium is also new. Means of transportation are not explicitly mentioned in the space-time model, but are a prerequisite for the functioning of primary media. In principle, the reach of primary media therefore already includes the means of transportation used or the mobility of users. Mobility can therefore be seen as the equivalent of interactions in media communication.

Figure \ref{fig20} ``Reach and spatial density'' clearly shows the simultaneous multilocality of media events in hybrid space. We experience this multilocality today not only in our dealings with the media. Many authors even relate the phenomenon of multilocality closely to real mobility \cite{b60,b61}. This investigation of the integration of means of transportation and mobility would be the task of further research.\\

\begin{quote}
"Multilocality means vita activa in several places: the active everyday life is distributed in its entirety over several places, which are visited in more or less large periods of time and used with a more or less large degree of functional diversity." \cite{b61}\\
\end{quote}

The dynamic topologies of hybrid spaces, as shown in the figures \ref{fig21}, \ref{fig22} and \ref{fig23}, show certain patterns. They represent events. Events in time form certain rhythms. These sequences and rhythms can be analyzed, evaluated and measured. Further work can clarify whether we can arrive at a new and deeper understanding of hybrid spaces in the future by comparing different patterns, sequences or rhythms. It is interesting to ask to what extent this also opens up a new understanding of architecture and urban space.

\newpage
\section*{Acknowledgments}

My gratitude goes to the municipality of Herrsching and to the cultural officer Hans-Hermann Weinen for the invitation to this art exhibition and all organizational matters. I would also like to thank Mr. Maximilian Pfertner from the Chair of Urban Structure and Transport Planning, TUM School of Engineering and Design at the Technical University of Munich (TUM). He made it possible for me to use the online survey tool "Lime Survey" and was always available to answer technical questions. This work would not have been possible without the extensive and valuable collection and the useful online services of the University Library of the Technical University of Munich.

\section*{Disclosure Statement}

The entire project was freely financed. No other financial resources were applied for or used. This work is therefore also free of potential other and competing interests.

\section*{Disclaimer}

The author wishes to exclude discrimination on the reasons of ethnic origin, gender, religion or ideology, disability, age or sexual identity in the content of this work. For the sake of correct linguistic syntax and grammar, for better readability and to promote textual comprehension, no form of gender-inclusive language is used in this work. All generic formulations used in this work address all genders in an equal way. Should the contents of this work be perceived as discriminatory, this is not intended by the author.


© 2024 by the author. This publication is licensed under the terms and conditions of the Creative Commons Attribution  license (CC BY-NC-ND). \url{https://creativecommons.org/licenses/by-nc-nd/4.0/}

\begin{wrapfigure}{l}{2.1cm}
\includegraphics[width=0.35\columnwidth]{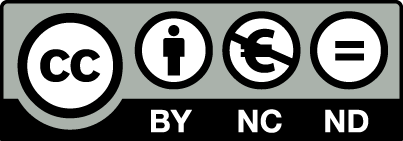}\\
\end{wrapfigure}

\cleardoublepage
\begin{appendices}

\section{Raw data on questionnaires and results}
\label{appendix:questionnaire}

On the following pages you will find the original documents for three raw data sets:\\

\begin{itemize}
\item Blank copy of the questionnaire (analog/online) (5 pages)
\item Results of the online survey on Lime Survey with graphical representations as a polar diagram (22 pages) 
\item All completed questionnaires of the online survey with the original answers (22 pages) \\
\end{itemize}

On the following five pages you will find the raw data sets for the questionnaires. The first raw data set is a blank copy of the questionnaire (analog/online). On 22 further pages you will find a second raw data set containing all the results of the survey in a graphical representation as a polar diagram. This is a complete presentation of the results of the online survey on Lime Survey. The third raw data set at the end of the appendix contains all completed original questionnaires with all responses from the 14 participants, also on 22 pages. All raw data on traffic on the social media wall was documented separately in the months of September 2022 to August 2023 via the author's blog. The author also has a pictorial documentation of the 28 posts under the Instagram hashtag \#kunstwirklichkeiten and the other social media activities.


\end{appendices}
\cleardoublepage



\section*{Supplementary material}

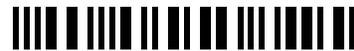

# Teil A: Angaben zur Person

**A1.** **Persönliche Anrede**

Herr ☐

Frau ☐

keine Angabe ☐

**A2.** **Wie alt sind Sie?**

0-9 Jahre ☐

10-19 Jahre ☐

20-29 Jahre ☐

30-39 Jahre ☐

40-49 Jahre ☐

50+ Jahre ☐

**A3.** **Was ist Ihr höchster Bildungsabschluß?**

Grund- / Hauptschule ☐

Mittlere Reife ☐

Fachabitur / Abitur ☐

Berufsausbildung / Duale Ausbildung ☐

Fachhochschule / Universität ☐

Promotion ☐

**A4.** **Wie ist Ihr Bezug zur Kunst?**

Kunstinteressiert ☐

Amateur ☐

Professional ☐

**A5.** **Welche Kunstwerke können Sie am Besten verstehen?**

Landschaftsmalerei und Stilleben ☐

Gegenständliche Kunst ☐

Abstrakte Kunst ☐

Konkrete Kunst ☐

Medienkunst und Performance ☐

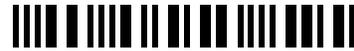

## Teil B: Ort und Situation

**B1.** **Wie viele Personen waren gleichzeitig mit Ihnen im Ausstellungsraum?**

- 0-2 ☐
- 3-5 ☐
- 5-10 ☐
- 10-15 ☐
- mehr als 15 ☐

**B2.** **Zu welcher Tageszeit haben Sie die Ausstellung besucht?**

- morgens ☐
- vormittags ☐
- mittags ☐
- nachmittags ☐
- abends ☐

**B3.** **Wie lange dauerte Ihr Besuch?**

- 0-15 Minuten ☐
- 15-30 Minuten ☐
- 30-45 Minuten ☐
- 45-60 Minuten ☐
- mehr als eine Stunde ☐

## Teil C: Künstler, Kunstwerk und Gesellschaft

**C1.** **Wie sehen Sie die gesellschaftliche Rolle des Künstlers?**

- Künstler lassen uns Dinge in einem anderen Licht sehen. ☐
- Künstler sind für andere Menschen schwer zu verstehen. ☐
- Jeder Mensch ist ein Künstler. ☐

**C2.** **Was ist der Gedanke hinter dem Werk 4´33'' von John Cage?**

- Es gibt ein Kunstwerk ohne einen Künstler. ☐
- Das Werk hat keinen Sinn, es ist absurd. ☐
- Absolute Schönheit und Erhabenheit. ☐

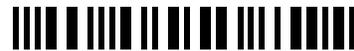

**C3.** **Der Wiener Architekt Josef Frank bezeichnet Kunst als „zwecklos", aber auch als „Endzweck aller übrigen Tätigkeit". Wie ist Ihre Meinung?**

- Kunst ist nur wertvoll, wenn sie handwerklich gut gemacht ist. ☐
- Manche Kunstwerke sehen für mich zwecklos aus. ☐
- Kunst gehört zu den wertvollsten kulturellen Errungenschaften. ☐

## Teil D: Grundlagen künstlerischer Praxis

**D1.** **Josef Albers beschreibt Farbe als „relativstes Medium in der Kunst". Welches Buch ist eines seiner Hauptwerke?**

- Kunst der Farbe ☐
- Interaction of Color ☐
- Farbe als Sprache ☐

**D2.** **Warum stellt sich Otl Aicher die Frage: „Wie frei ist die Freiheit?"**

- Die Kunst ist frei. Es gibt nur die freie Entscheidung des Künstlers. ☐
- Er beschäftigt sich mit nachvollziehbaren Gestaltungsprinzipien. ☐
- Kunst ist immer auch Politik. ☐

**D3.** **Frei Otto definierte den künstlerischen Gestaltungsprozeß neu. Er vertrat folgende Idee:**

- "Kunst mit dem alltäglichen Leben zu verbinden." ☐
- "Weniger ist mehr." ☐
- "Form finden" statt "Form geben" oder "Form wollen". ☐

## Teil E: Einflüsse anderer Wissensgebiete

**E1.** **Was behandelt das Modell der 3 Welten von Karl R. Popper?**

- den Unterschied zwischen Industrie- und Schwellenländern ☐
- die physische, psychische und geistige Wirklichkeit ☐
- die Darstellung des Universums in der Wissenschaft ☐

**E2.** **Max Bill ist ein Vertreter der „Konkreten Kunst". Worauf beziehen sich Werke dieser Kunstrichtung?**

- auf gestalterische Gesetzmäßigkeiten von Form, Farbe, Raum, Licht und Bewegung ☐
- auf Abstraktionen von anderen Dingen ☐
- auf eine konkrete Bedeutung, die nur der Künstler kennt ☐

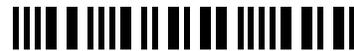

**E3.** **In seinem „Pädagogischen Skizzenbuch" beschäftigt sich Paul Klee mit der Richtungsänderung einer Linie. Was bewegt ihn dazu?**

- Nur die Übung macht den Meister. ☐
- Seine Idee ist nur für den Künstler selbst nachvollziehbar. ☐
- Es sind objektive, geometrisch-künstlerische Grundlagen. ☐

# Teil F: Individuelle Aktivitäten

**F1.** **Welche Veranstaltungen haben Sie besucht?**

- Vernissage ☐
- Künstlergespräch ☐
- Finissage ☐

**F2.** **Wie viele Kontakte hatten Sie im Rahmen der Ausstellung?**

- mit dem Künstler ☐
- mit 2-5 anderen Besuchern, Bekannten oder Freunden ☐
- mit mehr als 5 anderen Personen ☐

**F3.** **Welche Medien haben Sie zur Interaktion genutzt?**

- Persönliches Gespräch ☐
- Ausstellungskatalog ☐
- Analoger Fragebogen ☐
- Online-Fragebogen ☐
- Instagram / Social Media Wall ☐
- Andere Soziale Medien / Internet ☐
- Bibliothek / Bücherei ☐

**F4.** **Hat Sie die Ausstellung künstlerisch inspiriert? Haben Sie etwas Neues erfahren? Hinterlassen Sie uns hier eine Anmerkung oder einen Kommentar.**

[ ]

**F5.** **Hier können Sie uns eine eigene künstlerische Skizze oder ein Foto hinterlassen.**

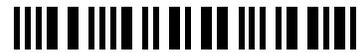

**Wirklichkeiten | Realities . Kunstausstellung | 1. Mai - 30. Juni 2023 Sitzungssaal im Rathaus . Herrsching am Ammersee Vielen Dank für Ihre Teilnahme! Sie können das Browserfenster jetzt schließen.**

## Zusammenfassung für A1

### Persönliche Anrede

| Antwort | Anzahl | Brutto-Prozentsatz |
|---|---|---|
| Herr (1) | 4 | 28.57% |
| Frau (2) | 3 | 21.43% |
| keine Angabe (3) | 1 | 7.14% |
| Keine Antwort | 0 | 0.00% |
| Nicht beendet oder nicht gezeigt | 6 | 42.86% |
| **Gesamt(Brutto)** | **14** | **100.00%** |

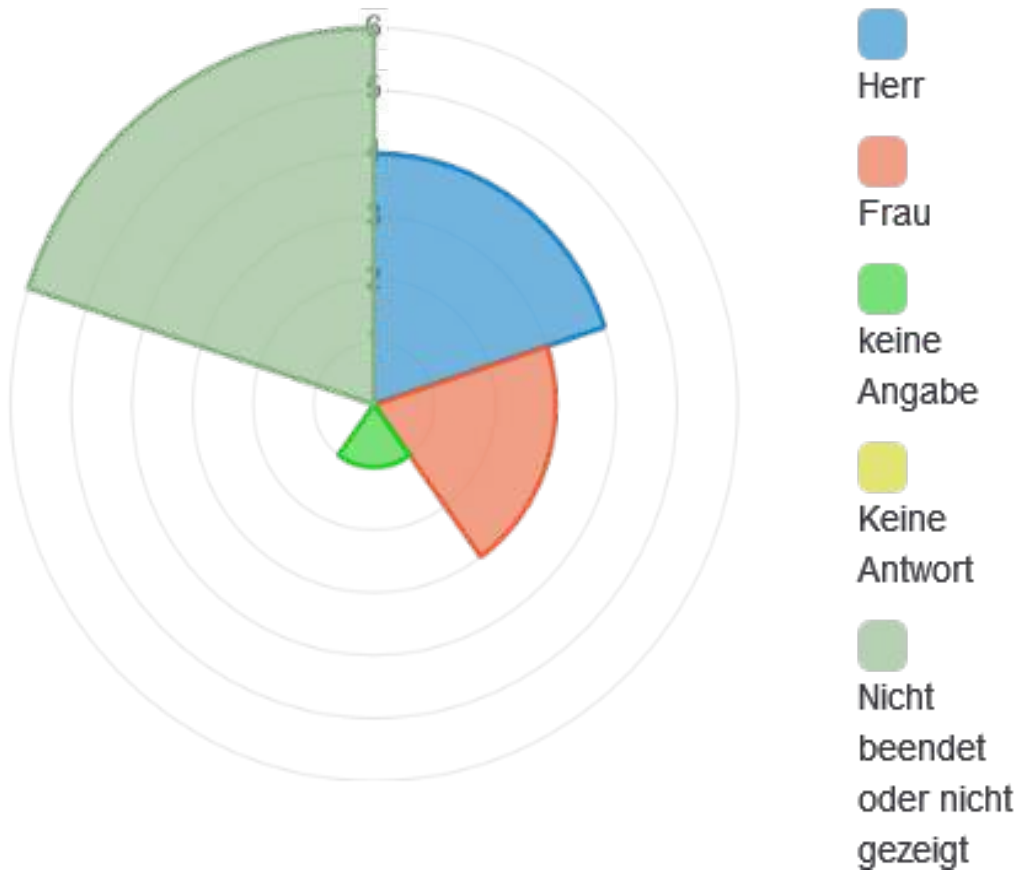

- Herr
- Frau
- keine Angabe
- Keine Antwort
- Nicht beendet oder nicht gezeigt

| Zusammenfassung für A2 |||
|---|---|---|
| Wie alt sind Sie? |||
| **Antwort** | **Anzahl** | **Brutto-Prozentsatz** |
| 0-9 Jahre (1) | 0 | 0.00% |
| 10-19 Jahre (2) | 0 | 0.00% |
| 20-29 Jahre (3) | 3 | 21.43% |
| 30-39 Jahre (4) | 1 | 7.14% |
| 40-49 Jahre (5) | 2 | 14.29% |
| 50+ Jahre (6) | 2 | 14.29% |
| Keine Antwort | 0 | 0.00% |
| Nicht beendet oder nicht gezeigt | 6 | 42.86% |
| **Gesamt(Brutto)** | **14** | **100.00%** |

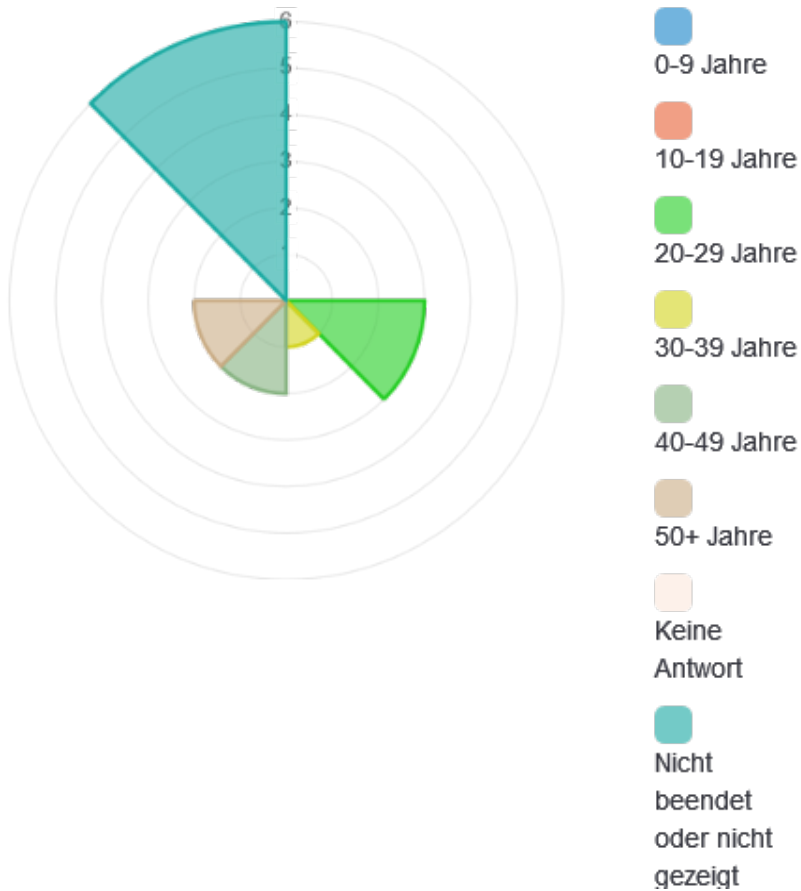

## Zusammenfassung für A3

### Was ist Ihr höchster Bildungsabschluß?

| Antwort | Anzahl | Brutto-Prozentsatz |
|---|---|---|
| Grund- / Hauptschule (1) | 0 | 0.00% |
| Mittlere Reife (2) | 0 | 0.00% |
| Fachabitur / Abitur (3) | 3 | 21.43% |
| Berufsausbildung / Duale Ausbildung (4) | 1 | 7.14% |
| Fachhochschule / Universität (5) | 4 | 28.57% |
| Promotion (6) | 0 | 0.00% |
| Keine Antwort | 0 | 0.00% |
| Nicht beendet oder nicht gezeigt | 6 | 42.86% |
| **Gesamt(Brutto)** | **14** | **100.00%** |

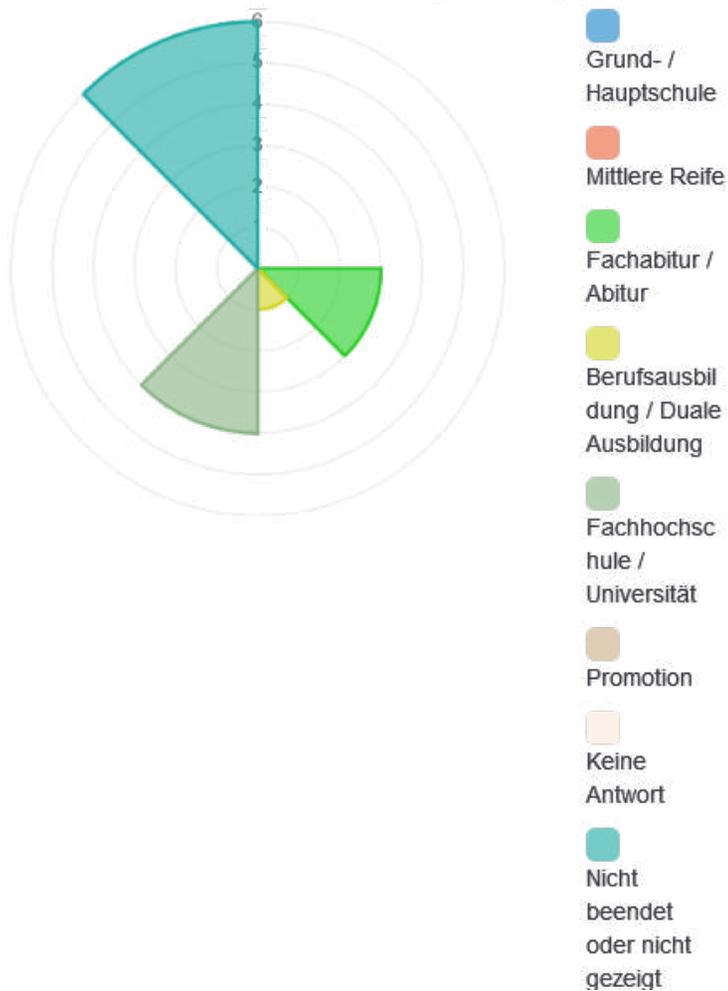

- Grund- / Hauptschule
- Mittlere Reife
- Fachabitur / Abitur
- Berufsausbildung / Duale Ausbildung
- Fachhochschule / Universität
- Promotion
- Keine Antwort
- Nicht beendet oder nicht gezeigt

## Zusammenfassung für A4

### Wie ist Ihr Bezug zur Kunst?

| Antwort | Anzahl | Brutto-Prozentsatz |
|---|---|---|
| Kunstinteressiert (1) | 5 | 35.71% |
| Amateur (2) | 1 | 7.14% |
| Professional (3) | 2 | 14.29% |
| Keine Antwort | 0 | 0.00% |
| Nicht beendet oder nicht gezeigt | 6 | 42.86% |
| **Gesamt(Brutto)** | **14** | **100.00%** |

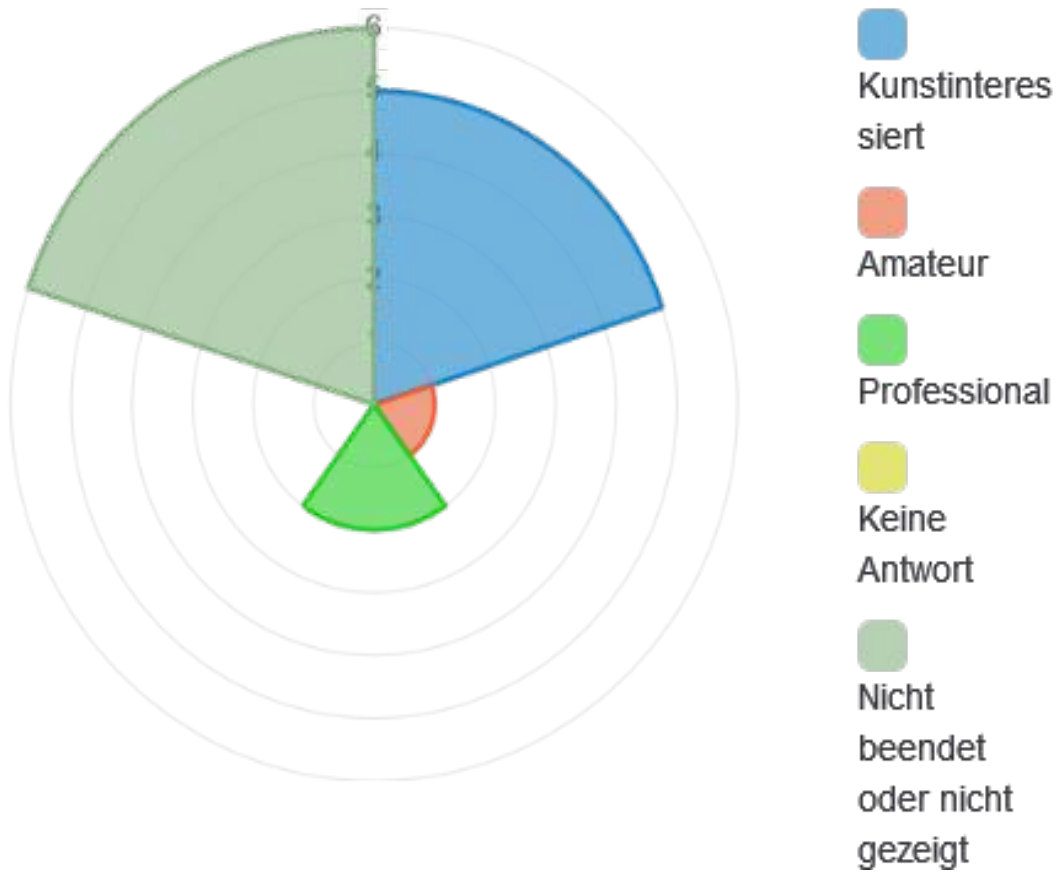

- Kunstinteressiert
- Amateur
- Professional
- Keine Antwort
- Nicht beendet oder nicht gezeigt

## Zusammenfassung für A5

### Welche Kunstwerke können Sie am Besten verstehen?

| Antwort | Anzahl | Brutto-Prozentsatz |
|---|---|---|
| Landschaftsmalerei und Stilleben (SQ001) | 5 | 35.71% |
| Gegenständliche Kunst (SQ002) | 4 | 28.57% |
| Abstrakte Kunst (SQ003) | 5 | 35.71% |
| Konkrete Kunst (SQ004) | 1 | 7.14% |
| Medienkunst und Performance (SQ005) | 5 | 35.71% |
| Nicht beendet oder nicht gezeigt | 6 | 42.86% |
| **Gesamt(Brutto)** | **26** | **100.00%** |

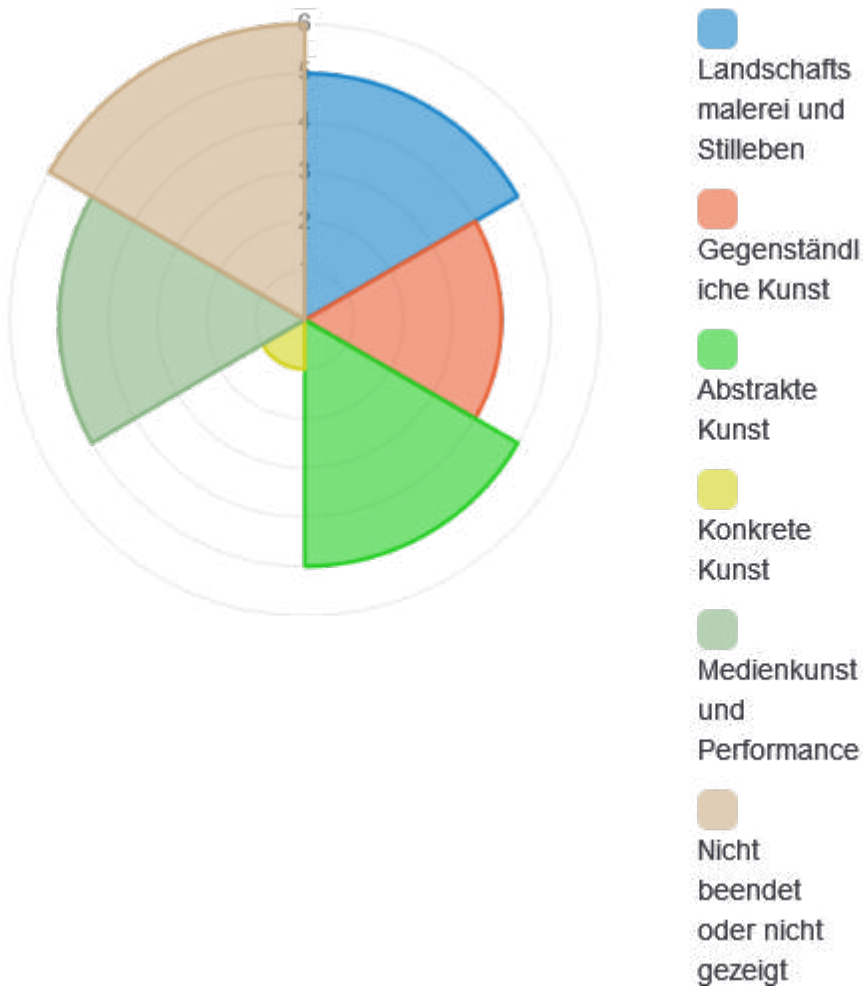

- Landschaftsmalerei und Stilleben
- Gegenständliche Kunst
- Abstrakte Kunst
- Konkrete Kunst
- Medienkunst und Performance
- Nicht beendet oder nicht gezeigt

## Zusammenfassung für B1

**Wie viele Personen waren gleichzeitig mit Ihnen im Ausstellungsraum?**

| Antwort | Anzahl | Brutto-Prozentsatz |
|---|---:|---:|
| 0-2 (1) | 1 | 7.14% |
| 3-5 (2) | 2 | 14.29% |
| 5-10 (3) | 0 | 0.00% |
| 10-15 (4) | 0 | 0.00% |
| mehr als 15 (5) | 0 | 0.00% |
| Keine Antwort | 0 | 0.00% |
| Nicht beendet oder nicht gezeigt | 11 | 78.57% |
| **Gesamt(Brutto)** | **14** | **100.00%** |

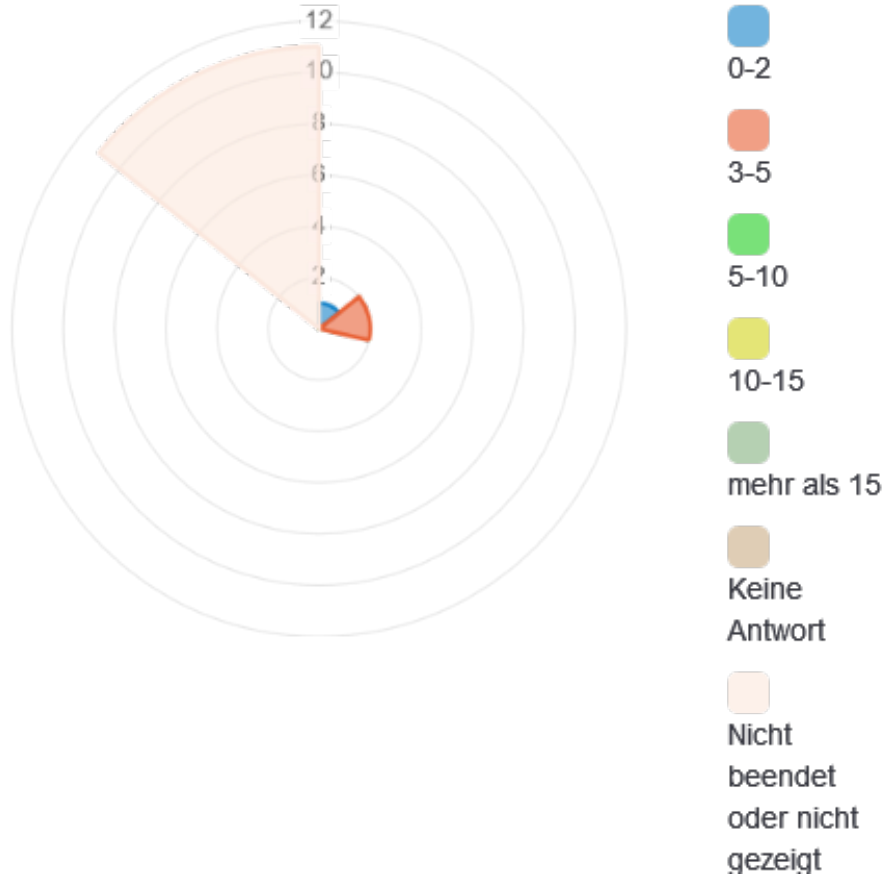

## Zusammenfassung für B2

**Zu welcher Tageszeit haben Sie die Ausstellung besucht?**

| Antwort | Anzahl | Brutto-Prozentsatz |
|---|---|---|
| morgens (1) | 0 | 0.00% |
| vormittags (2) | 0 | 0.00% |
| mittags (3) | 0 | 0.00% |
| nachmittags (4) | 1 | 7.14% |
| abends (5) | 2 | 14.29% |
| Keine Antwort | 0 | 0.00% |
| Nicht beendet oder nicht gezeigt | 11 | 78.57% |
| **Gesamt(Brutto)** | **14** | **100.00%** |

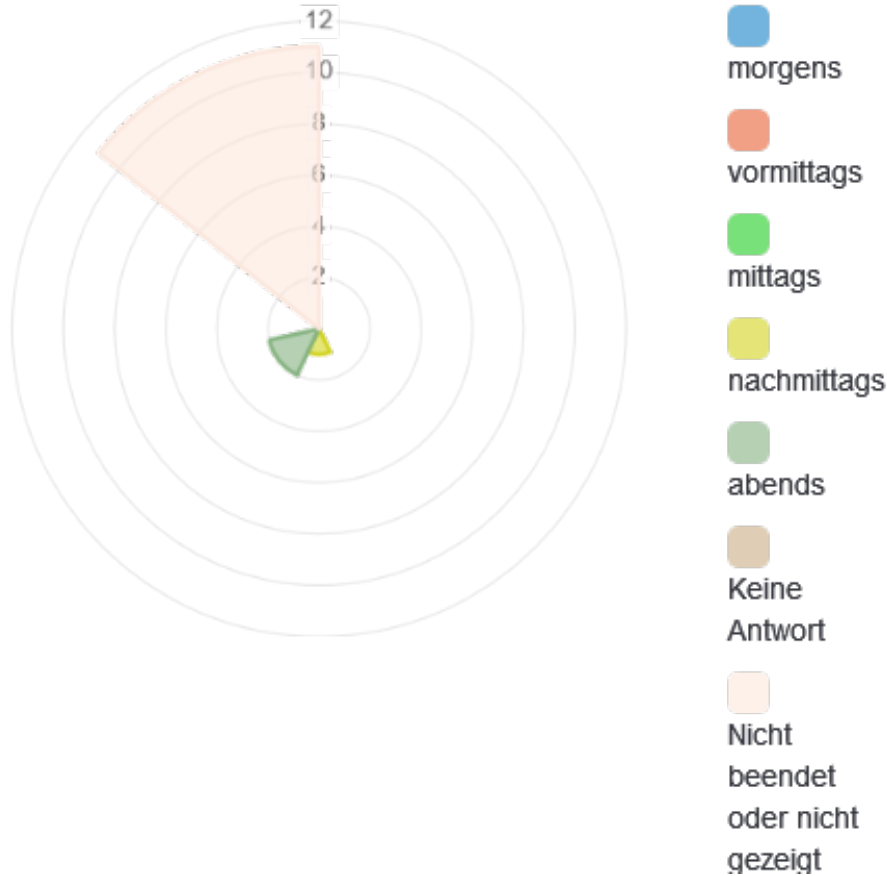

## Zusammenfassung für B3

### Wie lange dauerte Ihr Besuch?

| Antwort | Anzahl | Brutto-Prozentsatz |
|---|---|---|
| 0-15 Minuten (1) | 0 | 0.00% |
| 15-30 Minuten (2) | 0 | 0.00% |
| 30-45 Minuten (3) | 1 | 7.14% |
| 45-60 Minuten (4) | 1 | 7.14% |
| mehr als eine Stunde (5) | 1 | 7.14% |
| Keine Antwort | 0 | 0.00% |
| Nicht beendet oder nicht gezeigt | 11 | 78.57% |
| **Gesamt(Brutto)** | **14** | **100.00%** |

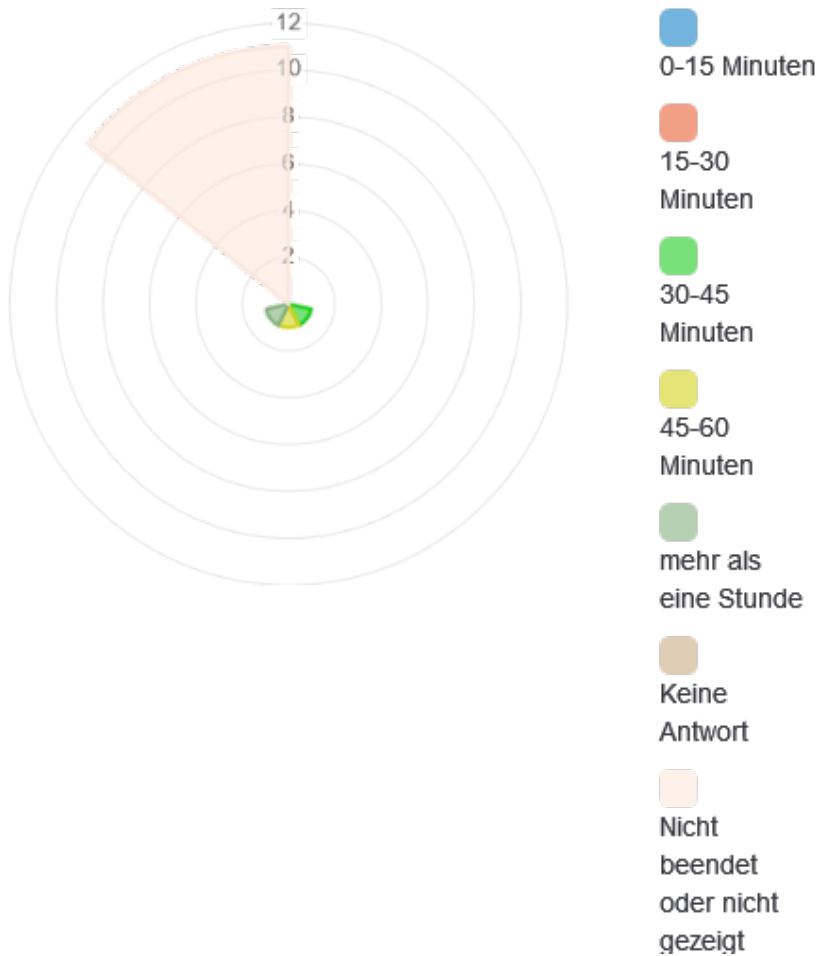

## Zusammenfassung für C1

### Wie sehen Sie die gesellschaftliche Rolle des Künstlers?

| Antwort | Anzahl | Brutto-Prozentsatz |
|---|---|---|
| Künstler lassen uns Dinge in einem anderen Licht sehen. (1) | 3 | 21.43% |
| Künstler sind für andere Menschen schwer zu verstehen. (2) | 0 | 0.00% |
| Jeder Mensch ist ein Künstler. (3) | 0 | 0.00% |
| Keine Antwort | 0 | 0.00% |
| Nicht beendet oder nicht gezeigt | 11 | 78.57% |
| **Gesamt(Brutto)** | **14** | **100.00%** |

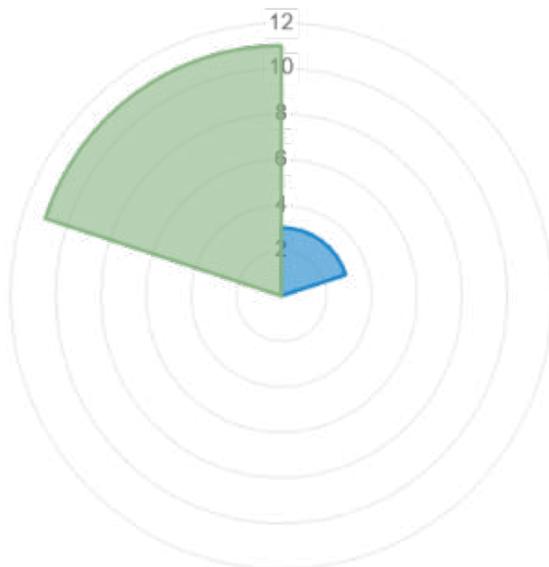

- Künstler lassen uns Dinge in einem anderen Licht sehen.
- Künstler sind für andere Menschen schwer zu verstehen.
- Jeder Mensch ist ein Künstler.
- Keine Antwort
- Nicht beendet oder nicht gezeigt

## Zusammenfassung für C2

**Was ist der Gedanke hinter dem Werk 4´33" von John Cage?**

| Antwort | Anzahl | Brutto-Prozentsatz |
|---|---:|---:|
| Es gibt ein Kunstwerk ohne einen Künstler. (1) | 1 | 7.14% |
| Das Werk hat keinen Sinn, es ist absurd. (2) | 1 | 7.14% |
| Absolute Schönheit und Erhabenheit. (3) | 1 | 7.14% |
| Keine Antwort | 0 | 0.00% |
| Nicht beendet oder nicht gezeigt | 11 | 78.57% |
| **Gesamt(Brutto)** | **14** | **100.00%** |

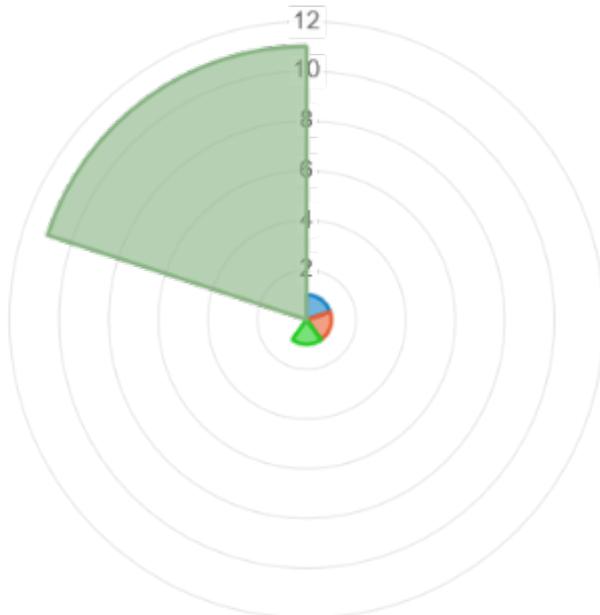

- Es gibt ein Kunstwerk ohne einen Künstler.
- Das Werk hat keinen Sinn, es ist absurd.
- Absolute Schönheit und Erhabenheit.
- Keine Antwort
- Nicht beendet oder nicht

## Zusammenfassung für C3

**Der Wiener Architekt Josef Frank bezeichnet Kunst als „zwecklos", aber auch als „Endzweck aller übrigen Tätigkeit". Wie ist Ihre Meinung?**

| Antwort | Anzahl | Brutto-Prozentsatz |
|---|---|---|
| Kunst ist nur wertvoll, wenn sie handwerklich gut gemacht ist. (1) | 0 | 0.00% |
| Manche Kunstwerke sehen für mich zwecklos aus. (2) | 0 | 0.00% |
| Kunst gehört zu den wertvollsten kulturellen Errungenschaften. (3) | 3 | 21.43% |
| Keine Antwort | 0 | 0.00% |
| Nicht beendet oder nicht gezeigt | 11 | 78.57% |
| **Gesamt(Brutto)** | **14** | **100.00%** |

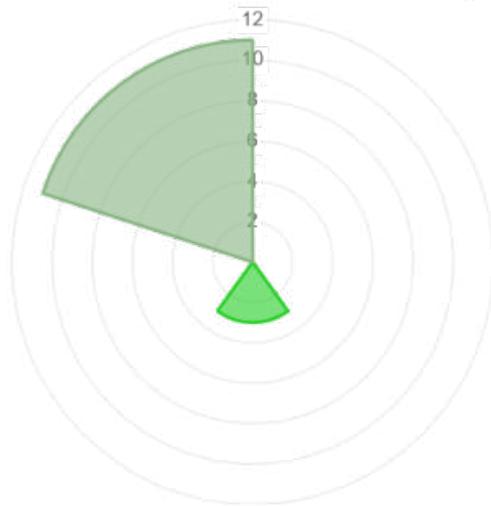

- Kunst ist nur wertvoll, wenn sie handwerklich gut gemacht ist.
- Manche Kunstwerke sehen für mich zwecklos aus.
- Kunst gehört zu den wertvollsten kulturellen Errungenschaften.
- Keine Antwort
- Nicht beendet oder nicht

## Zusammenfassung für D1

**Josef Albers beschreibt Farbe als „relativstes Medium in der Kunst". Welches Buch ist eines seiner Hauptwerke?**

| Antwort | Anzahl | Brutto-Prozentsatz |
|---|---|---|
| Kunst der Farbe (1) | 1 | 7.14% |
| Interaction of Color (2) | 1 | 7.14% |
| Farbe als Sprache (3) | 1 | 7.14% |
| Keine Antwort | 0 | 0.00% |
| Nicht beendet oder nicht gezeigt | 11 | 78.57% |
| **Gesamt(Brutto)** | **14** | **100.00%** |

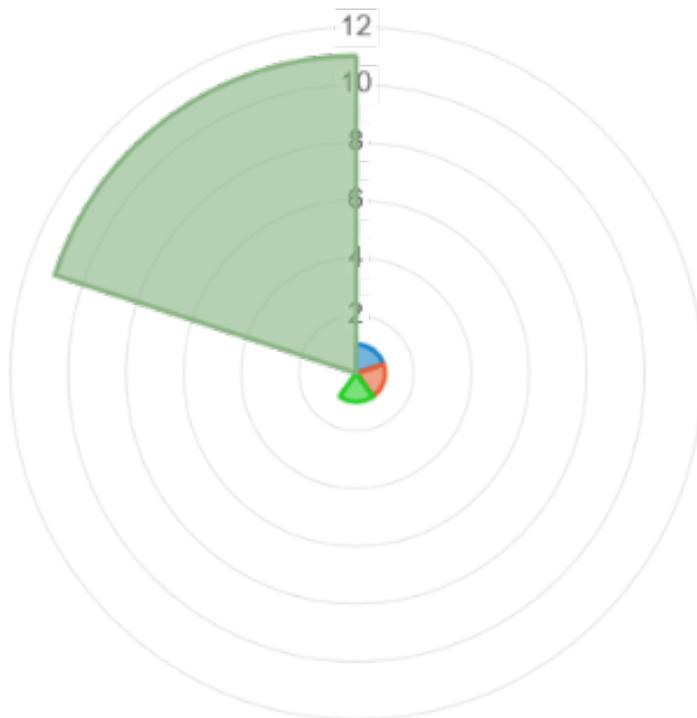

- Kunst der Farbe
- Interaction of Color
- Farbe als Sprache
- Keine Antwort
- Nicht beendet oder nicht gezeigt

## Zusammenfassung für D2

**Warum stellt sich Otl Aicher die Frage: „Wie frei ist die Freiheit?"**

| Antwort | Anzahl | Brutto-Prozentsatz |
|---|---|---|
| Die Kunst ist frei. Es gibt nur die freie Entscheidung des Künstlers. (1) | 1 | 7.14% |
| Er beschäftigt sich mit nachvollziehbaren Gestaltungsprinzipien. (2) | 1 | 7.14% |
| Kunst ist immer auch Politik. (3) | 1 | 7.14% |
| Keine Antwort | 0 | 0.00% |
| Nicht beendet oder nicht gezeigt | 11 | 78.57% |
| **Gesamt(Brutto)** | **14** | **100.00%** |

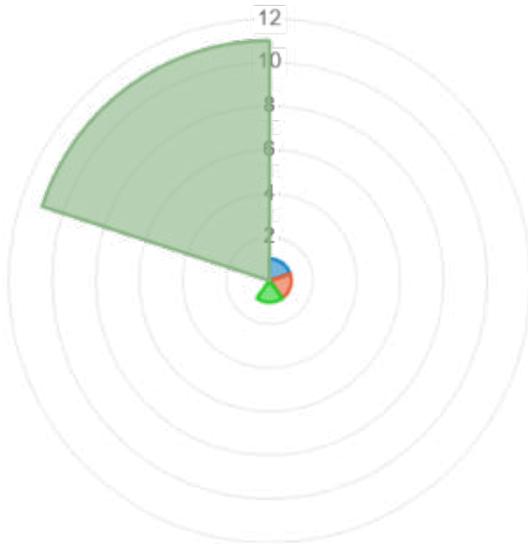

- Die Kunst ist frei. Es gibt nur die freie Entscheidung des Künstlers.
- Er beschäftigt sich mit nachvollziehbaren Gestaltungsprinzipien.
- Kunst ist immer auch Politik.
- Keine Antwort
- Nicht beendet oder nicht gezeigt

## Zusammenfassung für D3

**Frei Otto definierte den künstlerischen Gestaltungsprozeß neu. Er vertrat folgende Idee:**

| Antwort | Anzahl | Brutto-Prozentsatz |
|---|---|---|
| "Kunst mit dem alltäglichen Leben zu verbinden." (1) | 1 | 7.14% |
| "Weniger ist mehr." (2) | 0 | 0.00% |
| "Form finden" statt "Form geben" oder "Form wollen". (3) | 2 | 14.29% |
| Keine Antwort | 0 | 0.00% |
| Nicht beendet oder nicht gezeigt | 11 | 78.57% |
| **Gesamt(Brutto)** | **14** | **100.00%** |

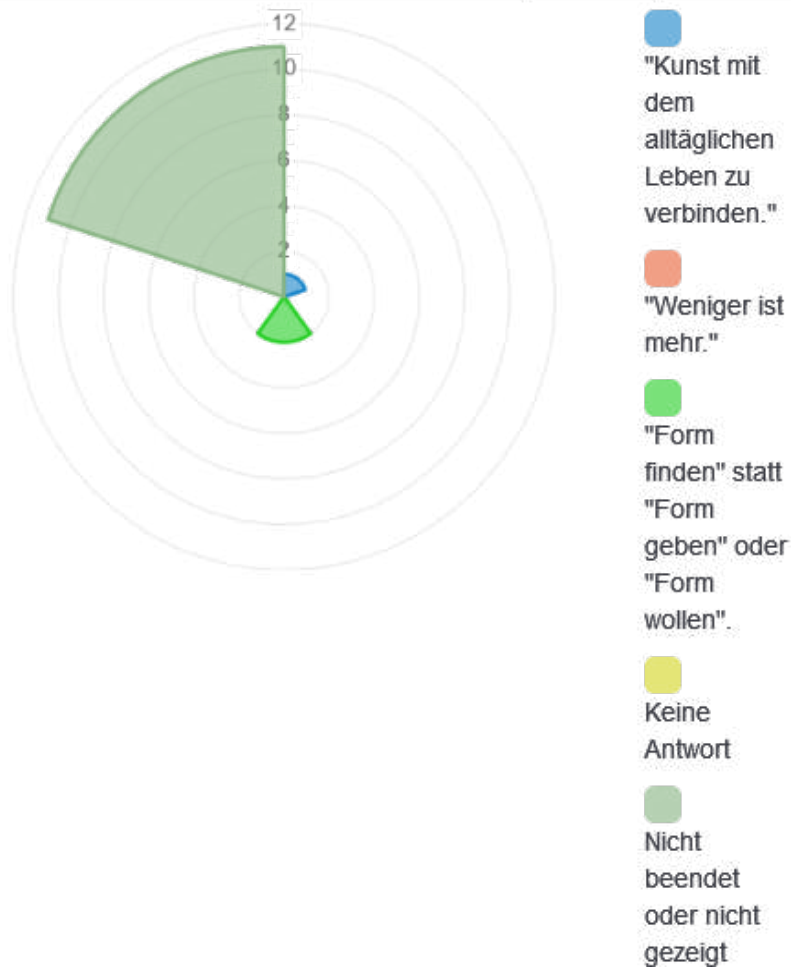

## Zusammenfassung für E1

### Was behandelt das Modell der 3 Welten von Karl R. Popper?

| Antwort | Anzahl | Brutto-Prozentsatz |
|---|---|---|
| den Unterschied zwischen Industrie- und Schwellenländern (1) | 0 | 0.00% |
| die physische, psychische und geistige Wirklichkeit (2) | 3 | 21.43% |
| die Darstellung des Universums in der Wissenschaft (3) | 0 | 0.00% |
| Keine Antwort | 0 | 0.00% |
| Nicht beendet oder nicht gezeigt | 11 | 78.57% |
| **Gesamt(Brutto)** | **14** | **100.00%** |

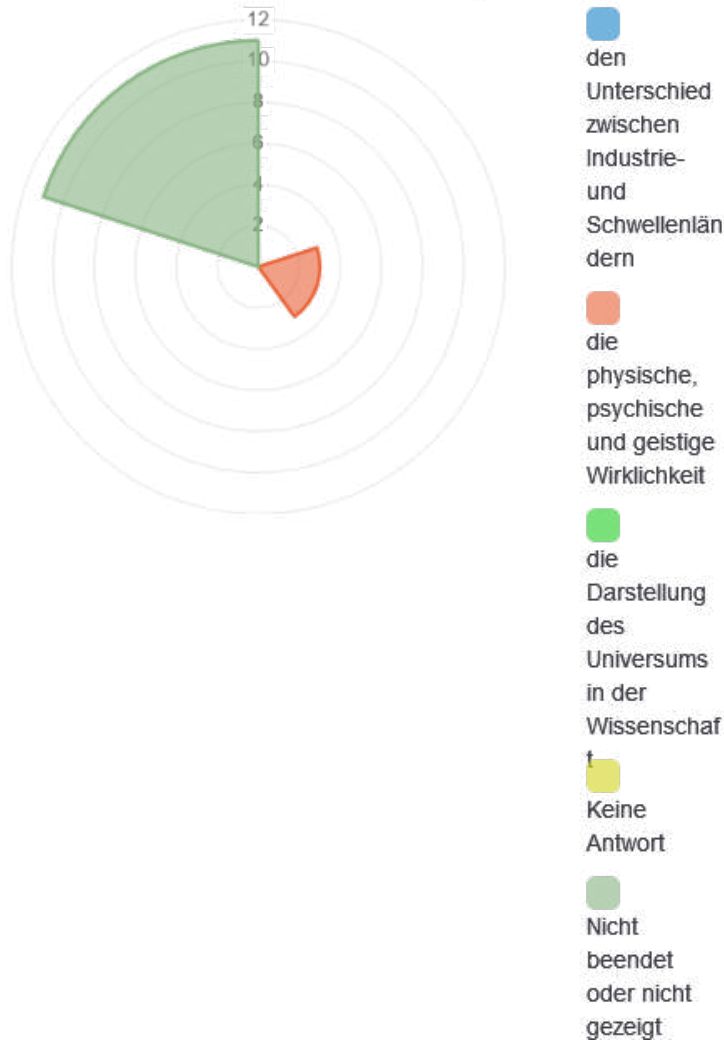

- den Unterschied zwischen Industrie- und Schwellenländern
- die physische, psychische und geistige Wirklichkeit
- die Darstellung des Universums in der Wissenschaft
- Keine Antwort
- Nicht beendet oder nicht gezeigt

## Zusammenfassung für E2

**Max Bill ist ein Vertreter der „Konkreten Kunst". Worauf beziehen sich Werke dieser Kunstrichtung?**

| Antwort | Anzahl | Brutto-Prozentsatz |
|---|---|---|
| auf gestalterische Gesetzmäßigkeiten von Form, Farbe, Raum, Licht und Bewegung (1) | 3 | 21.43% |
| auf Abstraktionen von anderen Dingen (2) | 0 | 0.00% |
| auf eine konkrete Bedeutung, die nur der Künstler kennt (3) | 0 | 0.00% |
| Keine Antwort | 0 | 0.00% |
| Nicht beendet oder nicht gezeigt | 11 | 78.57% |
| **Gesamt(Brutto)** | **14** | **100.00%** |

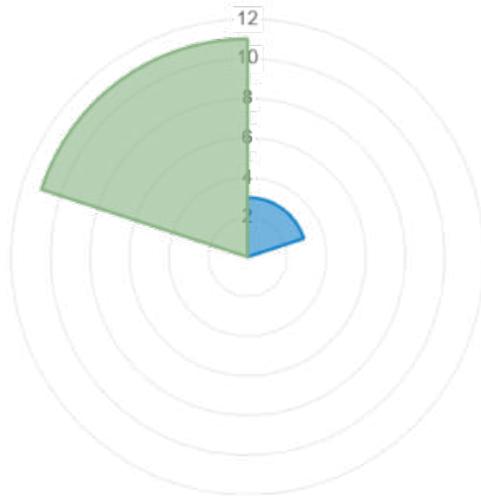

- auf gestalterische Gesetzmäßigkeiten von Form, Farbe, Raum, Licht und Bewegung
- auf Abstraktionen von anderen Dingen
- auf eine konkrete Bedeutung, die nur der Künstler kennt
- Keine Antwort
- Nicht beendet oder nicht gezeigt

## Zusammenfassung für E3

**In seinem „Pädagogischen Skizzenbuch" beschäftigt sich Paul Klee mit der Richtungsänderung einer Linie. Was bewegt ihn dazu?**

| Antwort | Anzahl | Brutto-Prozentsatz |
|---|---|---|
| Nur die Übung macht den Meister. (1) | 0 | 0.00% |
| Seine Idee ist nur für den Künstler selbst nachvollziehbar. (2) | 1 | 7.14% |
| Es sind objektive, geometrisch-künstlerische Grundlagen. (3) | 2 | 14.29% |
| Keine Antwort | 0 | 0.00% |
| Nicht beendet oder nicht gezeigt | 11 | 78.57% |
| **Gesamt(Brutto)** | **14** | **100.00%** |

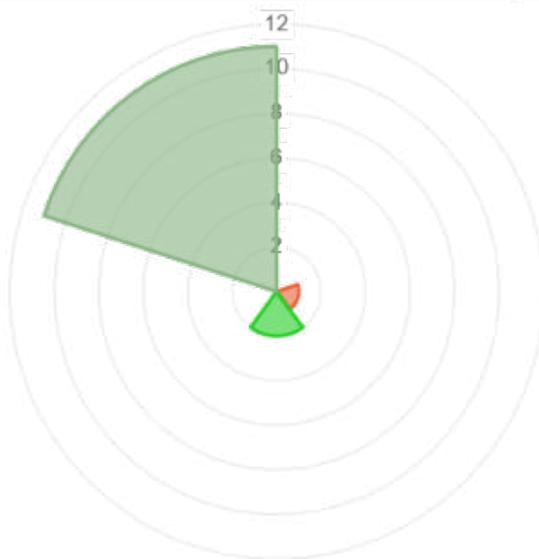

- Nur die Übung macht den Meister.
- Seine Idee ist nur für den Künstler selbst nachvollziehbar.
- Es sind objektive, geometrisch-künstlerische Grundlagen.
- Keine Antwort
- Nicht beendet oder nicht gezeigt

## Zusammenfassung für F1

### Welche Veranstaltungen haben Sie besucht?

| Antwort | Anzahl | Brutto-Prozentsatz |
|---|---|---|
| Vernissage (SQ001) | 1 | 7.14% |
| Künstlergespräch (SQ002) | 2 | 14.29% |
| Finissage (SQ003) | 1 | 7.14% |
| Nicht beendet oder nicht gezeigt | 11 | 78.57% |
| **Gesamt(Brutto)** | **15** | **100.00%** |

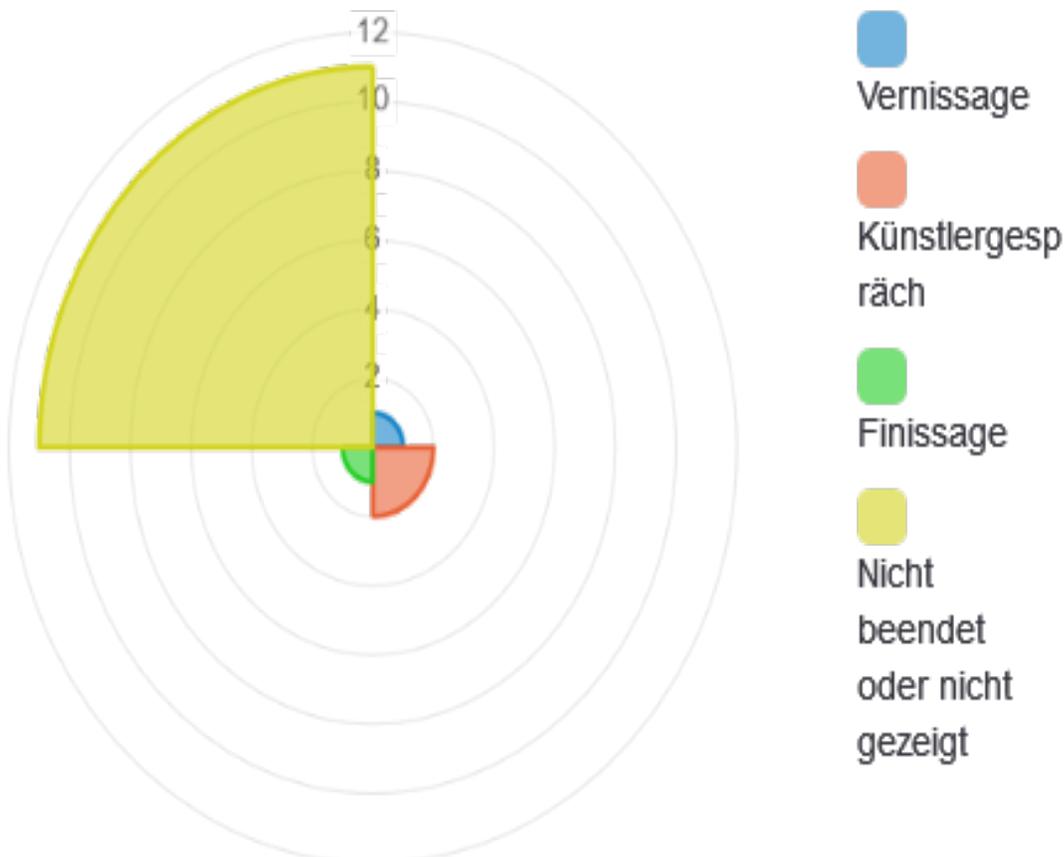

- Vernissage
- Künstlergespräch
- Finissage
- Nicht beendet oder nicht gezeigt

## Zusammenfassung für F2

**Wie viele Kontakte hatten Sie im Rahmen der Ausstellung?**

| Antwort | Anzahl | Brutto-Prozentsatz |
|---|---|---|
| mit dem Künstler (SQ001) | 3 | 21.43% |
| mit 2-5 anderen Besuchern, Bekannten oder Freunden (SQ002) | 3 | 21.43% |
| mit mehr als 5 anderen Personen (SQ003) | 0 | 0.00% |
| Nicht beendet oder nicht gezeigt | 11 | 78.57% |
| **Gesamt(Brutto)** | **17** | **100.00%** |

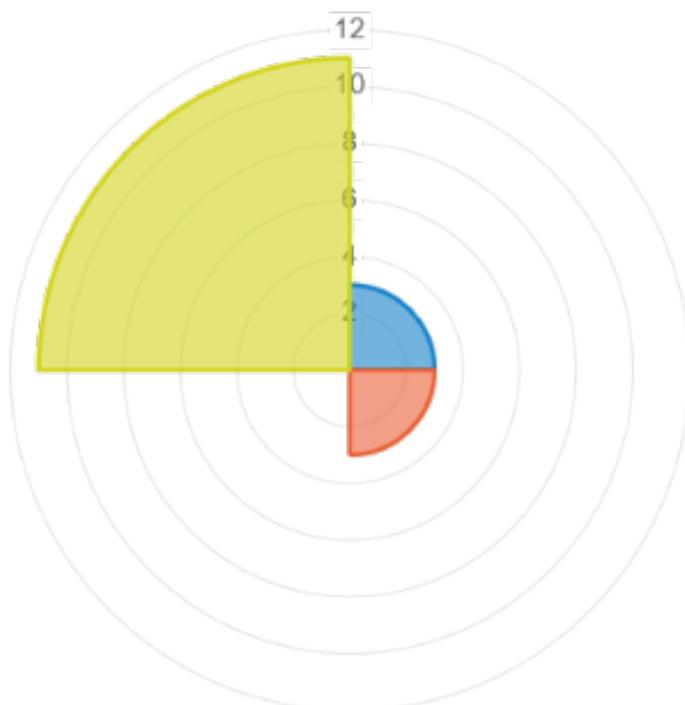

- mit dem Künstler
- mit 2-5 anderen Besuchern, Bekannten oder Freunden
- mit mehr als 5 anderen Personen
- Nicht beendet oder nicht gezeigt

## Zusammenfassung für F3

### Welche Medien haben Sie zur Interaktion genutzt?

| Antwort | Anzahl | Brutto-Prozentsatz |
|---|---|---|
| Persönliches Gespräch (SQ001) | 3 | 21.43% |
| Ausstellungskatalog (SQ002) | 1 | 7.14% |
| Analoger Fragebogen (SQ003) | 0 | 0.00% |
| Online-Fragebogen (SQ004) | 2 | 14.29% |
| Instagram / Social Media Wall (SQ005) | 1 | 7.14% |
| Andere Soziale Medien / Internet (SQ006) | 2 | 14.29% |
| Bibliothek / Bücherei (SQ007) | 0 | 0.00% |
| Nicht beendet oder nicht gezeigt | 11 | 78.57% |
| **Gesamt(Brutto)** | **20** | **100.00%** |

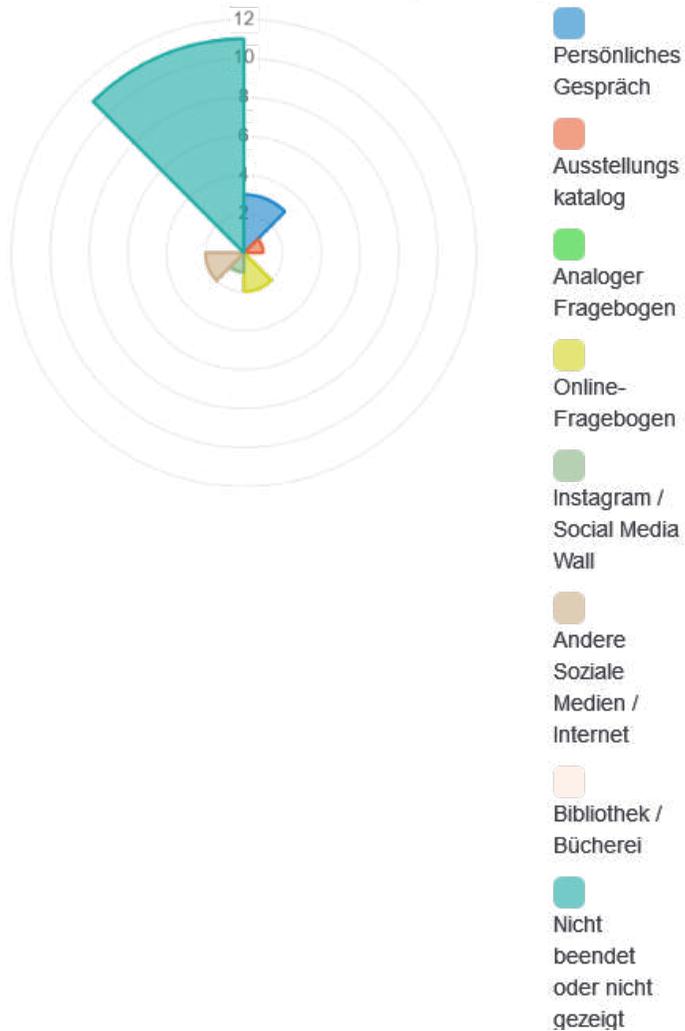

## Zusammenfassung für F4

**Hat Sie die Ausstellung künstlerisch inspiriert? Haben Sie etwas Neues erfahren? Hinterlassen Sie uns hier eine Anmerkung oder einen Kommentar.**

| Antwort | Anzahl | Brutto-Prozentsatz |
|---|---:|---:|
| Antwort [Ansehen] | 2 | 14.29% |
| Keine Antwort | 1 | 7.14% |
| Nicht beendet oder nicht gezeigt | 11 | 78.57% |
| **Gesamt(Brutto)** | **14** | **100.00%** |

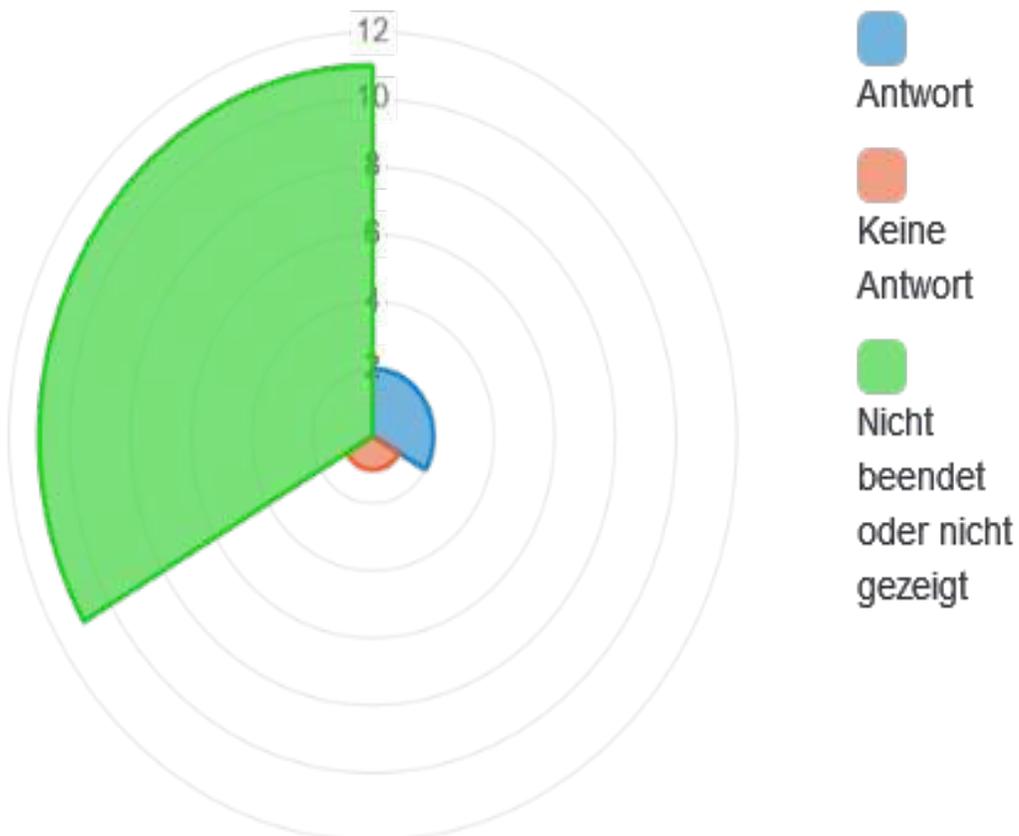

## Zusammenfassung für F5

**Hier können Sie uns eine eigene künstlerische Skizze oder ein Foto hinterlassen.**

| Antwort | Anzahl | Brutto-Prozentsatz |
|---|---|---|
| Keine Antwort | 3 | 21.43% |
| Nicht beendet oder nicht gezeigt | 11 | 78.57% |
| **Gesamt(Brutto)** | **14** | **100.00%** |

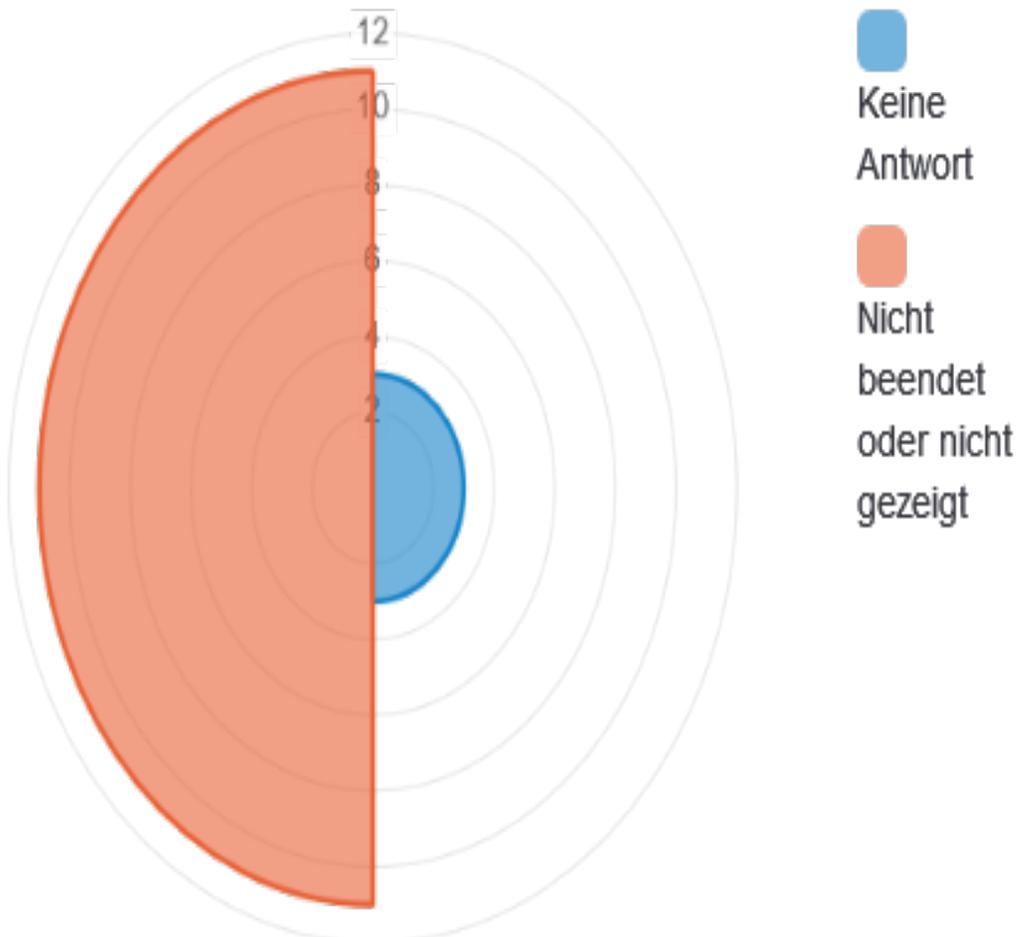

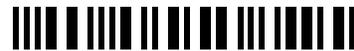

# Teil A: Angaben zur Person

**A1.** **Persönliche Anrede**

- Herr ☒
- Frau ☐
- keine Angabe ☐

**A2.** **Wie alt sind Sie?**

- 0-9 Jahre ☐
- 10-19 Jahre ☐
- 20-29 Jahre ☐
- 30-39 Jahre ☐
- 40-49 Jahre ☒
- 50+ Jahre ☐

**A3.** **Was ist Ihr höchster Bildungsabschluß?**

- Grund- / Hauptschule ☐
- Mittlere Reife ☐
- Fachabitur / Abitur ☒
- Berufsausbildung / Duale Ausbildung ☐
- Fachhochschule / Universität ☐
- Promotion ☐

**A4.** **Wie ist Ihr Bezug zur Kunst?**

- Kunstinteressiert ☒
- Amateur ☐
- Professional ☐

**A5.** **Welche Kunstwerke können Sie am Besten verstehen?**

- Landschaftsmalerei und Stillleben ☐
- Gegenständliche Kunst ☒
- Abstrakte Kunst ☐
- Konkrete Kunst ☐
- Medienkunst und Performance ☐

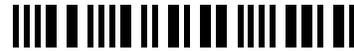

## Teil B: Ort und Situation

**B1.** **Wie viele Personen waren gleichzeitig mit Ihnen im Ausstellungsraum?**

- [ ] 0-2
- [x] 3-5
- [ ] 5-10
- [ ] 10-15
- [ ] mehr als 15

**B2.** **Zu welcher Tageszeit haben Sie die Ausstellung besucht?**

- [ ] morgens
- [ ] vormittags
- [ ] mittags
- [ ] nachmittags
- [x] abends

**B3.** **Wie lange dauerte Ihr Besuch?**

- [ ] 0-15 Minuten
- [ ] 15-30 Minuten
- [x] 30-45 Minuten
- [ ] 45-60 Minuten
- [ ] mehr als eine Stunde

## Teil C: Künstler, Kunstwerk und Gesellschaft

**C1.** **Wie sehen Sie die gesellschaftliche Rolle des Künstlers?**

- [x] Künstler lassen uns Dinge in einem anderen Licht sehen.
- [ ] Künstler sind für andere Menschen schwer zu verstehen.
- [ ] Jeder Mensch ist ein Künstler.

**C2.** **Was ist der Gedanke hinter dem Werk 4´33" von John Cage?**

- [ ] Es gibt ein Kunstwerk ohne einen Künstler.
- [ ] Das Werk hat keinen Sinn, es ist absurd.
- [x] Absolute Schönheit und Erhabenheit.

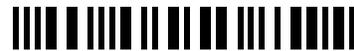

**C3.** Der Wiener Architekt Josef Frank bezeichnet Kunst als „zwecklos", aber auch als „Endzweck aller übrigen Tätigkeit". Wie ist Ihre Meinung?

- Kunst ist nur wertvoll, wenn sie handwerklich gut gemacht ist. ☐
- Manche Kunstwerke sehen für mich zwecklos aus. ☐
- Kunst gehört zu den wertvollsten kulturellen Errungenschaften. ☒

## Teil D: Grundlagen künstlerischer Praxis

**D1.** Josef Albers beschreibt Farbe als „relativstes Medium in der Kunst". Welches Buch ist eines seiner Hauptwerke?

- Kunst der Farbe ☒
- Interaction of Color ☐
- Farbe als Sprache ☐

**D2.** Warum stellt sich Otl Aicher die Frage: „Wie frei ist die Freiheit?"

- Die Kunst ist frei. Es gibt nur die freie Entscheidung des Künstlers. ☐
- Er beschäftigt sich mit nachvollziehbaren Gestaltungsprinzipien. ☐
- Kunst ist immer auch Politik. ☒

**D3.** Frei Otto definierte den künstlerischen Gestaltungsprozeß neu. Er vertrat folgende Idee:

- "Kunst mit dem alltäglichen Leben zu verbinden." ☒
- "Weniger ist mehr." ☐
- "Form finden" statt "Form geben" oder "Form wollen". ☐

## Teil E: Einflüsse anderer Wissensgebiete

**E1.** Was behandelt das Modell der 3 Welten von Karl R. Popper?

- den Unterschied zwischen Industrie- und Schwellenländern ☐
- die physische, psychische und geistige Wirklichkeit ☒
- die Darstellung des Universums in der Wissenschaft ☐

**E2.** Max Bill ist ein Vertreter der „Konkreten Kunst". Worauf beziehen sich Werke dieser Kunstrichtung?

- auf gestalterische Gesetzmäßigkeiten von Form, Farbe, Raum, Licht und Bewegung ☒
- auf Abstraktionen von anderen Dingen ☐
- auf eine konkrete Bedeutung, die nur der Künstler kennt ☐

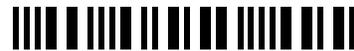

| E3. | In seinem „Pädagogischen Skizzenbuch" beschäftigt sich Paul Klee mit der Richtungsänderung einer Linie. Was bewegt ihn dazu? | |
|---|---|---|
| | Nur die Übung macht den Meister. | ☐ |
| | Seine Idee ist nur für den Künstler selbst nachvollziehbar. | ☒ |
| | Es sind objektive, geometrisch-künstlerische Grundlagen. | ☐ |

# Teil F: Individuelle Aktivitäten

| F1. | Welche Veranstaltungen haben Sie besucht? | |
|---|---|---|
| | Vernissage | ☒ |
| | Künstlergespräch | ☐ |
| | Finissage | ☒ |

| F2. | Wie viele Kontakte hatten Sie im Rahmen der Ausstellung? | |
|---|---|---|
| | mit dem Künstler | ☒ |
| | mit 2-5 anderen Besuchern, Bekannten oder Freunden | ☒ |
| | mit mehr als 5 anderen Personen | ☐ |

| F3. | Welche Medien haben Sie zur Interaktion genutzt? | |
|---|---|---|
| | Persönliches Gespräch | ☒ |
| | Ausstellungskatalog | ☐ |
| | Analoger Fragebogen | ☐ |
| | Online-Fragebogen | ☐ |
| | Instagram / Social Media Wall | ☐ |
| | Andere Soziale Medien / Internet | ☐ |
| | Bibliothek / Bücherei | ☐ |

**F4.** Hat Sie die Ausstellung künstlerisch inspiriert? Haben Sie etwas Neues erfahren? Hinterlassen Sie uns hier eine Anmerkung oder einen Kommentar.

**F5.** Hier können Sie uns eine eigene künstlerische Skizze oder ein Foto hinterlassen.

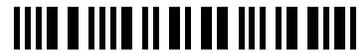

**Wirklichkeiten | Realities . Kunstausstellung | 1. Mai - 30. Juni 2023 Sitzungssaal im Rathaus . Herrsching am Ammersee Vielen Dank für Ihre Teilnahme! Sie können das Browserfenster jetzt schließen.**

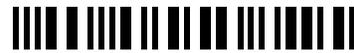

# Teil A: Angaben zur Person

**A1.** **Persönliche Anrede**

- ☐ Herr
- ☐ Frau
- ☒ keine Angabe

**A2.** **Wie alt sind Sie?**

- ☐ 0-9 Jahre
- ☐ 10-19 Jahre
- ☐ 20-29 Jahre
- ☐ 30-39 Jahre
- ☐ 40-49 Jahre
- ☒ 50+ Jahre

**A3.** **Was ist Ihr höchster Bildungsabschluß?**

- ☐ Grund- / Hauptschule
- ☐ Mittlere Reife
- ☒ Fachabitur / Abitur
- ☐ Berufsausbildung / Duale Ausbildung
- ☐ Fachhochschule / Universität
- ☐ Promotion

**A4.** **Wie ist Ihr Bezug zur Kunst?**

- ☐ Kunstinteressiert
- ☒ Amateur
- ☐ Professional

**A5.** **Welche Kunstwerke können Sie am Besten verstehen?**

- ☒ Landschaftsmalerei und Stilleben
- ☒ Gegenständliche Kunst
- ☒ Abstrakte Kunst
- ☐ Konkrete Kunst
- ☐ Medienkunst und Performance

# Teil B: Ort und Situation

**B1.** **Wie viele Personen waren gleichzeitig mit Ihnen im Ausstellungsraum?**

- 0-2 ☐
- 3-5 ☐
- 5-10 ☐
- 10-15 ☐
- mehr als 15 ☐

**B2.** **Zu welcher Tageszeit haben Sie die Ausstellung besucht?**

- morgens ☐
- vormittags ☐
- mittags ☐
- nachmittags ☐
- abends ☐

**B3.** **Wie lange dauerte Ihr Besuch?**

- 0-15 Minuten ☐
- 15-30 Minuten ☐
- 30-45 Minuten ☐
- 45-60 Minuten ☐
- mehr als eine Stunde ☐

# Teil C: Künstler, Kunstwerk und Gesellschaft

**C1.** **Wie sehen Sie die gesellschaftliche Rolle des Künstlers?**

- Künstler lassen uns Dinge in einem anderen Licht sehen. ☐
- Künstler sind für andere Menschen schwer zu verstehen. ☐
- Jeder Mensch ist ein Künstler. ☐

**C2.** **Was ist der Gedanke hinter dem Werk 4´33" von John Cage?**

- Es gibt ein Kunstwerk ohne einen Künstler. ☐
- Das Werk hat keinen Sinn, es ist absurd. ☐
- Absolute Schönheit und Erhabenheit. ☐

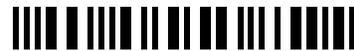

**C3. Der Wiener Architekt Josef Frank bezeichnet Kunst als „zwecklos", aber auch als „Endzweck aller übrigen Tätigkeit". Wie ist Ihre Meinung?**

- Kunst ist nur wertvoll, wenn sie handwerklich gut gemacht ist. ☐
- Manche Kunstwerke sehen für mich zwecklos aus. ☐
- Kunst gehört zu den wertvollsten kulturellen Errungenschaften. ☐

## Teil D: Grundlagen künstlerischer Praxis

**D1. Josef Albers beschreibt Farbe als „relativstes Medium in der Kunst". Welches Buch ist eines seiner Hauptwerke?**

- Kunst der Farbe ☐
- Interaction of Color ☐
- Farbe als Sprache ☐

**D2. Warum stellt sich Otl Aicher die Frage: „Wie frei ist die Freiheit?"**

- Die Kunst ist frei. Es gibt nur die freie Entscheidung des Künstlers. ☐
- Er beschäftigt sich mit nachvollziehbaren Gestaltungsprinzipien. ☐
- Kunst ist immer auch Politik. ☐

**D3. Frei Otto definierte den künstlerischen Gestaltungsprozeß neu. Er vertrat folgende Idee:**

- "Kunst mit dem alltäglichen Leben zu verbinden." ☐
- "Weniger ist mehr." ☐
- "Form finden" statt "Form geben" oder "Form wollen". ☐

## Teil E: Einflüsse anderer Wissensgebiete

**E1. Was behandelt das Modell der 3 Welten von Karl R. Popper?**

- den Unterschied zwischen Industrie- und Schwellenländern ☐
- die physische, psychische und geistige Wirklichkeit ☐
- die Darstellung des Universums in der Wissenschaft ☐

**E2. Max Bill ist ein Vertreter der „Konkreten Kunst". Worauf beziehen sich Werke dieser Kunstrichtung?**

- auf gestalterische Gesetzmäßigkeiten von Form, Farbe, Raum, Licht und Bewegung ☐
- auf Abstraktionen von anderen Dingen ☐
- auf eine konkrete Bedeutung, die nur der Künstler kennt ☐

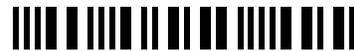

**E3.** In seinem „Pädagogischen Skizzenbuch" beschäftigt sich Paul Klee mit der Richtungsänderung einer Linie. Was bewegt ihn dazu?

- [ ] Nur die Übung macht den Meister.
- [ ] Seine Idee ist nur für den Künstler selbst nachvollziehbar.
- [ ] Es sind objektive, geometrisch-künstlerische Grundlagen.

# Teil F: Individuelle Aktivitäten

**F1.** Welche Veranstaltungen haben Sie besucht?

- [ ] Vernissage
- [ ] Künstlergespräch
- [ ] Finissage

**F2.** Wie viele Kontakte hatten Sie im Rahmen der Ausstellung?

- [ ] mit dem Künstler
- [ ] mit 2-5 anderen Besuchern, Bekannten oder Freunden
- [ ] mit mehr als 5 anderen Personen

**F3.** Welche Medien haben Sie zur Interaktion genutzt?

- [ ] Persönliches Gespräch
- [ ] Ausstellungskatalog
- [ ] Analoger Fragebogen
- [ ] Online-Fragebogen
- [ ] Instagram / Social Media Wall
- [ ] Andere Soziale Medien / Internet
- [ ] Bibliothek / Bücherei

**F4.** Hat Sie die Ausstellung künstlerisch inspiriert? Haben Sie etwas Neues erfahren? Hinterlassen Sie uns hier eine Anmerkung oder einen Kommentar.

**F5.** Hier können Sie uns eine eigene künstlerische Skizze oder ein Foto hinterlassen.

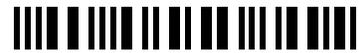

**Wirklichkeiten | Realities . Kunstausstellung | 1. Mai - 30. Juni 2023 Sitzungssaal im Rathaus . Herrsching am Ammersee Vielen Dank für Ihre Teilnahme! Sie können das Browserfenster jetzt schließen.**

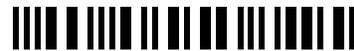

# Teil A: Angaben zur Person

**A1.** **Persönliche Anrede**

- Herr ☐
- Frau ☐
- keine Angabe ☐

**A2.** **Wie alt sind Sie?**

- 0-9 Jahre ☐
- 10-19 Jahre ☐
- 20-29 Jahre ☐
- 30-39 Jahre ☐
- 40-49 Jahre ☐
- 50+ Jahre ☐

**A3.** **Was ist Ihr höchster Bildungsabschluß?**

- Grund- / Hauptschule ☐
- Mittlere Reife ☐
- Fachabitur / Abitur ☐
- Berufsausbildung / Duale Ausbildung ☐
- Fachhochschule / Universität ☐
- Promotion ☐

**A4.** **Wie ist Ihr Bezug zur Kunst?**

- Kunstinteressiert ☐
- Amateur ☐
- Professional ☐

**A5.** **Welche Kunstwerke können Sie am Besten verstehen?**

- Landschaftsmalerei und Stilleben ☐
- Gegenständliche Kunst ☐
- Abstrakte Kunst ☐
- Konkrete Kunst ☐
- Medienkunst und Performance ☐

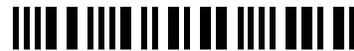

# Teil B: Ort und Situation

**B1.** **Wie viele Personen waren gleichzeitig mit Ihnen im Ausstellungsraum?**

- 0-2 ☐
- 3-5 ☐
- 5-10 ☐
- 10-15 ☐
- mehr als 15 ☐

**B2.** **Zu welcher Tageszeit haben Sie die Ausstellung besucht?**

- morgens ☐
- vormittags ☐
- mittags ☐
- nachmittags ☐
- abends ☐

**B3.** **Wie lange dauerte Ihr Besuch?**

- 0-15 Minuten ☐
- 15-30 Minuten ☐
- 30-45 Minuten ☐
- 45-60 Minuten ☐
- mehr als eine Stunde ☐

# Teil C: Künstler, Kunstwerk und Gesellschaft

**C1.** **Wie sehen Sie die gesellschaftliche Rolle des Künstlers?**

- Künstler lassen uns Dinge in einem anderen Licht sehen. ☐
- Künstler sind für andere Menschen schwer zu verstehen. ☐
- Jeder Mensch ist ein Künstler. ☐

**C2.** **Was ist der Gedanke hinter dem Werk 4´33" von John Cage?**

- Es gibt ein Kunstwerk ohne einen Künstler. ☐
- Das Werk hat keinen Sinn, es ist absurd. ☐
- Absolute Schönheit und Erhabenheit. ☐

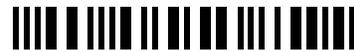

**C3.** Der Wiener Architekt Josef Frank bezeichnet Kunst als „zwecklos", aber auch als „Endzweck aller übrigen Tätigkeit". Wie ist Ihre Meinung?

- Kunst ist nur wertvoll, wenn sie handwerklich gut gemacht ist. ☐
- Manche Kunstwerke sehen für mich zwecklos aus. ☐
- Kunst gehört zu den wertvollsten kulturellen Errungenschaften. ☐

## Teil D: Grundlagen künstlerischer Praxis

**D1.** Josef Albers beschreibt Farbe als „relativstes Medium in der Kunst". Welches Buch ist eines seiner Hauptwerke?

- Kunst der Farbe ☐
- Interaction of Color ☐
- Farbe als Sprache ☐

**D2.** Warum stellt sich Otl Aicher die Frage: „Wie frei ist die Freiheit?"

- Die Kunst ist frei. Es gibt nur die freie Entscheidung des Künstlers. ☐
- Er beschäftigt sich mit nachvollziehbaren Gestaltungsprinzipien. ☐
- Kunst ist immer auch Politik. ☐

**D3.** Frei Otto definierte den künstlerischen Gestaltungsprozeß neu. Er vertrat folgende Idee:

- "Kunst mit dem alltäglichen Leben zu verbinden." ☐
- "Weniger ist mehr." ☐
- "Form finden" statt "Form geben" oder "Form wollen". ☐

## Teil E: Einflüsse anderer Wissensgebiete

**E1.** Was behandelt das Modell der 3 Welten von Karl R. Popper?

- den Unterschied zwischen Industrie- und Schwellenländern ☐
- die physische, psychische und geistige Wirklichkeit ☐
- die Darstellung des Universums in der Wissenschaft ☐

**E2.** Max Bill ist ein Vertreter der „Konkreten Kunst". Worauf beziehen sich Werke dieser Kunstrichtung?

- auf gestalterische Gesetzmäßigkeiten von Form, Farbe, Raum, Licht und Bewegung ☐
- auf Abstraktionen von anderen Dingen ☐
- auf eine konkrete Bedeutung, die nur der Künstler kennt ☐

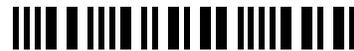

**E3.** **In seinem „Pädagogischen Skizzenbuch" beschäftigt sich Paul Klee mit der Richtungsänderung einer Linie. Was bewegt ihn dazu?**

- Nur die Übung macht den Meister. ☐
- Seine Idee ist nur für den Künstler selbst nachvollziehbar. ☐
- Es sind objektive, geometrisch-künstlerische Grundlagen. ☐

# Teil F: Individuelle Aktivitäten

**F1.** **Welche Veranstaltungen haben Sie besucht?**

- Vernissage ☐
- Künstlergespräch ☐
- Finissage ☐

**F2.** **Wie viele Kontakte hatten Sie im Rahmen der Ausstellung?**

- mit dem Künstler ☐
- mit 2-5 anderen Besuchern, Bekannten oder Freunden ☐
- mit mehr als 5 anderen Personen ☐

**F3.** **Welche Medien haben Sie zur Interaktion genutzt?**

- Persönliches Gespräch ☐
- Ausstellungskatalog ☐
- Analoger Fragebogen ☐
- Online-Fragebogen ☐
- Instagram / Social Media Wall ☐
- Andere Soziale Medien / Internet ☐
- Bibliothek / Bücherei ☐

**F4.** **Hat Sie die Ausstellung künstlerisch inspiriert? Haben Sie etwas Neues erfahren? Hinterlassen Sie uns hier eine Anmerkung oder einen Kommentar.**

[                                                          ]

**F5.** **Hier können Sie uns eine eigene künstlerische Skizze oder ein Foto hinterlassen.**

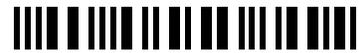

**Wirklichkeiten | Realities . Kunstausstellung | 1. Mai - 30. Juni 2023 Sitzungssaal im Rathaus . Herrsching am Ammersee Vielen Dank für Ihre Teilnahme! Sie können das Browserfenster jetzt schließen.**

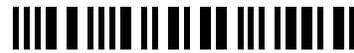

# Teil A: Angaben zur Person

**A1.** **Persönliche Anrede**

Herr ☒
Frau ☐
keine Angabe ☐

**A2.** **Wie alt sind Sie?**

0-9 Jahre ☐
10-19 Jahre ☐
20-29 Jahre ☒
30-39 Jahre ☐
40-49 Jahre ☐
50+ Jahre ☐

**A3.** **Was ist Ihr höchster Bildungsabschluß?**

Grund- / Hauptschule ☐
Mittlere Reife ☐
Fachabitur / Abitur ☐
Berufsausbildung / Duale Ausbildung ☐
Fachhochschule / Universität ☒
Promotion ☐

**A4.** **Wie ist Ihr Bezug zur Kunst?**

Kunstinteressiert ☐
Amateur ☐
Professional ☒

**A5.** **Welche Kunstwerke können Sie am Besten verstehen?**

Landschaftsmalerei und Stillleben ☒
Gegenständliche Kunst ☒
Abstrakte Kunst ☒
Konkrete Kunst ☒
Medienkunst und Performance ☒

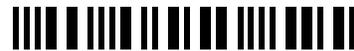

# Teil B: Ort und Situation

**B1.** **Wie viele Personen waren gleichzeitig mit Ihnen im Ausstellungsraum?**

- 0-2 ☐
- 3-5 ☐
- 5-10 ☐
- 10-15 ☐
- mehr als 15 ☐

**B2.** **Zu welcher Tageszeit haben Sie die Ausstellung besucht?**

- morgens ☐
- vormittags ☐
- mittags ☐
- nachmittags ☐
- abends ☐

**B3.** **Wie lange dauerte Ihr Besuch?**

- 0-15 Minuten ☐
- 15-30 Minuten ☐
- 30-45 Minuten ☐
- 45-60 Minuten ☐
- mehr als eine Stunde ☐

# Teil C: Künstler, Kunstwerk und Gesellschaft

**C1.** **Wie sehen Sie die gesellschaftliche Rolle des Künstlers?**

- Künstler lassen uns Dinge in einem anderen Licht sehen. ☐
- Künstler sind für andere Menschen schwer zu verstehen. ☐
- Jeder Mensch ist ein Künstler. ☐

**C2.** **Was ist der Gedanke hinter dem Werk 4´33" von John Cage?**

- Es gibt ein Kunstwerk ohne einen Künstler. ☐
- Das Werk hat keinen Sinn, es ist absurd. ☐
- Absolute Schönheit und Erhabenheit. ☐

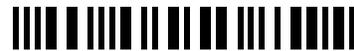

**C3.** Der Wiener Architekt Josef Frank bezeichnet Kunst als „zwecklos", aber auch als „Endzweck aller übrigen Tätigkeit". Wie ist Ihre Meinung?

- [ ] Kunst ist nur wertvoll, wenn sie handwerklich gut gemacht ist.
- [ ] Manche Kunstwerke sehen für mich zwecklos aus.
- [ ] Kunst gehört zu den wertvollsten kulturellen Errungenschaften.

## Teil D: Grundlagen künstlerischer Praxis

**D1.** Josef Albers beschreibt Farbe als „relativstes Medium in der Kunst". Welches Buch ist eines seiner Hauptwerke?

- [ ] Kunst der Farbe
- [ ] Interaction of Color
- [ ] Farbe als Sprache

**D2.** Warum stellt sich Otl Aicher die Frage: „Wie frei ist die Freiheit?"

- [ ] Die Kunst ist frei. Es gibt nur die freie Entscheidung des Künstlers.
- [ ] Er beschäftigt sich mit nachvollziehbaren Gestaltungsprinzipien.
- [ ] Kunst ist immer auch Politik.

**D3.** Frei Otto definierte den künstlerischen Gestaltungsprozeß neu. Er vertrat folgende Idee:

- [ ] "Kunst mit dem alltäglichen Leben zu verbinden."
- [ ] "Weniger ist mehr."
- [ ] "Form finden" statt "Form geben" oder "Form wollen".

## Teil E: Einflüsse anderer Wissensgebiete

**E1.** Was behandelt das Modell der 3 Welten von Karl R. Popper?

- [ ] den Unterschied zwischen Industrie- und Schwellenländern
- [ ] die physische, psychische und geistige Wirklichkeit
- [ ] die Darstellung des Universums in der Wissenschaft

**E2.** Max Bill ist ein Vertreter der „Konkreten Kunst". Worauf beziehen sich Werke dieser Kunstrichtung?

- [ ] auf gestalterische Gesetzmäßigkeiten von Form, Farbe, Raum, Licht und Bewegung
- [ ] auf Abstraktionen von anderen Dingen
- [ ] auf eine konkrete Bedeutung, die nur der Künstler kennt

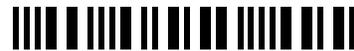

| E3. | In seinem „Pädagogischen Skizzenbuch" beschäftigt sich Paul Klee mit der Richtungsänderung einer Linie. Was bewegt ihn dazu? |
|---|---|

- Nur die Übung macht den Meister. ☐
- Seine Idee ist nur für den Künstler selbst nachvollziehbar. ☐
- Es sind objektive, geometrisch-künstlerische Grundlagen. ☐

# Teil F: Individuelle Aktivitäten

**F1. Welche Veranstaltungen haben Sie besucht?**

- Vernissage ☐
- Künstlergespräch ☐
- Finissage ☐

**F2. Wie viele Kontakte hatten Sie im Rahmen der Ausstellung?**

- mit dem Künstler ☐
- mit 2-5 anderen Besuchern, Bekannten oder Freunden ☐
- mit mehr als 5 anderen Personen ☐

**F3. Welche Medien haben Sie zur Interaktion genutzt?**

- Persönliches Gespräch ☐
- Ausstellungskatalog ☐
- Analoger Fragebogen ☐
- Online-Fragebogen ☐
- Instagram / Social Media Wall ☐
- Andere Soziale Medien / Internet ☐
- Bibliothek / Bücherei ☐

**F4. Hat Sie die Ausstellung künstlerisch inspiriert? Haben Sie etwas Neues erfahren? Hinterlassen Sie uns hier eine Anmerkung oder einen Kommentar.**

[                                                                 ]

**F5. Hier können Sie uns eine eigene künstlerische Skizze oder ein Foto hinterlassen.**

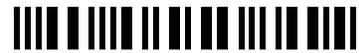

**Wirklichkeiten | Realities . Kunstausstellung | 1. Mai - 30. Juni 2023 Sitzungssaal im Rathaus . Herrsching am Ammersee Vielen Dank für Ihre Teilnahme! Sie können das Browserfenster jetzt schließen.**

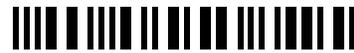

# Teil A: Angaben zur Person

**A1.** **Persönliche Anrede**

- [ ] Herr
- [x] Frau
- [ ] keine Angabe

**A2.** **Wie alt sind Sie?**

- [ ] 0-9 Jahre
- [ ] 10-19 Jahre
- [ ] 20-29 Jahre
- [ ] 30-39 Jahre
- [ ] 40-49 Jahre
- [x] 50+ Jahre

**A3.** **Was ist Ihr höchster Bildungsabschluß?**

- [ ] Grund- / Hauptschule
- [ ] Mittlere Reife
- [ ] Fachabitur / Abitur
- [x] Berufsausbildung / Duale Ausbildung
- [ ] Fachhochschule / Universität
- [ ] Promotion

**A4.** **Wie ist Ihr Bezug zur Kunst?**

- [x] Kunstinteressiert
- [ ] Amateur
- [ ] Professional

**A5.** **Welche Kunstwerke können Sie am Besten verstehen?**

- [x] Landschaftsmalerei und Stilleben
- [ ] Gegenständliche Kunst
- [x] Abstrakte Kunst
- [ ] Konkrete Kunst
- [x] Medienkunst und Performance

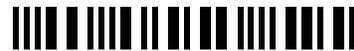

## Teil B: Ort und Situation

**B1.** **Wie viele Personen waren gleichzeitig mit Ihnen im Ausstellungsraum?**

- 0-2 ☐
- 3-5 ☒
- 5-10 ☐
- 10-15 ☐
- mehr als 15 ☐

**B2.** **Zu welcher Tageszeit haben Sie die Ausstellung besucht?**

- morgens ☐
- vormittags ☐
- mittags ☐
- nachmittags ☐
- abends ☒

**B3.** **Wie lange dauerte Ihr Besuch?**

- 0-15 Minuten ☐
- 15-30 Minuten ☐
- 30-45 Minuten ☐
- 45-60 Minuten ☐
- mehr als eine Stunde ☒

## Teil C: Künstler, Kunstwerk und Gesellschaft

**C1.** **Wie sehen Sie die gesellschaftliche Rolle des Künstlers?**

- Künstler lassen uns Dinge in einem anderen Licht sehen. ☒
- Künstler sind für andere Menschen schwer zu verstehen. ☐
- Jeder Mensch ist ein Künstler. ☐

**C2.** **Was ist der Gedanke hinter dem Werk 4´33" von John Cage?**

- Es gibt ein Kunstwerk ohne einen Künstler. ☒
- Das Werk hat keinen Sinn, es ist absurd. ☐
- Absolute Schönheit und Erhabenheit. ☐




\end{document}